\documentclass[]{aa-package/aa} 
\usepackage{amsmath,amsfonts,latexsym,amscd, tabularx}
\usepackage{graphicx}
\usepackage{graphbox}
\usepackage{threeparttablex, longtable, multirow} 
\usepackage{lscape}
\usepackage{array}
\usepackage{rotating} 

\begin{document}

\title{Sampling Molecular Gas in the Helix Planetary Nebula: Variation in HNC/HCN with UV Flux}

\author{
J.~Bublitz \inst{\ref{instGBO}, \ref{instSOPA}, \ref{instLAMA}, \ref{instFR}}
\and J.H. Kastner \inst{\ref{instSOPA}, \ref{instLAMA}, \ref{instCIS}}
\and P. Hily-Blant \inst{\ref{instFR}}
\and T. Forveille \inst{\ref{instFR}}
\and M. Santander-Garc\'{i}a \inst{\ref{instES}}
\and J. Alcolea \inst{\ref{instES}}
\and V. Bujarrabal \inst{\ref{instES}}
}

\institute{
Green Bank Observatory, 155 Observatory Road, P.O. Box 2, Green Bank, WV 24944, USA \label{instGBO}
\and School of Physics and Astronomy, Rochester Institute of Technology, Rochester NY 14623, USA \label{instSOPA}
\and Laboratory for Multiwavelength Astrophysics, Rochester Institute of Technology, USA \label{instLAMA}
\and Center for Imaging Science, Rochester Institute of Technology, USA \label{instCIS}
\and Institut de Plan\'{e}tologie et d'Astrophysique de Grenoble (IPAG) UMR 5274, F-38041, Grenoble, France \label{instFR}
\and Observatorio Astron\'{o}mico Nacional, Alfonso XII, 3, 28014, Madrid, Spain \label{instES}
} 


\abstract
{Observations of molecular clouds, prestellar cores, and protoplanetary disks have established that the HNC/HCN ratio may be a potent diagnostic of molecular gas physical conditions. The processes that govern the relative abundances of these molecules nevertheless remain poorly understood.} 
{We seek to exploit the wide range of UV irradiation strengths within the $\sim$1 pc diameter Helix planetary nebula to explore the potential role of UV radiation in driving HNC/HCN.}
{We performed IRAM 30~m and APEX 12~m radio line observations across six positions within the Helix Nebula, making use of radiative transfer and photodissociation modeling codes to interpret the results for line intensities and line ratios in terms of the molecular gas properties.} 
{We have obtained the first detections of the plasma-embedded Helix molecular knots (globules) in HCN, HNC, HCO$^+$, and other trace molecules. Analysis of the HNC/HCN integrated line intensity ratio reveals an increase with radial distance from the Helix central star. In the context of molecular line ratios of other planetary nebulae from the literature, the HNC/HCN ratio appears to be anticorrelated with UV emission over four orders of magnitude in incident flux. 
Models of the photodissociation regions within the Helix using the RADEX and Meudon codes reveal strong constraints on column density (1.5--2.5$\times$10$^{12}$ cm$^{-2}$) of the molecular gas, as well as pressure and temperature. 
Analysis of the molecular ion HCO$^+$ across the Helix indicates that X-ray irradiation is likely driving HCO$^+$ production in the outer regions of planetary nebulae, where photodissociation is limited, yet cold gas and ionized molecules are abundant.}
{Although the observational results clearly indicate that UV irradiation is important in determining the HNC/HCN ratio, our PDR modeling indicates that the UV flux gradient alone cannot reproduce the observed variation of HNC/HCN across the Helix Nebula. Instead, HNC/HCN appears to be dependent on both UV irradiation and gas pressure and density. }


\maketitle

\section{Introduction}

Planetary nebulae (PNe) form through rapid mass loss during the asymptotic giant branch (AGB) stages of low-to-intermediate mass stars (0.8-8 M$_{\odot}$). As the newly revealed, inert stellar core reaches sufficiently hot temperatures to produce far UV photons ($\ge 13.6$ eV), the ensuing photodissociation and ionization of material ejected during the AGB phase establishes the PN. 
Most of the central stars of PNe (CSPNe) have large UV fluxes as a consequence of their high T$_{eff}$ (often exceeding 100 kK) and large L$_{bol}$ (100--1000 L$_{\odot}$) \citep[e.g.,][]{Frew08}. Many PNe are also luminous X-ray sources, where X-ray photons originate with hot CSPN photospheres, X-ray-bright CSPN companions, and/or energetic wind shocks \citep{Leahy96, Kastner12, Freeman14, Montez15}.

Despite harboring these intense internal sources of high-energy radiation, some PNe retain sizable masses of cold ($\sim$10$^2$~K) molecular gas \citep[e.g.,][]{Huggins89, Huggins05, Schmidt16, Schmidt17}. This UV- and X-irradiated molecular gas and dust emits in the millimeter regime \citep[][and references therein]{Bublitz19}, providing rich radio sources well-suited to improving our understanding of the chemistry within photo-dissociation regions (PDRs).

In this study, we focus on what PNe can teach us about the molecules HCN and HNC, which form primarily through dissociative recombination of HCNH$^+$ with electrons. 
The ongoing investigations of HNC/HCN in OMC-1 make clear the ratio's potential as a temperature probe for cold molecular gas \citep{Schilke92, Graninger14, Hacar19}. 
Studies of dark cloud cores \citep{Hirota98, Nickerson21}, prestellar cores \citep{Padovani11}, star-forming H{\sc ii} regions \citep{Jin15} and galaxies \citep{Canameras21}, the protoplanetary disk orbiting TW Hya \citep{Graninger15, Long21}, and even low-mass protostars and young brown dwarfs \citep{Riaz18} reveal the utility of HNC/HCN ratio measurements across environments ranging from molecular clouds through early stages of stellar evolution. The interplay between the two molecules appears to be governed primarily by the reaction HNC + H $\rightarrow$ HCN + H. At temperatures above $\sim$100~K, this process favors conversion of HNC to HCN, thus decreasing the HNC/HCN ratio as protostellar evolution proceeds \citep{Jin15}. 
Modeling by \cite{Graninger14} suggests that a temperature range of 25--40~K is sufficient to prompt conversion of HNC to HCN, suggesting kinetic-driven reactions, despite the higher barrier temperature predicted by theory.

Selective photodissociation of HNC may also produce a gradient in the HNC/HCN ratio within molecular clouds. 
\cite{Aguado17} predicted that the photodissociation rate of HNC can reach 2--10 times that of HCN in the presence of strong high-energy radiation fields. 
In the case of PNe, photodissociation should be most important in the exposed layers of molecular gas, but may be less effective in the densest gas or outer reaches of a PN. We would then expect to see pre-PN levels of HNC and HCN where the gas is shielded from UV, while HCN abundance rises with respect to HNC at the surface layers of the molecular gas due to selective photodissociation.

A recent survey of molecule-rich PNe \citep{Bublitz19} established that the HNC/HCN ratio is anticorrelated with PN CSPN UV luminosity, indicating that UV irradiation could be driving the conversion of HNC into HCN. 
However, these results suggest that HNC/HCN should also vary within an individual PN due to the gradient of incident UV photons across the molecular gas regions. This provides motivation for followup studies focused on variation in the irradiation of a geometrically well-characterized distribution of molecular gas. The Helix Nebula provides such a powerful testbed for HNC/HCN evolution due to its highly extended molecular envelope, which presents gas parcels exposed to a wide range of UV flux across the nebula.

NGC 7293 is one of the nearest known \citep[$199.6 \pm 0.6$ pc;][]{GaiaEDR3} and most extended (0.46 pc radius) PNe. Its massive molecular envelope has been mapped in mm-wave CO \citep{Young99}, HCO$^+$ \citep{Zeigler13}, and near-IR H$_2$ emission \citep{Kastner96, Hora06}. 
The inner ionized cavity of the nebula also contains many embedded, compact regions of molecular gas and dust that have been identified by a variety of names such as knots, globules, and filaments \citep[e.g.,][]{Huggins92, ODell05, ODell07}. Throughout this paper, such features will be referred to as `globules'. Estimates of the number of globules range from $\sim$3500 to tens of thousands \citep{ODell07, Hora06}. They reside primarily within the outer edges of the ionized core, though are also present along the inner ring of molecular gas \citep{ODell04}. A characteristic mass of $\sim$10$^{-5}$~M$_{\odot}$ has been attributed to individuals in the globule population \citep{Meaburn96, ODell07}, with a typical density of 10$^6$~cm$^{-3}$ \citep{Huggins92, Meaburn96}. A recent analysis of one such globule \cite[identified as C1,][]{Huggins92} uses ALMA maps of CO isotopologues to refine these estimates, finding a molecular mass of $\sim$2$\times10^{-4}$~M$_{\odot}$ \citep{GlobC19}. 
The internal structures of globules, such as their temperature and density gradients, have yet to be established.

Due to the immense span of the the Helix's molecular regions, from the dense globules scattered within 1.5--2.5$'$ (2.7--4.5$\times$10$^{17}$~cm) of the CSPN to the outer molecular plumes at distances of $\sim$7.5$'$ (1.3$\times$10$^{18}$~cm), there exists a significant (factor $\sim$20) gradient in the CSPN radiation incident on the molecular gas within this PN. The Helix central star emits strongly in the UV, with CSPN $T_{eff}=$ 104~kK and a luminosity $L_{bol}$ = 89 $L_\odot$ \citep{Napiwotzki99, Montez15}, as well as a point-like source of X-ray emission \citep{Guerrero01}. These properties make the Helix a prime testbed to study the effects of central star irradiation on PN molecular gas chemistry. 
\cite{Schmidt17} performed multi-position molecular line studies, probing eight regions of the extended Helix envelope with 3~mm observations that yielded detections of HCN and HNC. 
The eight positions they targeted provided good coverage of the Helix's primary molecular ring. However, both studies focus on points within the molecular shell at offsets of 3.7--7.4$'$ from the central star, given the $\sim$1.2$'$ beam size of these measurements, this coverage is insufficient to establish whether HNC/HCN varies systematically across the PN.

In this work, we present a targeted molecular line study of the Helix Nebula using the Institut de Radioastronomie Millim\'{e}trique (IRAM) 30 m telescope and Atacama Large Pathfinder Experiment (APEX) 12~m telescope. The primary goal is to further investigate the apparent anticorrelation between the HNC/HCN line ratio and UV radiation in PNe by measuring the ratio with respect to distance from the Helix central star.
In Section \ref{sect_observations}, we describe the single-dish molecular line observations. In Section \ref{sect_results}, we describe the line detection and mapping results. In Section \ref{sect_analysis} we present our analysis of these results, while in Section \ref{sect_models}, radiative transfer and photodissociation region modeling are discussed. Conclusions from this work are presented in Section \ref{sect_conclusions}.

	\begin{figure*}
		\centering
		\includegraphics[trim={5.9cm 2.5cm 6.5cm 2cm}, clip, height=0.38\linewidth, angle=90]{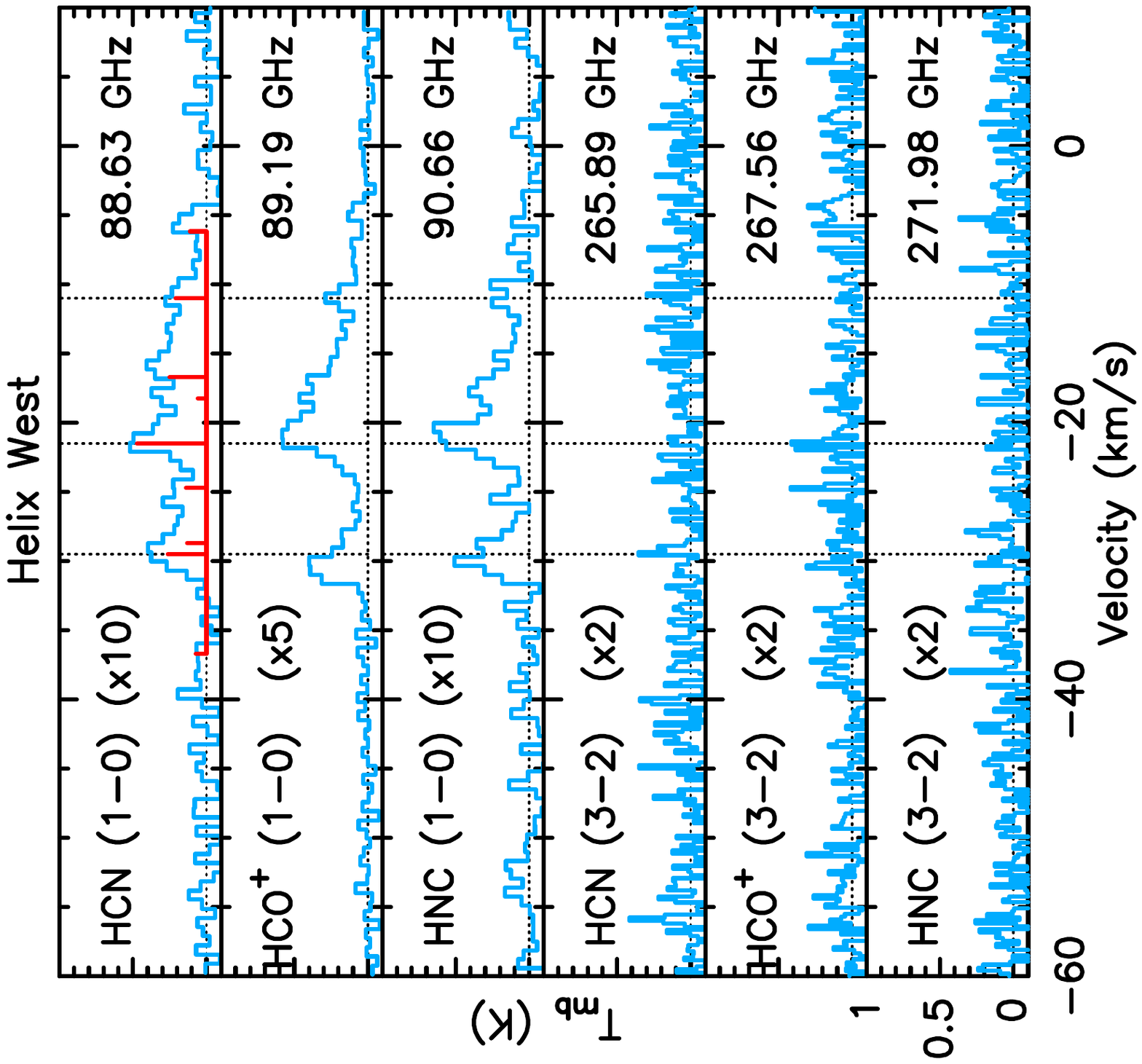}
		\includegraphics[trim={5.9cm 2.5cm 6.5cm 2cm}, clip, height=0.38\linewidth, angle=90]{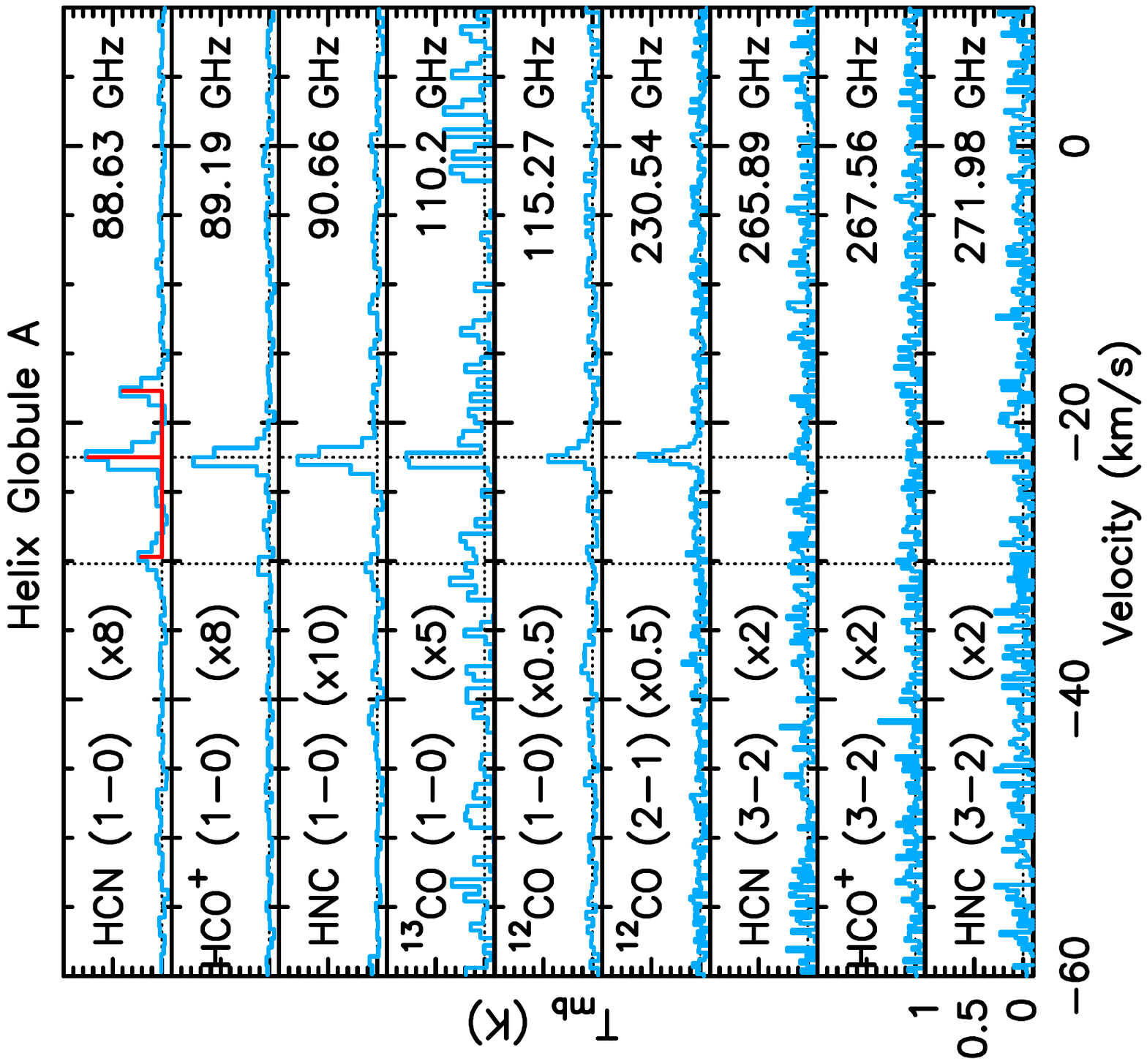} \\
		\includegraphics[trim={3cm 4cm 6.5cm 2cm}, clip, height=0.35\linewidth, angle=90, align=c]{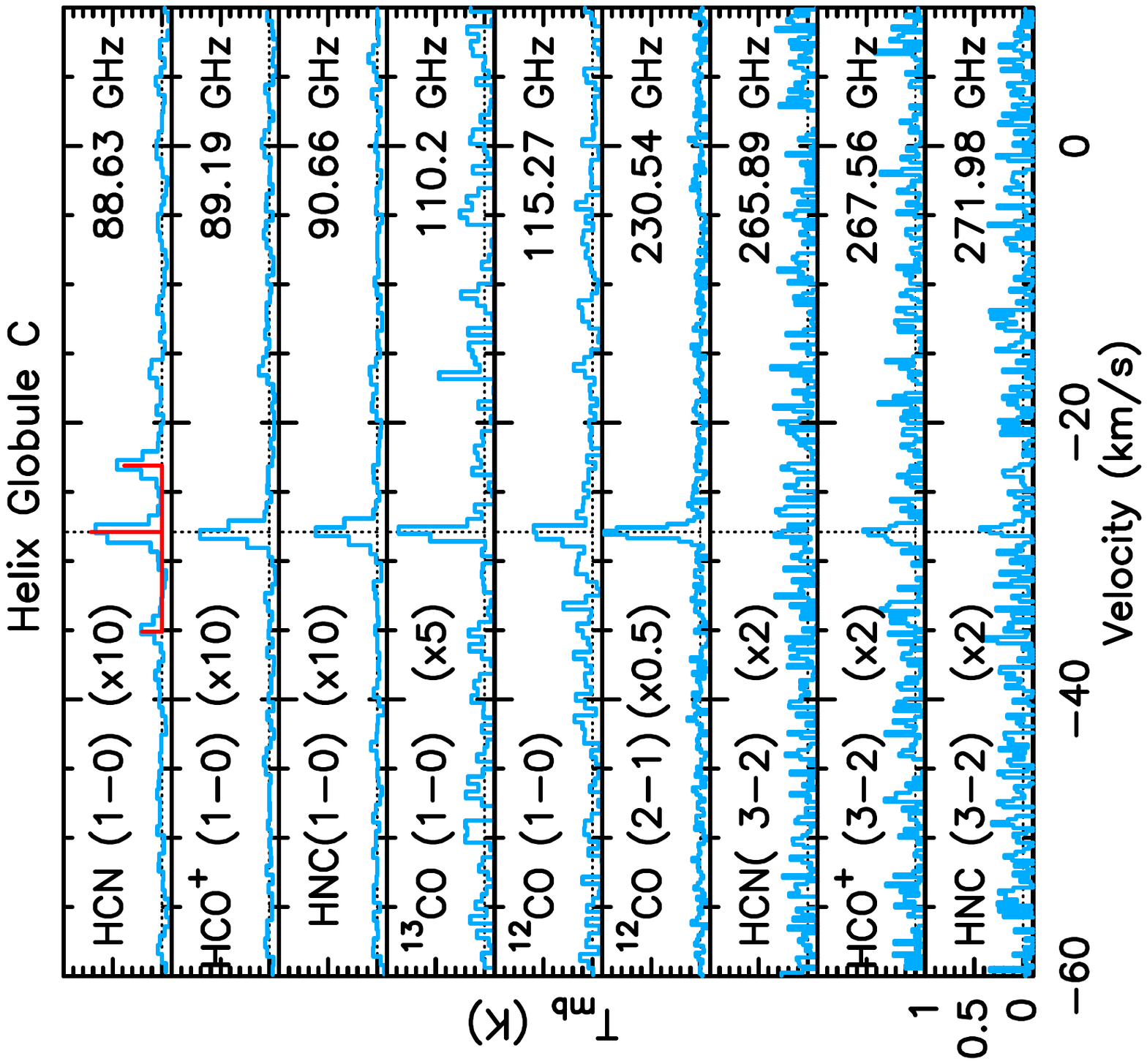}
		\includegraphics[trim={8cm 15cm 8cm 15cm}, clip, height=0.25\linewidth, angle=90]{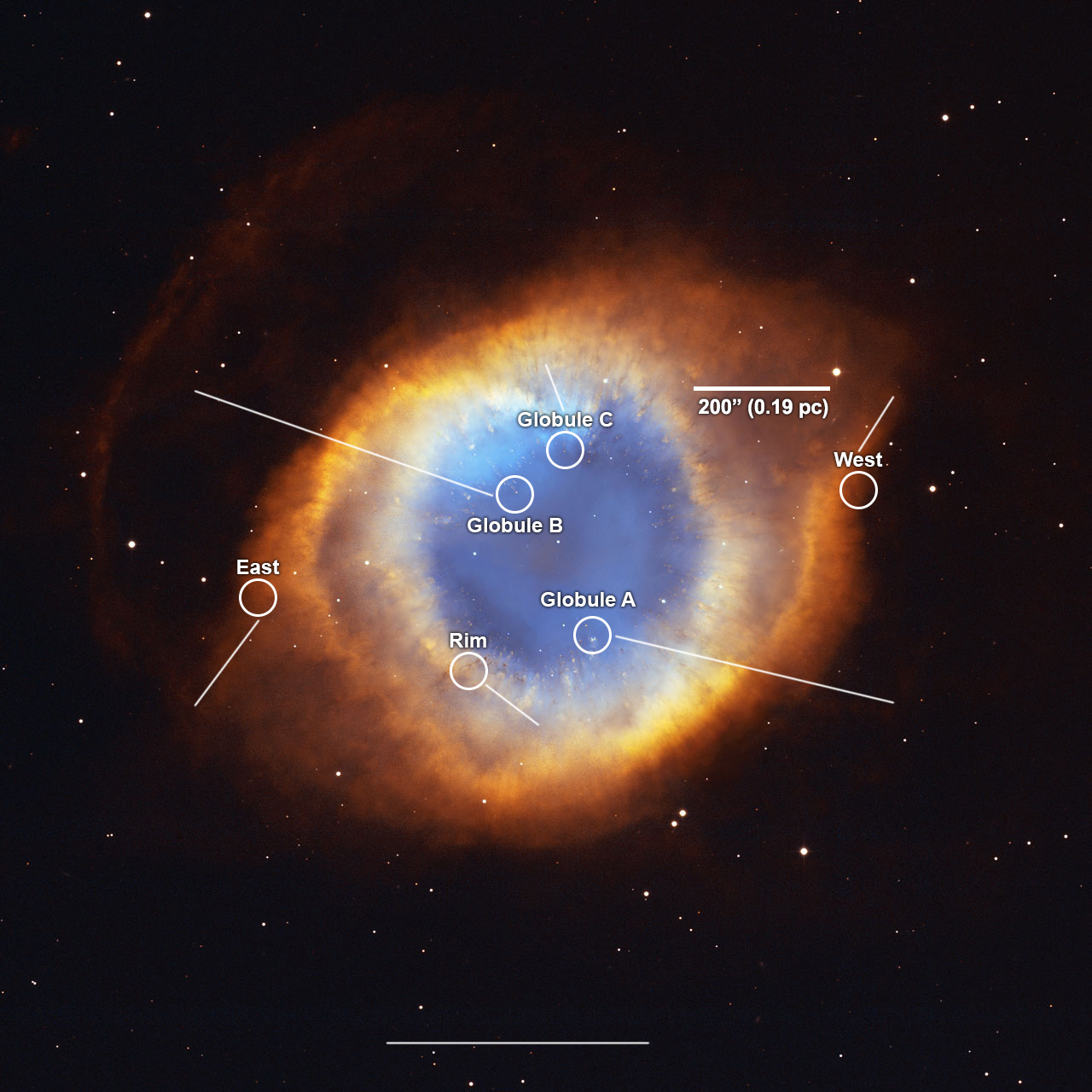}
		\includegraphics[trim={3cm 2.5cm 6.5cm 2cm}, clip, height=0.38\linewidth, angle=90, align=c]{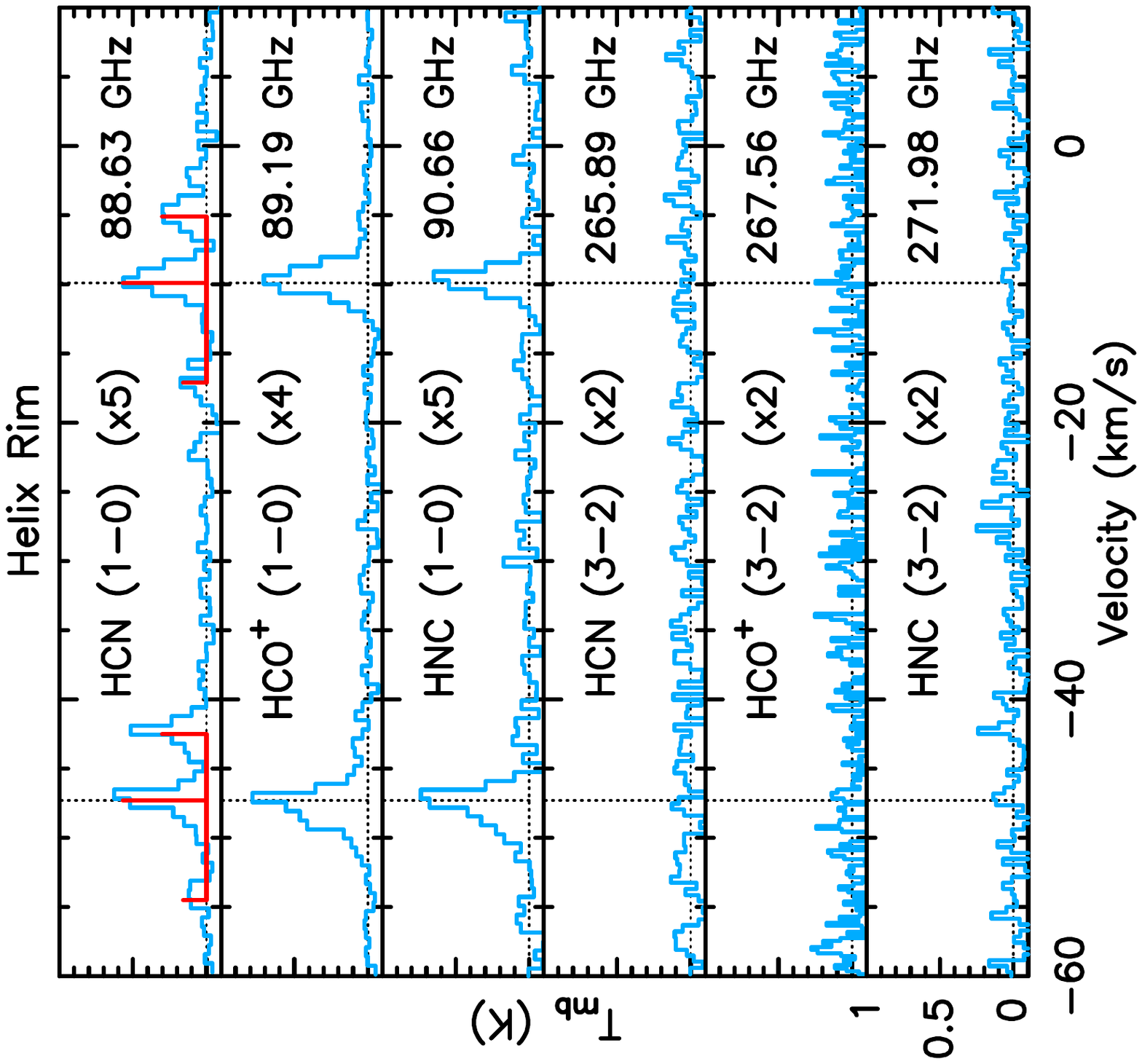} \\
		\includegraphics[trim={5.9cm 2.5cm 6.5cm 2cm}, clip, height=0.38\linewidth, angle=90]{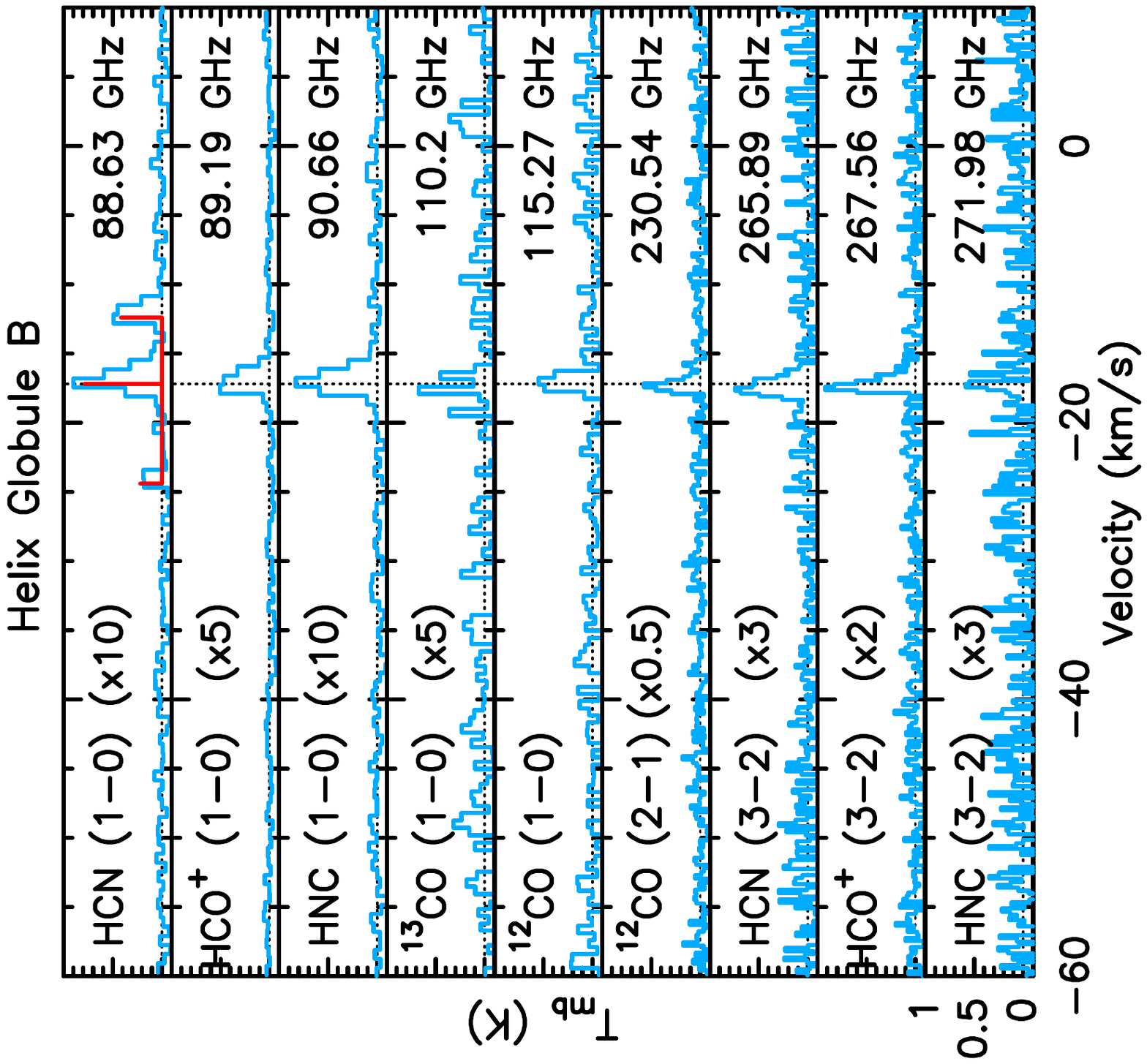}
		\includegraphics[trim={5.9cm 2.5cm 6.5cm 2cm}, clip, height=0.38\linewidth, angle=90]{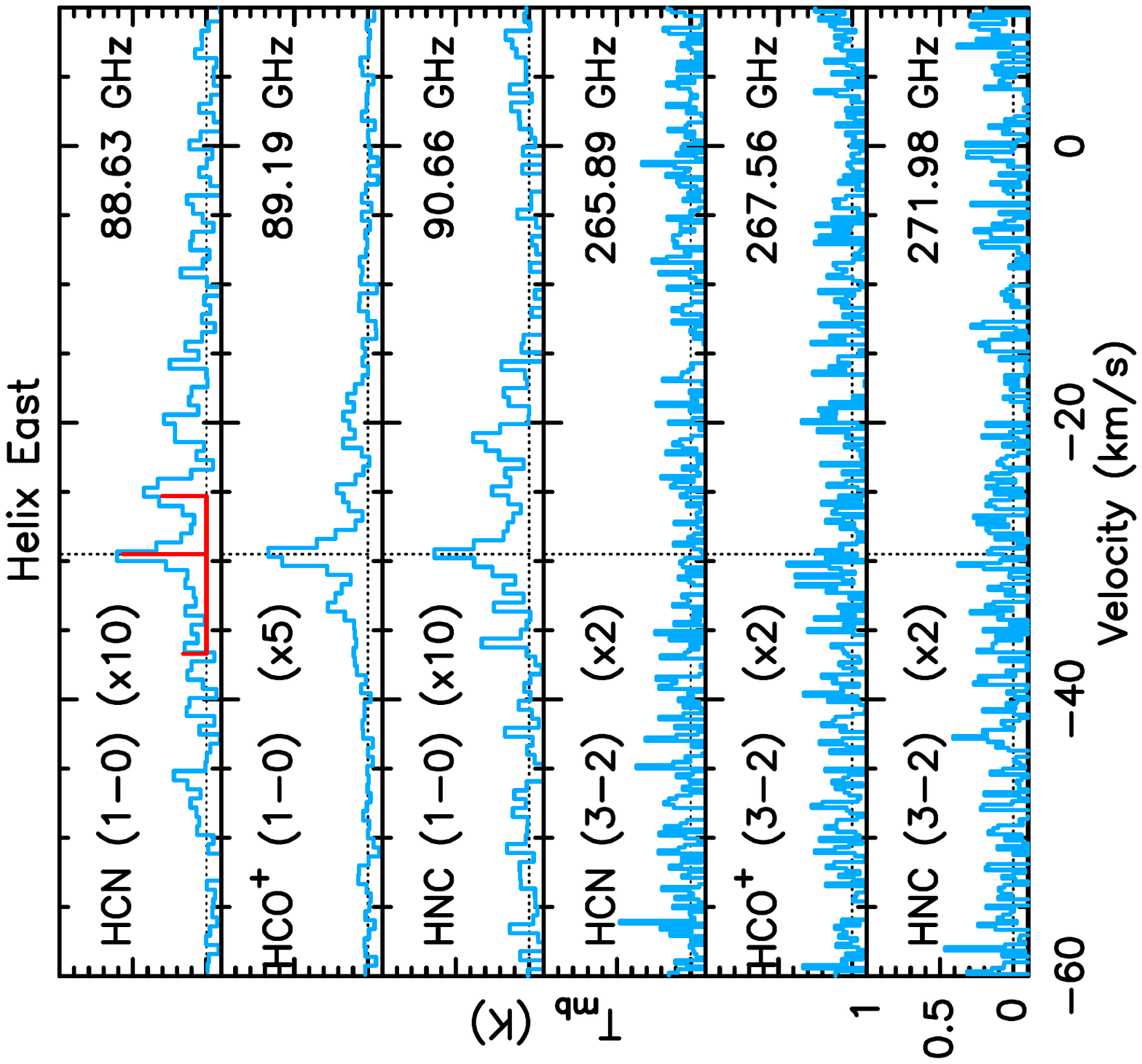}
		
		\caption{Spectra for transitions of molecules detected in the six Helix positions using IRAM 30m. Individual spectra have been scaled by a multiplier indicated for each line to facilitate comparison. The x-axis indicates velocity with respect to the local standard of rest (V$_{LSR}$) and the y-axis is the antenna main-beam temperature (K). Vertical dotted lines mark the systemic velocity of individual emission features with respect to the entire nebula. Hyperfine structure lines are indicated in red. The two separate velocity features identified towards the Globule A, Helix Rim, and West positions represent emission from nearer and more distant clumps of gas. Positions are indicated on an HST optical image of the Helix (NASA, NOAO, ESA, the Hubble Helix Nebula Team, M. Meixner (STScI), and T.A. Rector (NRAO)), with the HCN ($J=1\rightarrow0$) beamsize (27.6$''$) and positions of our line survey observations as circles (globules A, B, C -- from Huggins et al. 1992 \citep{Huggins92} -- molecular Rim, East, and West positions).}
		\label{Helix_SpectraMap}
	\end{figure*}

\begin{table}
	\caption{Observed Positions}
	\label{Positions} \label{Table_Positions}
	\centering
	\resizebox{\linewidth}{!}{%
	\begin{tabular}{l c c c}
\hline\hline
Position & $\alpha$ (J2000.0) & $\delta$ (J2000.0) & N, E Offset ($''$)\\
\hline
Globule A	&	22$^h$29$^m$35.0$^s$ & -20\degr52$'$30.0$''$ & (-136.4, -49.6)\\
Globule B	&	22$^h$29$^m$43.0$^s$ & -20\degr49$'$02.6$''$ & (+71.0, +62.5)\\
Globule C	&	22$^h$29$^m$37.8$^s$ & -20\degr47$'$58.0$''$ & (+135.6, -10.4)\\
Helix Rim	&	22$^h$29$^m$47.9$^s$ & -20\degr53$'$18.0$''$ & (-184.4, +131.2)\\
Helix East	&	22$^h$30$^m$09.6$^s$ & -20\degr51$'$29.0$''$ & (-75.4, +435.4)\\
Helix West &	22$^h$29$^m$07.5$^s$ & -20\degr48$'$59.0$''$ & (+74.6, -435.2)\\
\hline
	\end{tabular}
	}
	\tablefoot{RA and dec of the observed regions in the Helix, as well as their offsets from the Helix CSPN in arcseconds. For reference, the central star is located at 22$^h$29$^m$38.5$^s$, -20\degr50$'$13.6$''$.
}
\end{table}

\section{Observations} \label{sect_observations}
\subsection{IRAM 30m}
Six positions in the Helix Nebula were observed in 2018 February and May with the IRAM 30m telescope on Pico Veleta. The targeted positions were selected to reflect a range in distances from the CSPN (Figure \ref{Helix_SpectraMap} and Table \ref{Positions}). Three globule structures (hereafter Globules A, B, and C), previously the focus of IRAM 30~m CO observations \citep{Huggins92}, were targeted to exploit their proximity to the CSPN. The position labeled ``Helix West'' was the subject of the previously mentioned PN molecular line survey \citep{Bublitz19}. A position directly opposite the CSPN, labeled ``Helix East'', was chosen as a comparison for molecular and UV properties in the extended molecular envelope. The final ``Helix Rim'' position was selected from among the positions observed by \cite{Schmidt17}, so as to sample an intermediate distance from the CSPN.

Data were obtained with the Eight MIxer Receiver (EMIR) and Fast Fourier Transform Spectrometer (FTS) backend, which allowed for observing in both the 3~mm and 1~mm regimes simultaneously, with 8 GHz bandwidth per sideband and velocity resolutions of 0.66 and 0.22 km s$^{-1}$, respectively. Two frequency setups were devised, with central frequencies of 89 and 267 GHz for simultaneous acquisition of the $J=1\rightarrow0$ and $J=3\rightarrow2$ lines of HCN, HCO$^+$, and HNC, as well as the central frequencies of 111 and 227 GHz for lines of CO and neighboring molecular transitions.

Weather was typically clear with calm winds. Strong winds resulted in loss of observing time for two of the nine observing days allocated. Another day was lost to excessive snow and wind, while a fourth was lost to an electrical fault.

Globules A, B, and C were also targeted with 5-point maps, using observational offsets of 10$"$ in $J=1\rightarrow0$ $^{13}$CO, $^{12}$CO, and $J=2\rightarrow1$ $^{12}$CO lines (Figures \ref{Glob_5ptA}--\ref{Glob_5ptC}). The intent was to roughly ascertain the spatial extension of the globules with respect to the 30~m beam (Section \ref{5PointMaps}).
Calibration of data used the chopper wheel method with scaling in units of main-beam temperature (T$_{mb}$). 
Worst-case flux calibration uncertainties from ripples in the receiver (by coupling to hot and cold calibration loads) are given at 2.5\% \citep{Carter12}, however local fluctuation around individual molecular transitions contain stable baselines.

The beam sizes (half power beam widths) are 27.8$''$ and 9.25$''$, with beam efficiencies of 0.77 and 0.45, respectively, at the 88.6 GHz and 265.9 GHz frequencies of HCN $J=1\rightarrow0$ and $J=3\rightarrow2$. 
Venus and Mars were used to check telescope pointing and focus every hour. Pointing errors were corrected to an accuracy of $\sim$1.5$''$. Observations were made in position-switching mode. Offset positions that display no detected CO emission were chosen from \cite{Young99}: (+28.6$''$, +35.0$''$) from the CSPN for Globules B and C, ($-24.81''$, $-67.41''$) from the CSPN for Globule A and Rim position, and two positions well outside the molecular envelope (400$''$ east and 250$''$ west of positions East and West, respectively).

Integration times varied by globule and by molecular transition, as integrations for the receiver setting covering the $J=1\rightarrow0$ transitions of HCN, HCO$^+$, and HNC ranged from 2.8 hours (Rim) to 10.4 hours (Globule C), for a total of 38 hours across the six positions. HCN and HNC (3-2) transitions were observed for $\sim$1 hour at all positions. Select observations of the CO isotopologues were made during the followup observing run of the globules with integration times of $\sim$20 minutes to yield similar noise constraints.

\subsection{APEX}

\begin{figure*}
		\includegraphics[width=0.32\linewidth, trim={4.5cm 2.5cm 6cm 2cm}, clip]{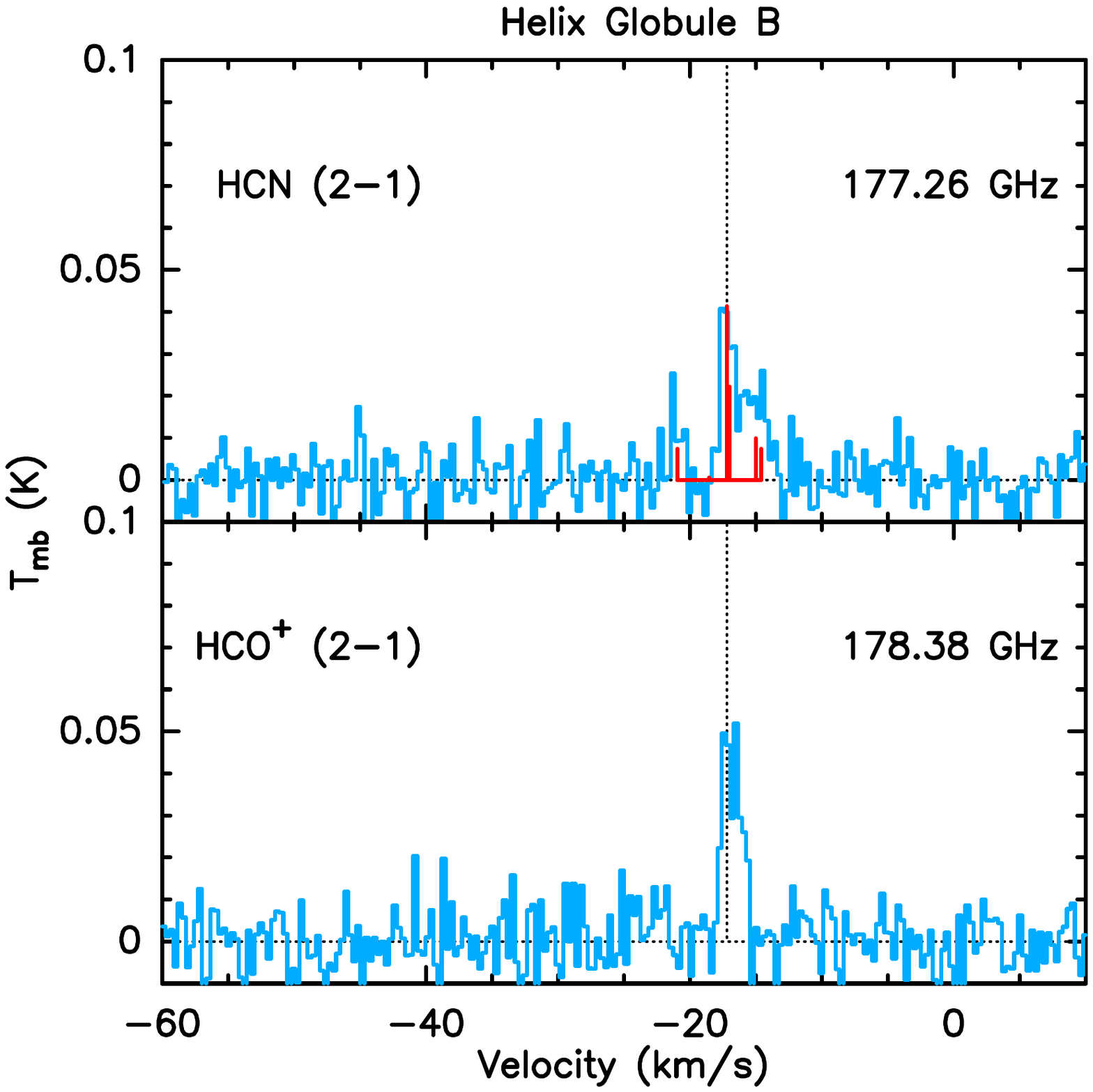}
		\includegraphics[width=0.32\linewidth, trim={4.5cm 2.5cm 6cm 2cm}, clip]{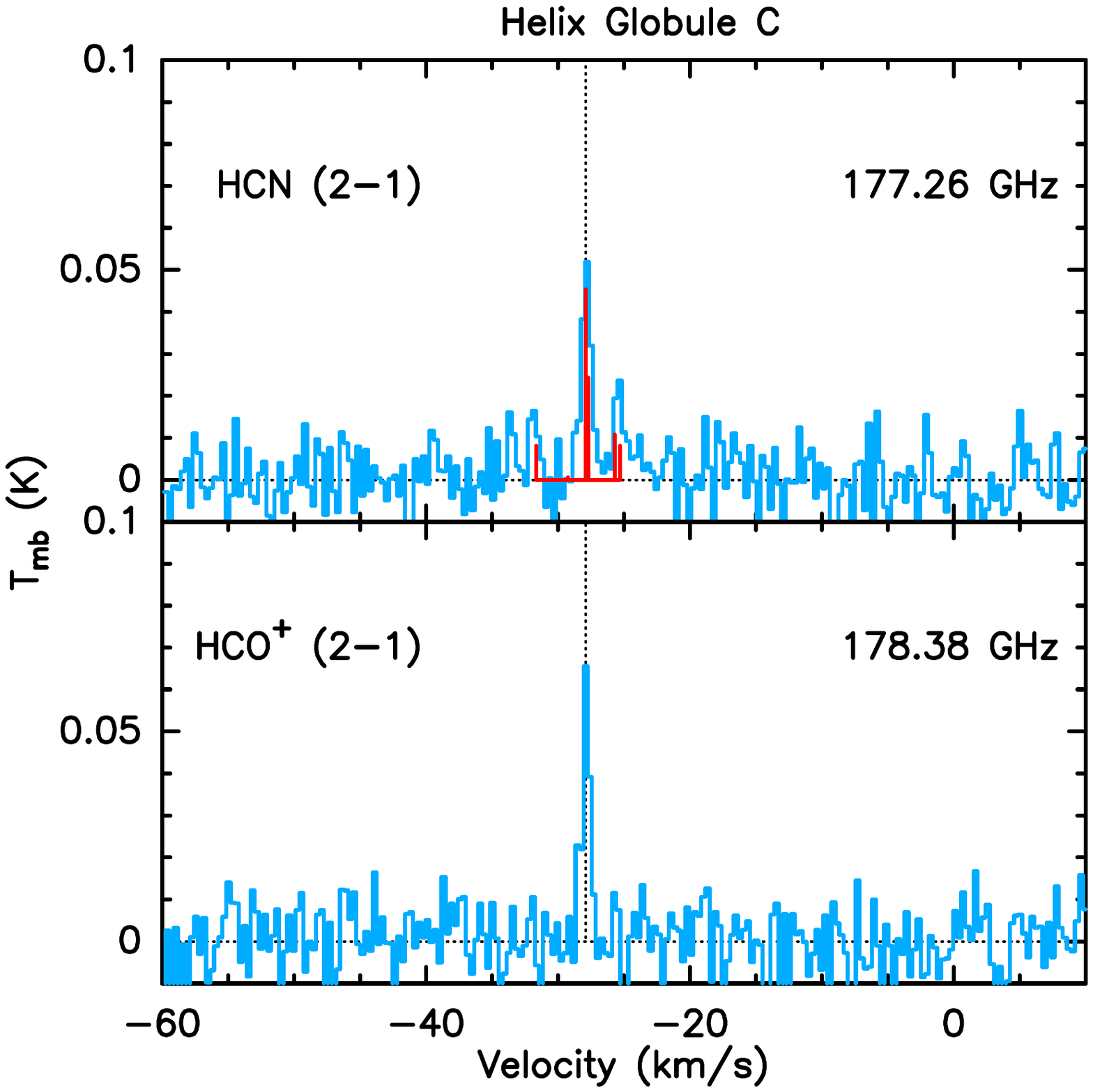}
		\includegraphics[width=0.32\linewidth, trim={4.5cm 2.5cm 6cm 2cm}, clip]{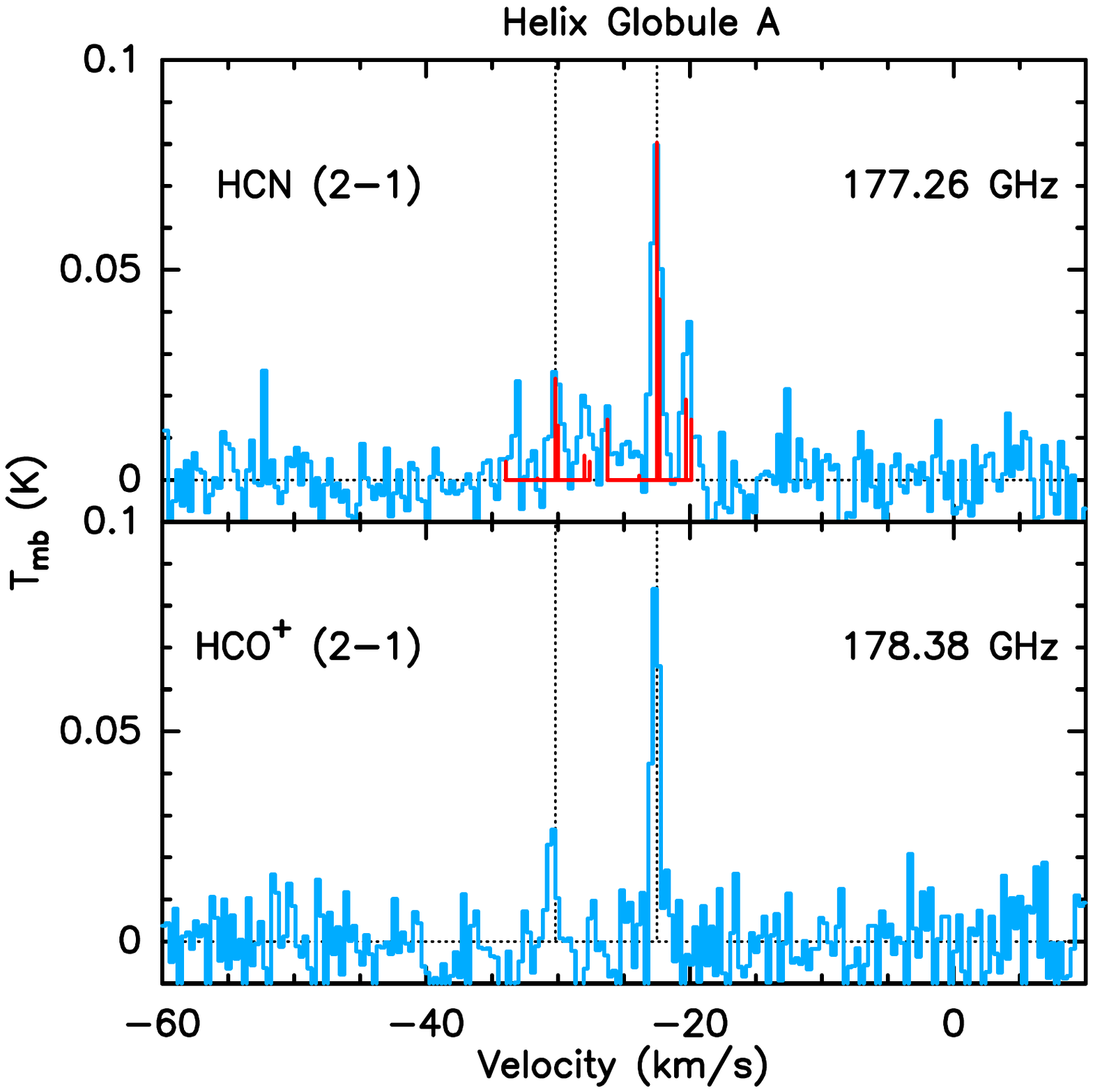}\\
		\includegraphics[width=0.32\linewidth, trim={4.5cm 2.5cm 6cm 2cm}, clip]{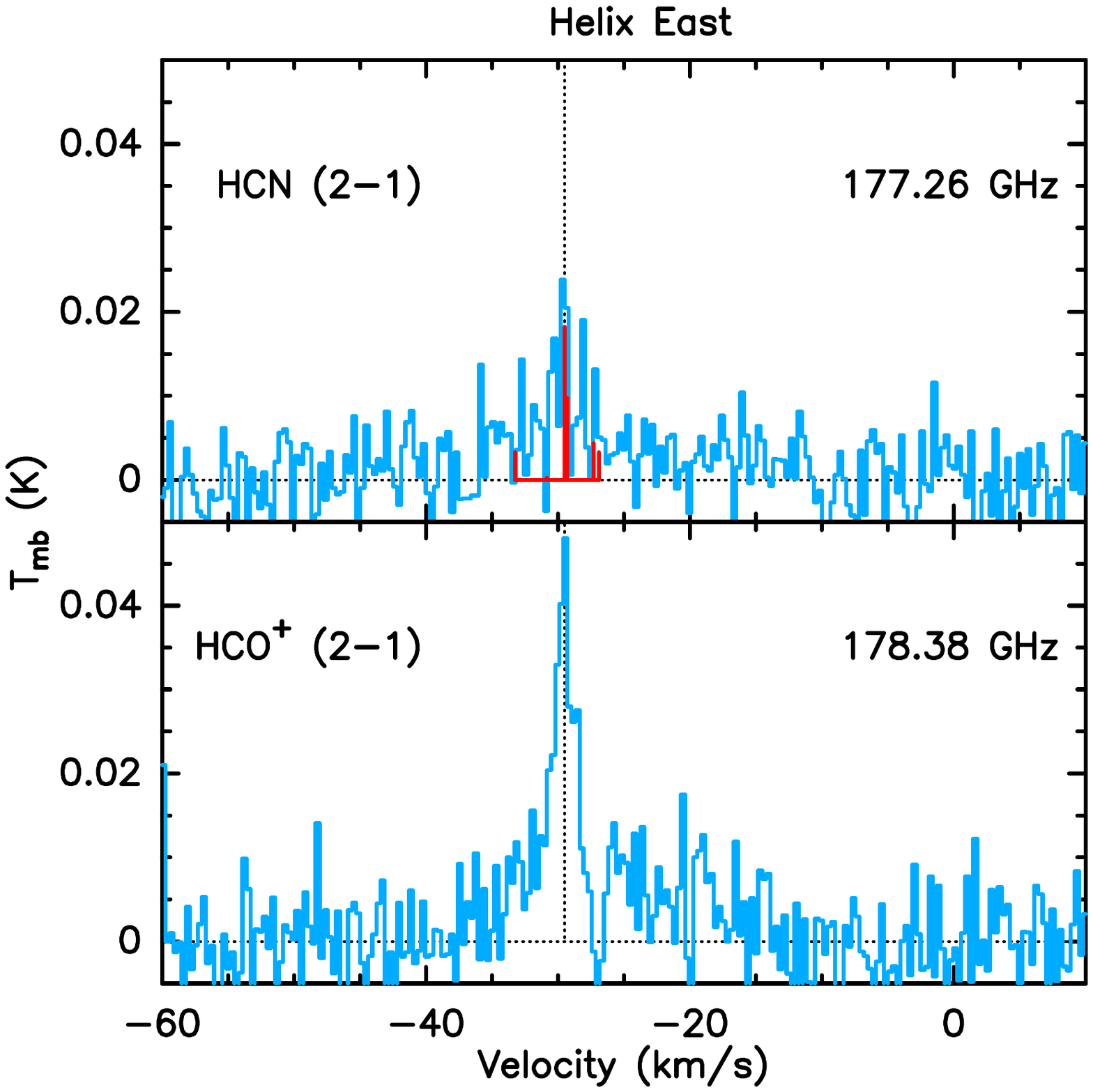}
		\includegraphics[width=0.32\linewidth, trim={4.5cm 2.5cm 6cm 2cm}, clip]{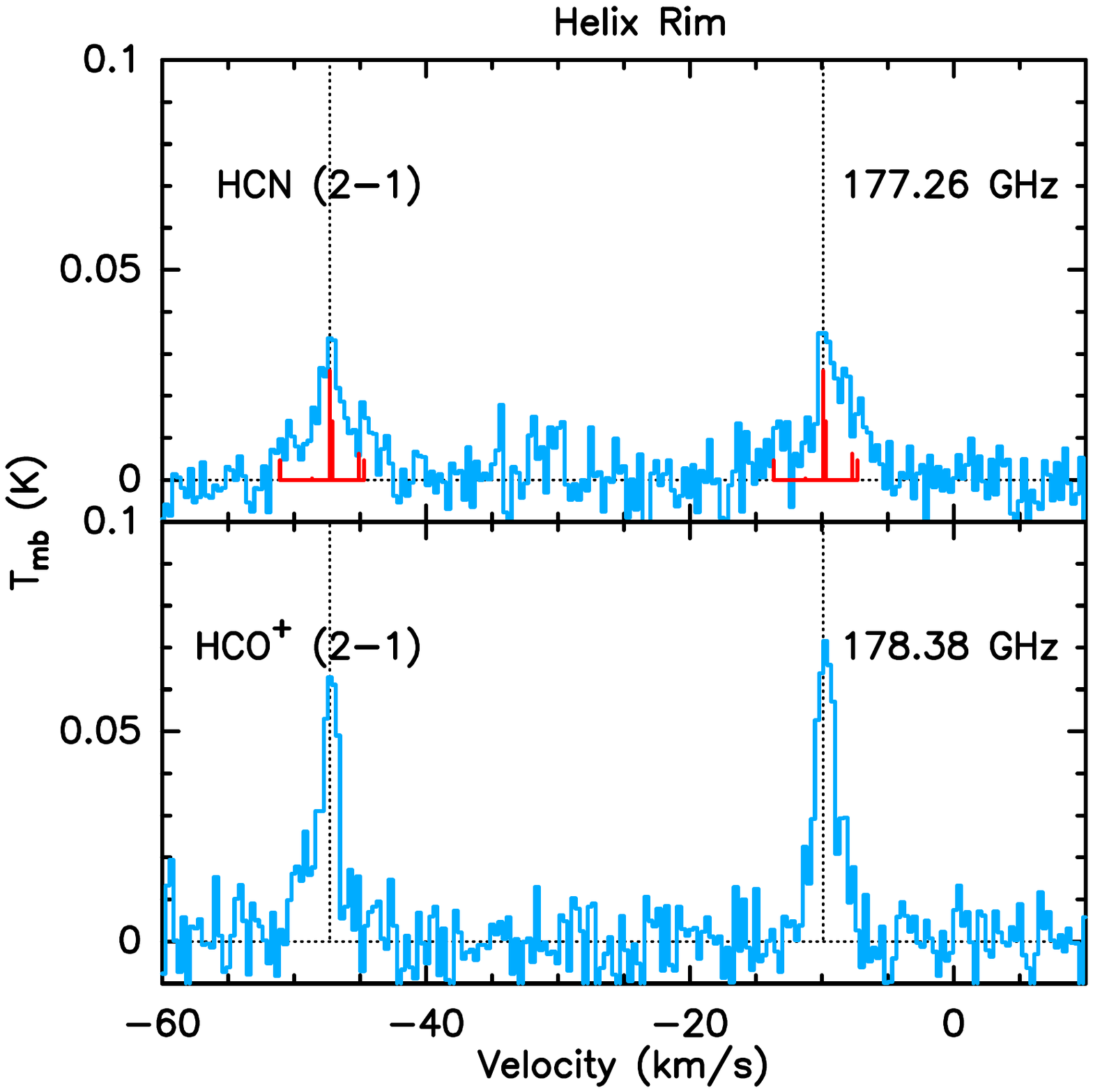}
		\includegraphics[width=0.32\linewidth, trim={4.5cm 2.5cm 6cm 2cm}, clip]{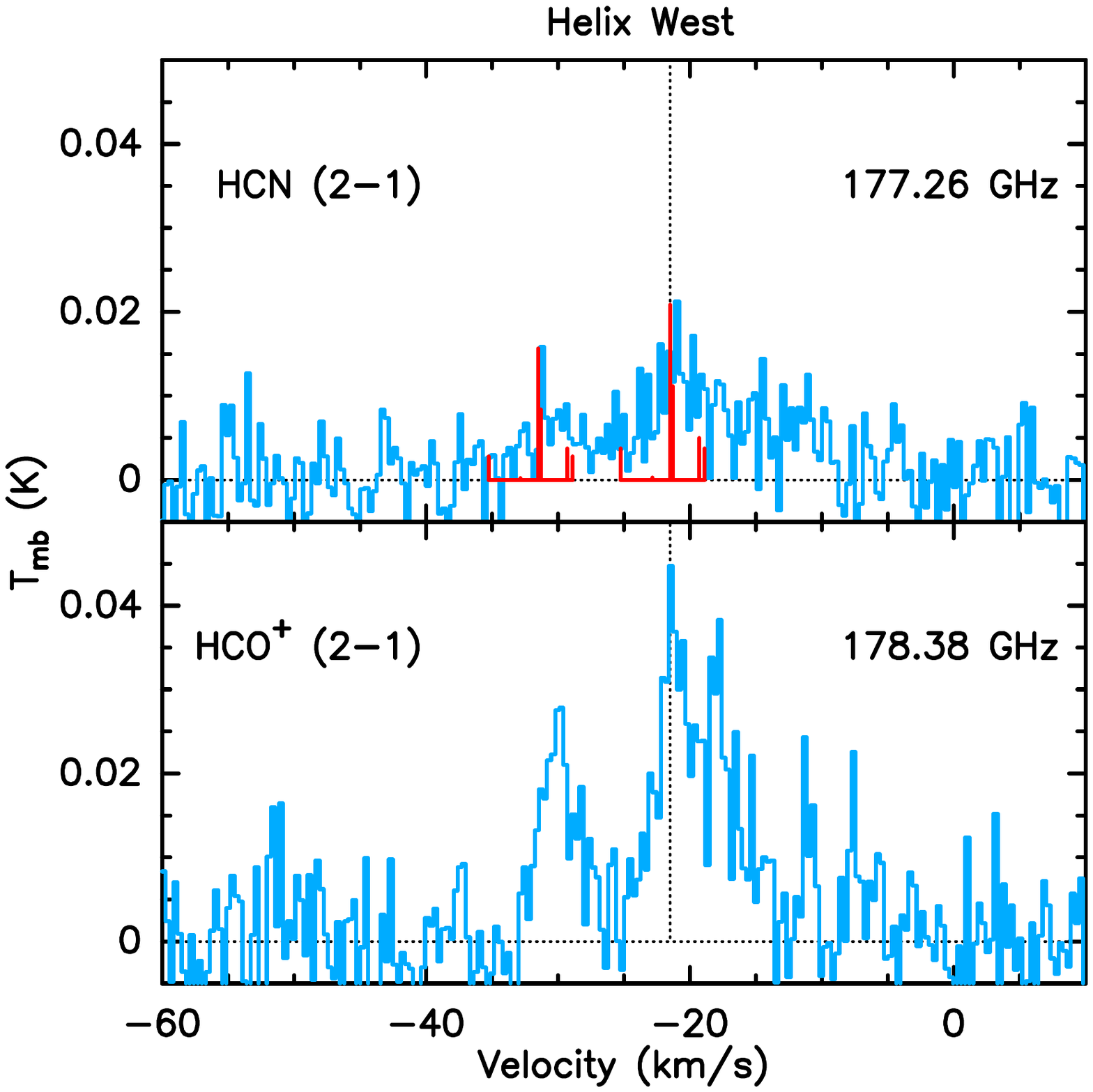}
	\caption{Spectra for $J=2\rightarrow1$ transitions of HCN and HCO$^+$ detected in the six Table \ref{Table_Positions} positions with the APEX 12m Telescope. X- and Y-axes are same as Figure \ref{Helix_SpectraMap}. Hyperfine lines of HCN --- where identified --- have been indicated in red.}
		\label{Fig_APEX}
	\end{figure*}

The six Table \ref{Table_Positions} positions in the Helix Nebula were also targeted with APEX in 2019 August using the SEPIA180 receiver in Band 5 (Figure \ref{Fig_APEX}). A baseband setup at 177.8 GHz was used, with resolution of 10 MHz and a velocity resolution of $\delta$v = 0.3 km s$^{-1}$. The backend consisted of the eXtended Fast Fourier Transform Spectrometer (XFFTS), whose bandwidth of 4 GHz allowed simultaneous observation of the $J=2\rightarrow1$ lines of HCN (177.26 GHz) and HCO$^+$ (178.38 GHz). Due to its close proximity to a 183 GHz telluric water line, the HNC $J=2\rightarrow1$ line at 181.33 GHz was not measurable. Data collection was performed in total power/position switching mode.

Beam width for this frequency range is 35$''$, with a main beam efficiency of 0.86. The brightness temperature and pointing calibrators were EP Aqr, RAqr, SGru, and $\pi^1$Gru. Focusing was performed on Jupiter before observations of each position. Integration time on-source ranged from 30-60 minutes per position, for a total observing time of 15.6 hours.

\subsection{Data Reduction}
Reduction of the IRAM 30~m and APEX 12~m spectra was performed using the Continuum and Line Analysis Single-dish Software (CLASS)\footnote{https://www.iram.fr/IRAMFR/GILDAS/}. A more detailed account of methods for data reduction used in this work are described in \cite{Bublitz19}. In brief, data reduction consisted of scan averaging, rescaling the spectra to main beam brightness temperature ($T_{mb}$), and baseline subtraction of a low-order polynomial.

\section{Results} \label{sect_results}

\subsection{Line Detections and Properties}
\label{LineDetections}

In total, seven molecular species across multiple transitions were identified throughout the six positions targeted. 
We have obtained detections of HCN, HCO$^+$, and HNC for the first time in the Helix globules, as well as detections of these molecules in the three PN envelope positions, with the IRAM 30~m telescope. The $J = 2 \rightarrow1$ transitions of HCN and HCO$^+$ were also detected with APEX across all six positions. The $J = 3 \rightarrow2$ transitions of HCN and HNC were only detected by the 30~m telescope toward Globule B, with upper limits obtained in the other regions. The $J = 3 \rightarrow2$ line of HCO$^+$ was also detected from Globules B and C, in addition to the Helix East and West positions.

Spectral maps in the $J = 1 \rightarrow0$ transitions have identified multiple emission line features in the Globule A, Rim, and West positions, indicative of parcels of emitting gas spanning a range of velocities. 
In Globule A, a secondary feature is observed at -30.2 km s$^{-1}$, while two features are identified at -29.5 and -11 km s$^{-1}$ towards the extended West envelope region, in addition to the dominant emission lines detailed in Table \ref{table_longtable}. 
These positions feature overlapping hyperfine structure lines in the molecule HCN. With insufficient information regarding the opacity and local thermodynamic equilibrium status of the regions, the hyperfine lines are not trivially disentangled and, as such, the analysis and discussion presented here is restricted to the brightest components. 
Towards the Helix Rim position, however, two velocity components are fully resolved, providing us with an additional data point, which translates into two different displacements from the CSPN (discussed further in Section \ref{FluxDetermination}). At -47.3 and -9.9 km s$^{-1}$, these features are hereafter referred to as `Outer Rim' and `Inner Rim', respectively, in order to differentiate their displacements from the CSPN.

Measured line intensities range from 20~mK in weak HCN hyperfine structure components to 2.09~K in CO towards Globule C. The corresponding FWHM line widths varied from 0.6-3.3 km~s$^{-1}$ for well-resolved lines, to upwards of 5.7 km~s$^{-1}$ for blended lines that required multi-gaussian fitting. 
For the 30~m observations, the average S/N for HCN $J=1\rightarrow0$ integrated line intensities in the globules is 23.0, while a S/N ratio of 5.6 is obtained for the molecular envelope positions. The resulting S/N ratios across the positions observed with the APEX Telescope are 6.8--10.9.

Properties of detected lines were measured with the Gaussian curve fitting tool available in CLASS, with individual fitting of all hyperfine lines when resolved (see Figure \ref{Fig_APEX}). 
Results of the spectral line analysis can be found in Tables \ref{table_longtable} and \ref{table_bonuslines}, with the individual spectra presented in Figures \ref{Helix_SpectraMap} and \ref{Fig_APEX} for lines discussed in this work, and \ref{BonusSpectra} for additional lines by source.

Our observations detected the hyperfine splitting of the rotational energy levels for several molecules. 
The hyperfine complex of HCN $J = 1 \rightarrow0$ was fully resolved, except in the case of the East and West positions where emission from multiple regions along the molecular envelope line-of-sight has blended and obscured any individual line components. This is consistent with observations of the Helix West position in \cite{Bublitz19}. The observed $J = 1 \rightarrow0$ hyperfine complex of HCN is consistent with the expected relative intensity levels of 3:5:1 for the $F = 0\rightarrow1, 2\rightarrow1, 1\rightarrow1$ transitions between quantum levels. For ratios of the integrated intensities of HCN hyperfine components I($F = 0\rightarrow1/2\rightarrow1$)=R$_{02}$ and I($F = 1\rightarrow1/2\rightarrow1$)=R$_{12}$, LTE gas will follow the theoretical intensity ratios of 0.2$<R_{02}<$1.0 and 0.6$<R_{12}<$1.0 for the case of optically thin gas \citep[e.g.,][]{Gottlieb75, Loughnane12}. Across the observed positions in the Helix Nebula with resolved hyperfine lines, we measured the ratios in the ranges R$_{02}$=0.25--0.40 and R$_{12}$=0.53--0.75. The similarity of hyperfine ratios in the Helix globules and Rim positions, within uncertainties of 4\%, indicate that the Helix is optically thin in HCN. 
While HNC $J = 1 \rightarrow0$ also exhibits hyperfine splitting, the line frequencies vary only by $\sim$0.1~MHz or $<$0.5 km~s$^{-1}$. Thus, the individual emission lines of HNC could not be resolved in these observations. 
Hyperfine lines of C$_2$H and CN were detected in the globules for the first time, with C$_2$H detected in the Rim position \citep[also previously detected by][]{Schmidt18} and marginal detections in Helix East (Table \ref{table_bonuslines}, Figure \ref{BonusSpectra}). 

Due to the complex nature of the line profiles, the results for the Helix East and West positions listed in Table \ref{table_longtable} only include those for the central peak gaussian fit. 
The integrated line intensity used for the analysis described in Section \ref{FluxDetermination} and Section \ref{sect_models}, however, includes the summation of all resolved velocity components. 

\begin{table*}
	\caption{Molecular Transitions Detected}
	\label{table_frequencies}
	\centering
	\begin{tabular}{l l l l r r}
\hline\hline
Molecule & \multicolumn{3}{c}{Resolved Quantum Numbers} & $\nu$ (GHz) & Rel. Intensity\\
\hline
CCH		& N=1$\rightarrow$0 & J=$\frac{3}{2}$$\rightarrow$$\frac{1}{2}$ & F=2$\rightarrow$1 & 87.317 & \\
		& 				& J=$\frac{1}{2}$$\rightarrow$$\frac{1}{2}$ & F=1$\rightarrow$1 & 87.402 & \\
HCN		&				& J=1$\rightarrow$0					& F=0$\rightarrow$1	& 88.630 & 0.3333 \\
		&  &				 								& F=2$\rightarrow$1	& 88.632 & 0.5555 \\
		&  &												& F=1$\rightarrow$1	& 88.634 & 0.1111 \\
HCO$^+$	&				& J=1$\rightarrow$0					&				& 89.189 & \\
HNC$^a$	&				&J=1$\rightarrow$0					&				& 90.664 & \\
$^{13}$CO &				& J=1$\rightarrow$0					&				& 110.201 & \\
CN		& N=1$\rightarrow$0 & J=$\frac{1}{2}$$\rightarrow$$\frac{1}{2}$ & F=$\frac{3}{2}$$\rightarrow$$\frac{1}{2}$ & 113.170 & \\
		& 				 & J=$\frac{1}{2}$$\rightarrow$$\frac{1}{2}$ & F=$\frac{3}{2}$$\rightarrow$$\frac{3}{2}$ & 113.191 & \\
		& 				 & J=$\frac{3}{2}$$\rightarrow$$\frac{1}{2}$ & F=$\frac{5}{2}$$\rightarrow$$\frac{3}{2}$ & 113.491 & \\
$^{12}$CO &				& J=1$\rightarrow$0					&				& 115.271 & \\
HCN		&				& J=2$\rightarrow$1					& F=2$\rightarrow$2	& 177.260 & 0.0833\\
		&				&				 				& F=1$\rightarrow$0	& 177.260 & 0.1111\\
		&				&								& F=2$\rightarrow$1	& 177.261 & 0.2500\\
		&				&								& F=3$\rightarrow$2	& 177.261 &0.4667 \\
		&				&								& F=1$\rightarrow$2	& 177.262 & 0.0056\\
		&				&				 				& F=1$\rightarrow$1	& 177.263 & 0.0833\\
HCO$^+$	&				& J=2$\rightarrow$1					& 				& 178.375 & \\
HNC		&				& J=2$\rightarrow$1					&				& 181.325 & \\
$^{12}$CO &				& J=2$\rightarrow$1					&				& 230.538 & \\
HCN		&				& J=3$\rightarrow$2					&				& 265.886 & \\
HCO$^+$	&				& J=3$\rightarrow$2					&				& 267.558 & \\
HNC		&				& J=3$\rightarrow$2					&				& 271.981 & \\
\hline
	\end{tabular}
\tablefoot{
		Line frequencies and relative intensities were measured in the laboratory and published in \cite{Muller01} and \cite{Muller05}. Frequencies of HCN transitions (and their hyperfine components) were established by \cite{Mullins16}. \\
		\tablefoottext{a}{The hyperfine structure of HNC (J=1$\rightarrow$0) is unresolved in these observations.}
}
\end{table*}

\subsection{Mapping the Globules in CO}
\label{5PointMaps}

Spectra obtained for the 5-point maps targeting the globules are presented in the Appendix for the $J=1 \rightarrow 0$ lines of $^{13}$CO and $^{12}$CO, as well as $J=2 \rightarrow 1$ for $^{12}$CO (Figures \ref{Glob_5ptA}--\ref{Glob_5ptC}). 
For Globule A, the western position ($-$10, 0) presents a weaker emission intensity at 29--47\% of its globule core value, which suggests a lack of centering on the emission source, or that there is a decrease in molecular gas along that side of the globule. 
Globules B and C display a similar drop-off in CO isotopologue intensities along their Southern sides (0, $-$10). Moreover, minimal emission is found at the western position of Globule B. When compared with the alignment of each globule observed in the optical, the decrease is consistent with gas facing the central star. The mapping might then be identifying the lack of molecular CO in the ionized shell that protects the cold molecule-rich gas from the CSPN's radiation.

Newly published ALMA maps ($J=2 \rightarrow 1$ transitions of $^{12}$CO, C$^{18}$O, and $^{13}$CO) have revealed that Helix Globule C consists of a N-S-oriented filamentary structure with a length of 11--16$''$ and average thickness of 2$''$ \citep{GlobC19}. The ALMA data yield a velocity- and spatially-integrated line flux of 14.4 Jy~km~s$^{-1}$. Within the two 30~m positions whose beams span the Globule's $^{12}$CO-emitting region detected in the ALMA data, we measure integrated line intensities of 1.70 and 1.54 K~km~s$^{-1}$, in the 10$''$ $J=2\rightarrow1$ beams, for a total beam intensity of 3.24 K~km~s$^{-1}$. Given the conversion factor of S/T$_{mb}$=4.95 J/K for the 30m telescope at 230 GHz)\footnote{https://www.iram.es/IRAMES/telescope/telescopeSummary/telescope\_summary.html}, we compute an integrated flux of 16.04$\pm$0.71 Jy~km~s$^{-1}$, in reasonable agreement with the total line flux measured by ALMA. 

\section{Analysis}
\label{sect_analysis}

In this section, we consider the variations of three key line ratios across the Helix Nebula, to identify whether and how these variations may be related to the local intensity of high-energy radiation from the CSPN.

\subsection{HNC/HCN in the Helix and Other PNe}
\label{FluxDetermination}

To calculate the UV flux incident on the targeted regions of molecular gas in the Helix, we have inferred their linear distances from the CSPN. 
We used the intrinsic Helix geometrical model presented in \cite{Meaburn98} to infer the 3D structure of the globules and envelope emission-line components from their velocities and projected positions. 
The resulting deprojected distances from the CSPN, along with the inferred incident CSPN UV fluxes, are listed in Table \ref{Ratio_Table}. Note that these incident UV flux estimates only account for geometrical dilution; we do not attempt to correct for potential extinction along the line of sight between gas parcel and CSPN.

\begin{figure}
		\includegraphics[width=\linewidth]{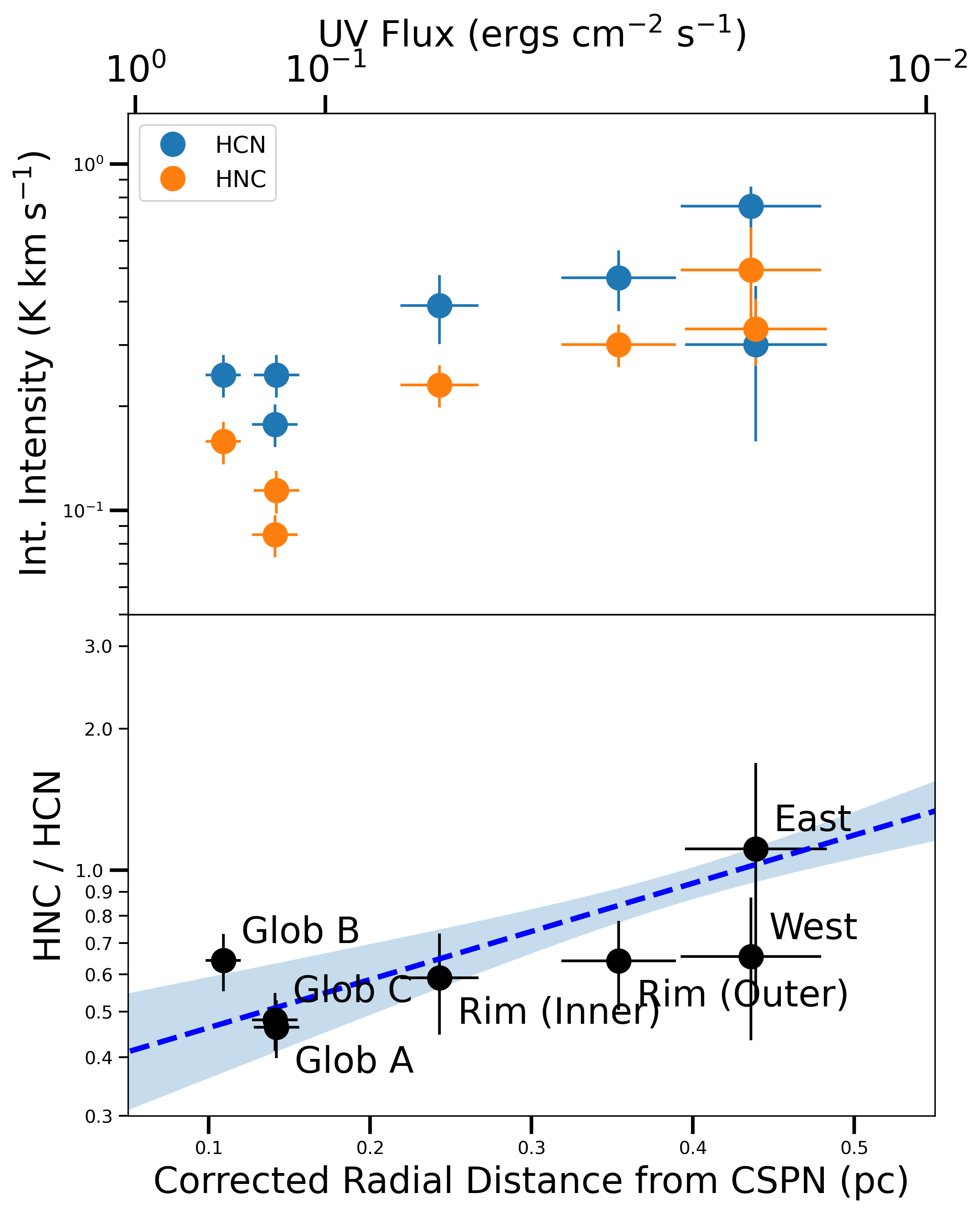}
		\caption{Top: Integrated line intensities of HCN and HNC for seven regions throughout the Helix Nebula at varying deprojected distance from the CSPN. Computed UV flux units are shown along the upper axis.
		Bottom: Line ratio of HNC/HCN against distance. The blue dashed line indicates a linear regression fit, with the shaded region representing 1$\sigma$ uncertainty.}
		\label{Helix_Ratio_Dist}
	\end{figure}

In Figure \ref{Helix_Ratio_Dist}, we plot the integrated HNC and HCN line intensities and their line ratio as functions of distance from CSPN and incident CSPN UV flux. Both molecular lines increase monotonically in intensity with increasing distance from the CSPN (Figure \ref{Helix_Ratio_Dist}, top panel). Such an increase in intensity may be indicative of decreasing beam dilution with the more distant positions, i.e., increasing emitting region solid angle as one goes from the compact inner globules to the more diffuse molecular rim. The lower panel of Figure \ref{Helix_Ratio_Dist} reveals that the HNC/HCN ratio increases as CSPN displacement increases and, hence, F$_{UV}$ declines.

\begin{table*}
	\caption{PNe used to compute HNC/HCN ratio}
	\label{Ratio_Table}
	\centering
\resizebox{\textwidth}{!}{%
	\begin{tabular}{l c c c c c}
\hline\hline
Object & 		HCN $\int{I dv}$ & 	HNC $\int{I dv}$ & 	HNC/HCN & 	Distance from 	& UV Flux$^c$ \\
Name & 		 (K km s$^{-1}$) &  	(K km s$^{-1}$) & 		Ratio & 	CSPN (pc) 	& (ergs cm$^{-2}$ s$^{-1}$)\\
\hline
Helix Globule A 		& 0.25 (0.01) 	& 0.11 (0.01) 	& 0.46 (0.01)	& 0.14 (0.02) 	& 0.15 (0.01) \\
Helix Globule B 		& 0.25 (0.02) 	& 0.16 (0.01) 	& 0.64 (0.06)	& 0.11 (0.01) 	& 0.25 (0.01) \\
Helix Globule C 		& 0.18 (0.01) 	& 0.09 (0.00) 	& 0.48 (0.04)	& 0.14 (0.02) 	& 0.15 (0.01) \\
Helix Rim (Outer) 		& 0.47 (0.10) 	& 0.30 (0.03) 	& 0.64 (0.14)	& 0.35 (0.08) 	& 0.02 (0.00) \\ 
Helix Rim (Inner) 		& 0.39 (0.09) 	& 0.23 (0.02) 	& 0.59 (0.14)	& 0.24 (0.04) 	& 0.05 (0.00) \\ 
Helix East 			& 0.30 (0.14) 	& 0.33 (0.07) 	& 1.11 (0.58)	& 0.44 (0.08) 	& 0.02 (0.00) \\
Helix West 			& 0.76 (0.06) 	& 0.49 (0.16) 	& 0.66 (0.22)	& 0.44 (0.08) 	& 0.02 (0.00) \\
\hline
NGC 7027	& 11.6 (0.03) 	& 0.31 (0.02) 	& 0.03 (0.001)	& 0.03$^a$ 	& 240.82 \\ 
Hb 5			& 0.63 (0.22) 	& 0.24 (0.06) 	& 0.38 (0.17)	& 0.09$^a$ 	& 18.33 \\ 
NGC 6720	& 1.83 (0.02) 	& 0.67 (0.02) 	& 0.37 (0.01)	& 0.17$^a$ 	& 0.44 \\ 
NGC 6445	& 4.54 (0.02) 	& 1.12 (0.02) 	& 0.25 (0.01)	& 0.19$^a$	& 0.85 \\ 
NGC 2346	& 1.0 (0.2) 	& 1.0 (0.2) 	& 1.0 (0.28)	& 0.32$^a$ 	& 0.01 \\ 
NGC 6720 Rim	& 2.08 (0.03) 	& 0.60 (0.03) 	& 0.29 (0.01)	& 0.16 (0.07) 	& 0.52 (0.03) \\ 
NGC 6853	& 0.64 (0.01) 	& 0.20 (0.07) 	& 0.32 (0.01)	& 0.16$^b$	& 0.42 \\ 
NGC 2440	& 0.37 (0.12) 	& 0.05 (0.0) 	& 0.14 (0.05)	& 0.12$^a$ 	& 4.83 \\ 
NGC 6772	& 0.42 (0.02) 	&  0.15 (0.01) 	& 0.37 (0.03)	& 0.20 (0.01) 	& 0.21 (0.02) \\ 
NGC 6781	& 3.83 (0.02) 	& 1.73 (0.02) 	& 0.45 (0.01)	& 0.32 (0.09) 	& 0.12 (0.04) \\ 
Helix-SZ 1 	& 0.13 (0.07) 	& 0.08 (0.03) 	& 0.63 (0.09)	& 0.22-0.29	& 0.047 (0.018) \\
Helix-SZ 2		& 0.17 (0.09) 	& 0.16 (0.06) 	& 1.00 (0.26)	& 0.34-0.36 	& 0.024 (0.007) \\
Helix-SZ 3		& 0.18 (0.07) 	& 0.12 (0.04) 	& 0.59 (0.10)	& 0.40-0.48 	& 0.015 (0.004) \\
\hline
	\end{tabular}
	}
\tablefoot{
	NGC 2346 line intensities from \cite{Bachiller89+}, while NGC 2440 and Hb 5 integrated intensities computed from \cite{Schmidt16, Schmidt17}. 
	Helix-SZ values from \cite{Schmidt17}.
	Other PN line intensities from \cite{Bublitz19}. \\
	\tablefoottext{a}{Distance estimated using H$\alpha$ radius \citep{Frew08}, assumed 50\% error in computed fluxes}
	\tablefoottext{b}{Distance estimated from H$_2$ map \citep{Kastner96}.}
	\tablefoottext{c}{Incident UV flux assuming CSPN L$_{UV}$ = 3.55$\times$10$^{35}$ erg s$^{-1}$ \citep{Montez15}.}\\
	All other distances computed from observed angular displacement and expansion velocities.
	UV Flux intensities calculated using deprojected distances of targeted positions from CSPN and CSPN UV luminosities \citep{Bachiller89+, Frew08, Montez15, Bublitz19}. Error estimated from 12~m beam.
	}
\end{table*}

\begin{figure}
		\includegraphics[width=\linewidth]{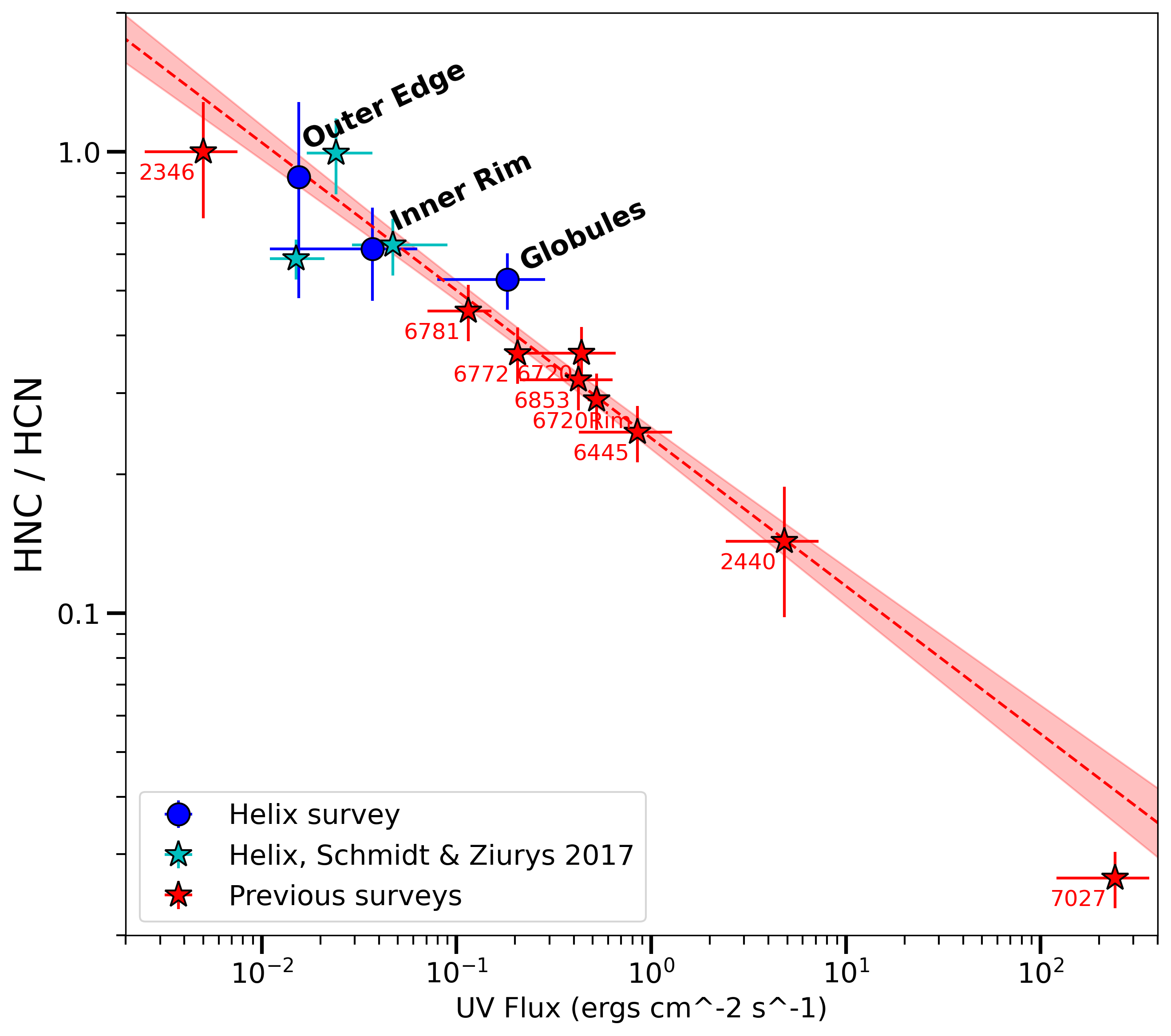} 
		\caption{HNC/HCN line ratio versus UV flux for the averaged representative Helix Nebula positions (blue symbols) and similar molecule-rich PNe (red symbols with associated NGC numbers). A linear regression fit to all data points is indicated as a red dashed line, with the range of 1$\sigma$ uncertainty shown as red shading.}
		\label{Extended_Helix_Ratio}
	\end{figure}

To place these results for the apparent dependence of Helix HNC/HCN on UV flux in context, we identified PNe with previous HNC and HCN observations and for which we could estimate the UV irradiation (Table \ref{Ratio_Table}). The Helix Nebula molecular envelope was previously observed in HNC and HCN by \cite{Schmidt17} across eight regions, including two locations close to those targeted in this work. Line intensities from these positions were divided into three distance bins (see Helix-SZ 1--3 in Table \ref{Ratio_Table}) and averaged into line ratios representative of the inner, middle, and outer Helix envelope. Other PNe studied in HNC and HCN \citep[and references therein]{Bublitz19} with molecular gas velocities distinct from their CSPN systemic velocity, such as NGC 6720 (rim position), NGC 6772, and NGC 6781, were also included in the sample, using the same method as used for the Helix to calculate the UV flux incident on the molecular gas. The remaining objects listed in Table \ref{Ratio_Table} were either unresolved or observed directly along the line-of-sight to their CSPN, thus sampling gas across a significant range of UV illumination. For these cases, displacement from CSPN to irradiated gas cannot be determined, so we adopt the H$\alpha$ radii of these PNe, with assumed 50\% uncertainties \citep{Frew16}. Due to its highly filamentary nature \citep{Huggins96, Bachiller2000}, the extent of NGC 6853's molecular envelope is poorly represented by the H$\alpha$ radius. We instead estimate the extent of molecular gas by selecting the midpoint to the outer edge of the nebula's H$_2$ region \citep{Kastner96}. 
The only Table \ref{Ratio_Table} PN not included in Figure \ref{Extended_Helix_Ratio} is the pinched-waist bipolar nebula Hb 5. Far infrared studies of Hb 5 and subsequent modeling have found dust mass levels higher than most PNe \citep{Pottasch07}, with significant quantities of molecular hydrogen in the core likely due to a high progenitor mass star. A nebular radius estimate from H$\alpha$ flux does not adequately represent the region where the molecular gas resides, as this PN has been shown to have highly filamentary lobes that stem from a dense core of material that is shaped by fast outflows \citep{Lopez12}.

The inclusion of these additional data points from PN literature sources extends the anticorrelation found between HNC/HCN and UV flux from the CSPN (Figure \ref{Extended_Helix_Ratio}). A linear regression fit over all PNe yields a power-law slope of $-0.32$. We thus broadly conclude that the relationship of HNC/HCN with UV flux takes the form
$$\frac{HNC}{HCN} \sim F_{UV}^{-1/3}.$$

That the power law slope of all PNe in this study is similar albeit steeper than $-0.29$ for the Helix alone may reflect the fact that most of these objects represent CSPNe with levels of UV irradiation 3--1600 times that of the UV incident on the Helix Nebula. 
When the seven regions of the Helix Nebula observed in this study are separated into three averaged bins, representing the globules of the ionized region, the inner rim positions, and the East and West positions of the outer envelope (dark blue, Figure \ref{Extended_Helix_Ratio}), the points are highly compatible with the general trend across all PNe. Despite the scatter in individual measurements, and the fact that the PNe in this study span a wide range of ages and morphologies, it is apparent that the anticorrelation of HNC/HCN with UV irradiation extends over 4 orders of magnitude in UV flux. We further explore the potential dependence of HNC/HCN on UV irradiation in Section \ref{PDR}.

\subsection{Variation of HCO$^+$ with Radius}
\begin{figure}[!h]
		\includegraphics[width=\linewidth]{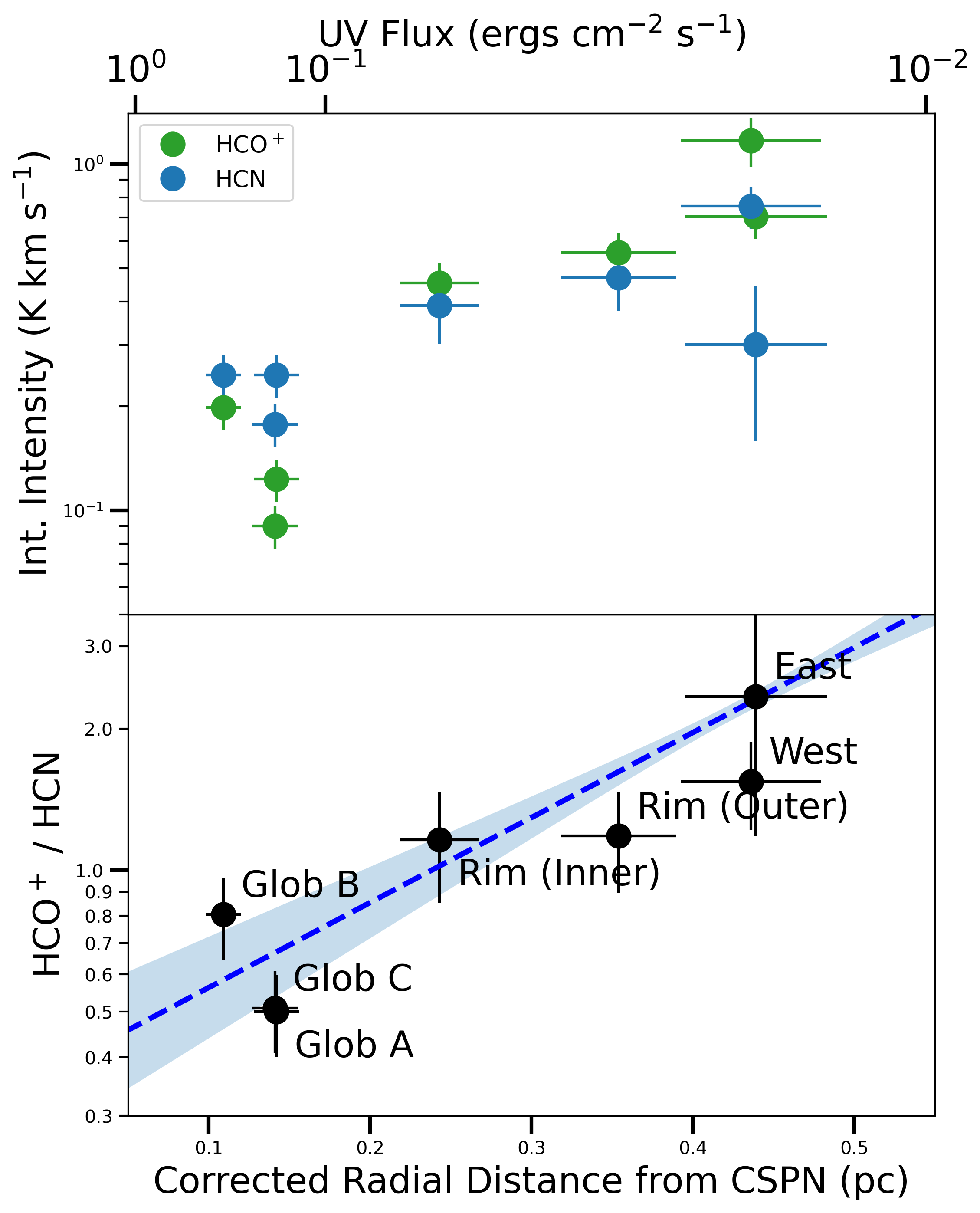}
		\caption{Top: Integrated line intensities of HCO$^+$ and HCN plotted against the deprojected distances from the CSPN. Computed UV flux emission at a given distance from the CSPN is included along the upper axis.
		Bottom: HCO$^+$/HCN line ratio against distance. Linear regression fit with 1$\sigma$ uncertainty is included in blue.}
		\label{HCO_HCN}
	\end{figure}

\begin{figure}[!h]
		\includegraphics[width=\linewidth]{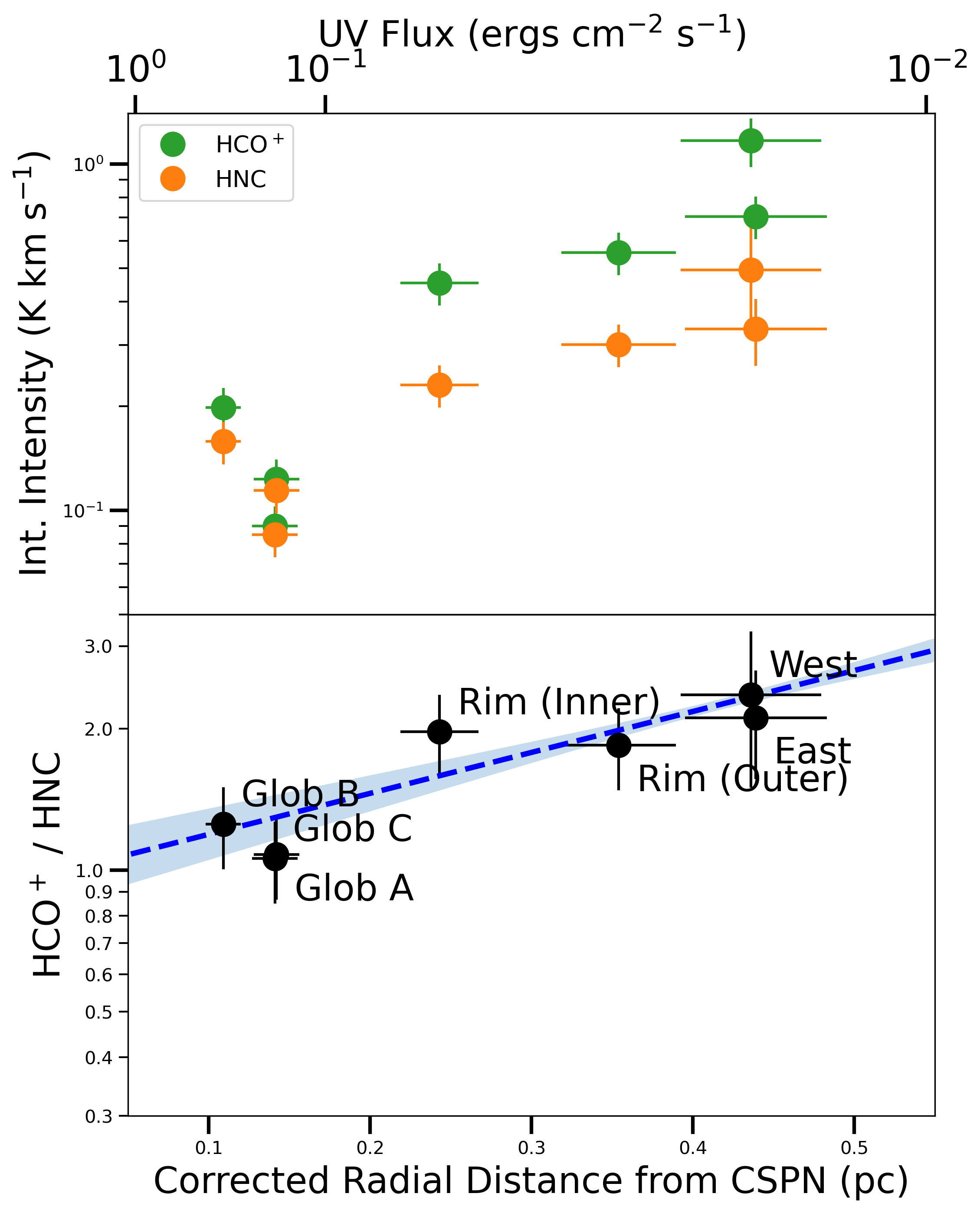}
		\caption{Top: Integrated line intensities of HCO$^+$ and HNC plotted against the deprojected distances from the CSPN. Computed UV flux emission at a given distance from the CSPN is included along the upper axis.
		Bottom: HCO$^+$/HNC line ratios against distance. Linear regression fit with 1$\sigma$ uncertainty is included in blue.}
		\label{HCO_HNC}
	\end{figure}

In Figures \ref{HCO_HCN} and \ref{HCO_HNC}, we plot the integrated intensities of HCO$^+$, and their ratio to HCN and HNC, as functions of projected distance from the CSPN and UV flux. These results demonstrate that HCO$^+$ increases in intensity with distance from the CSPN more rapidly and steadily than either HCN or HNC. HCO$^+$ has been well discussed as a likely diagnostic of X-ray emission incident on high-density gas, wherein the HCO$^+$ is enhanced via the production of H$_3^+$ and is shielded from UV-induced dissociative recombination \citep[e.g.,][]{Kimura12}. Cosmic rays may also play a role in HCO$^+$ production, though one might anticipate that CSPN-produced X-rays are more important within the Helix. HCO$^+$/HCN and HCO$^+$/HNC are found to rise with increased distance from the source of PN UV emission at slopes of 1.8 and 0.9 compared to the near unity increase of HNC/HCN with radius, respectively, with a correlation coefficient of $\sim$0.9 for both HCO$^+$ line ratios.

HCO$^+$ has previously been found to exhibit unusually high abundances in the molecular envelope of the Helix \citep{Zack13}. Figures \ref{HCO_HCN} and \ref{HCO_HNC} hint at an explanation for this enhanced HCO$^+$ production. 
As X-rays penetrate gas beyond the PDR, where dissociative UV photons are blocked, the extended molecular cloud can more readily produce and maintain heightened HCO$^+$ emission compared to regions that are highly UV-irradiated, such as the inner globules.

This correlation becomes apparent in the steep incline of the HCO$^+$/HCN ratio (Figure \ref{HCO_HCN}), wherein the outer envelope positions display $\sim$4 times the line ratio value of those positions closest to the CSPN. The HCO$^+$/HNC ratio is a factor $\sim$2 larger in the outer envelope than in the globules. The implication is that the molecular gas forms and maintains HCO$^+$ more readily where UV-irradiation is low and the molecule is not heavily dissociated, yet X-irradiation persists.

\section{Modeling} \label{sect_models}

\subsection{HCN and HNC Column Densities}
\label{Radex}

To interpret our molecular line data in terms of Helix globule properties such as density and kinetic temperature, we used RADEX, a publicly available, non-LTE radiative transfer code \citep{vandertak07}. RADEX takes in a set of initial conditions for a one-dimensional slab model of gas and uses an escape probability formulation to calculate the emergent line intensities within the modeled medium. The RADEX outputs are further refined through comparison with multiple transitions of the target molecular species. By varying parameters such as gas temperature and density and then comparing the outputs with the observed transition strengths sampled with the 30m and APEX, we have attempted to constrain the abundances of HCN and HNC. 

As the only position with resolved $J=1\rightarrow0$ and $J=3\rightarrow2$ transitions for HCN and HNC, as well as HCN $J=2\rightarrow1$, Globule B provides good constraints on the molecular excitation. We hence use Globule B as the subject of the modeling discussed here. A range of kinetic temperatures consistent with PDR gas (10--100~K) were explored, along with a range of H$_2$ number densities and HCN and HNC column densities typical of molecular cloud environments \citep[10$^4$--10$^7$ cm$^{-3}$ and 10$^{10}$--10$^{14}$ cm$^{-2}$, respectively;][]{Meaburn98, Huggins92}. Background radiation temperature was set to 2.73~K, while the linewidths were fixed to 1.0~km s$^{-1}$ based on our measurements of emission lines from Globule B (Table \ref{table_longtable}).
  
Preliminary analysis of ALMA interferometric mapping of Globule B (Bublitz et al. 2021, in prep.) indicates that its structure is similar to that of Globule C observed with ALMA \citep{GlobC19}. Thus, the structure of Globule B is represented by an elongated rectangle of dimensions 12$''$ by 2$''$ pointing away from the CSPN, for the purposes of beam dilution correction. Given the beamsizes of the IRAM and APEX telescopes (27.8$''$ and 35.2$''$, respectively), the beam dilution factors are hence 25 and 40.

In Figure \ref{Radex}, we compare the observed line intensities for $J=1\rightarrow0, 2\rightarrow1,$ and $3\rightarrow2$ molecular transitions in Globule B, corrected for beam dilution effects, with those predicted by the model through contour maps in log-density and log-column density space. 
Contours correspond to the ranges of predicted intensities that agree with the observations to within 1$\sigma$. 
The confluence of these curves mark the acceptable solution space for the range of density and column density explored at each value of $T_{kin}$.

The results of incrementally varying gas kinetic temperature in the model indicate that, up to $T_{kin}$ = 60 K, tight constraints exist on the column densities of HCN and HNC. The measurement of three low-lying J transitions of HCN provide the narrow constraints on N(HCN) and n(H$_2$), as demonstrated in Figure \ref{Radex} for $T_{kin}$ = 20 and 40~K (black star).
These comparisons of the RADEX models to the data yield estimated column densities of $\sim$2.5$\times10^{12}$ cm$^{-2}$ and $\sim$1.5$\times10^{12}$ cm$^{-2}$ for HCN and HNC, respectively. The models indicate densities of 10$^{6.5}$--10$^{5.4}$ cm$^{-3}$ across the temperature range of 10--40~K. For HNC, however, the H$_2$ density remains degenerate with kinetic temperature. We find no clear trend in the HNC/HCN column density ratio with temperature from the RADEX modeling, suggesting that the line ratio indeed serves as a proxy for the abundance ratio in the T$_{kin}$ regime of interest.

	\begin{figure*}
	\centering
	\begin{tabular}{c}
		\includegraphics[width=0.32\linewidth]{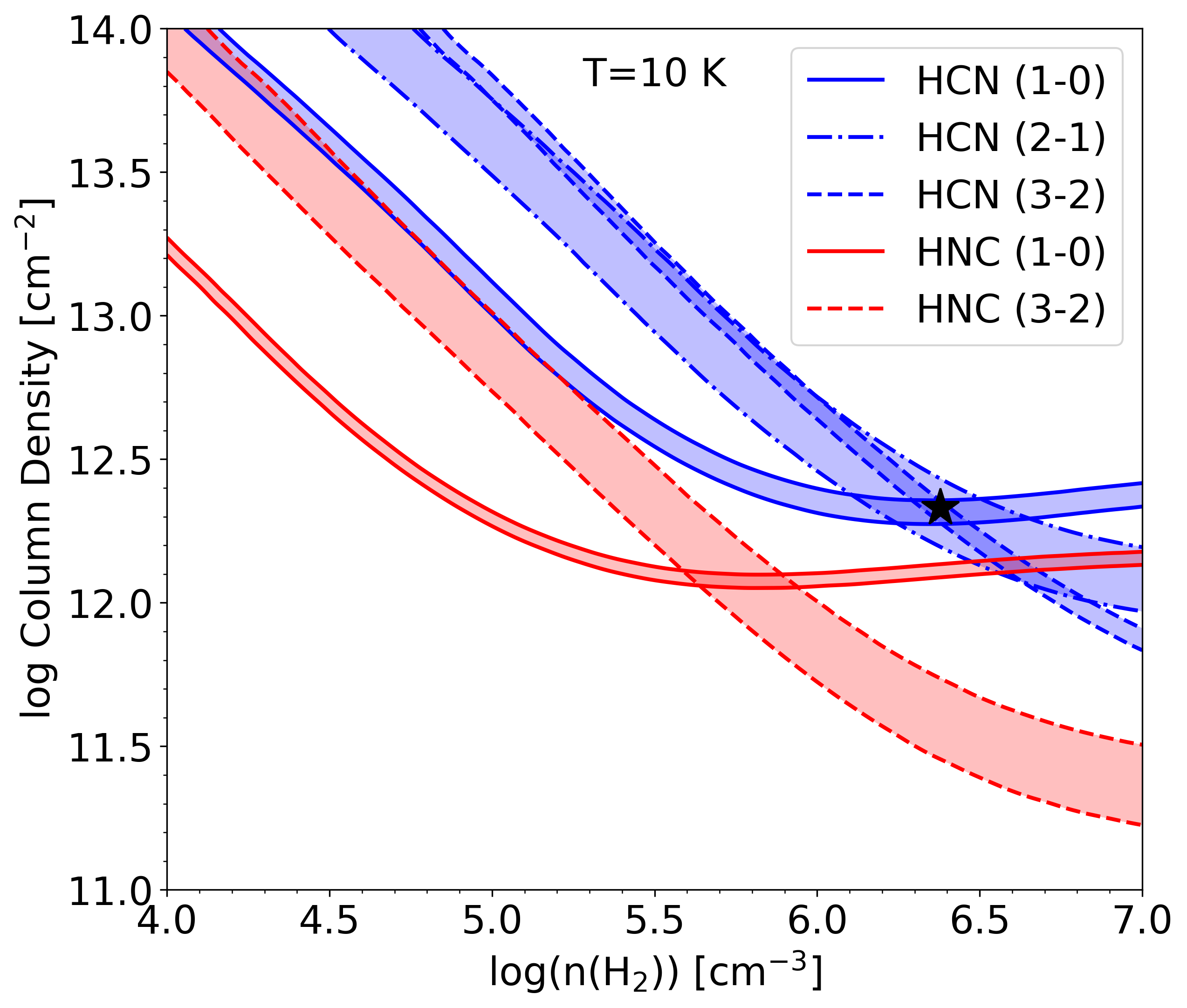}
		\includegraphics[width=0.32\linewidth]{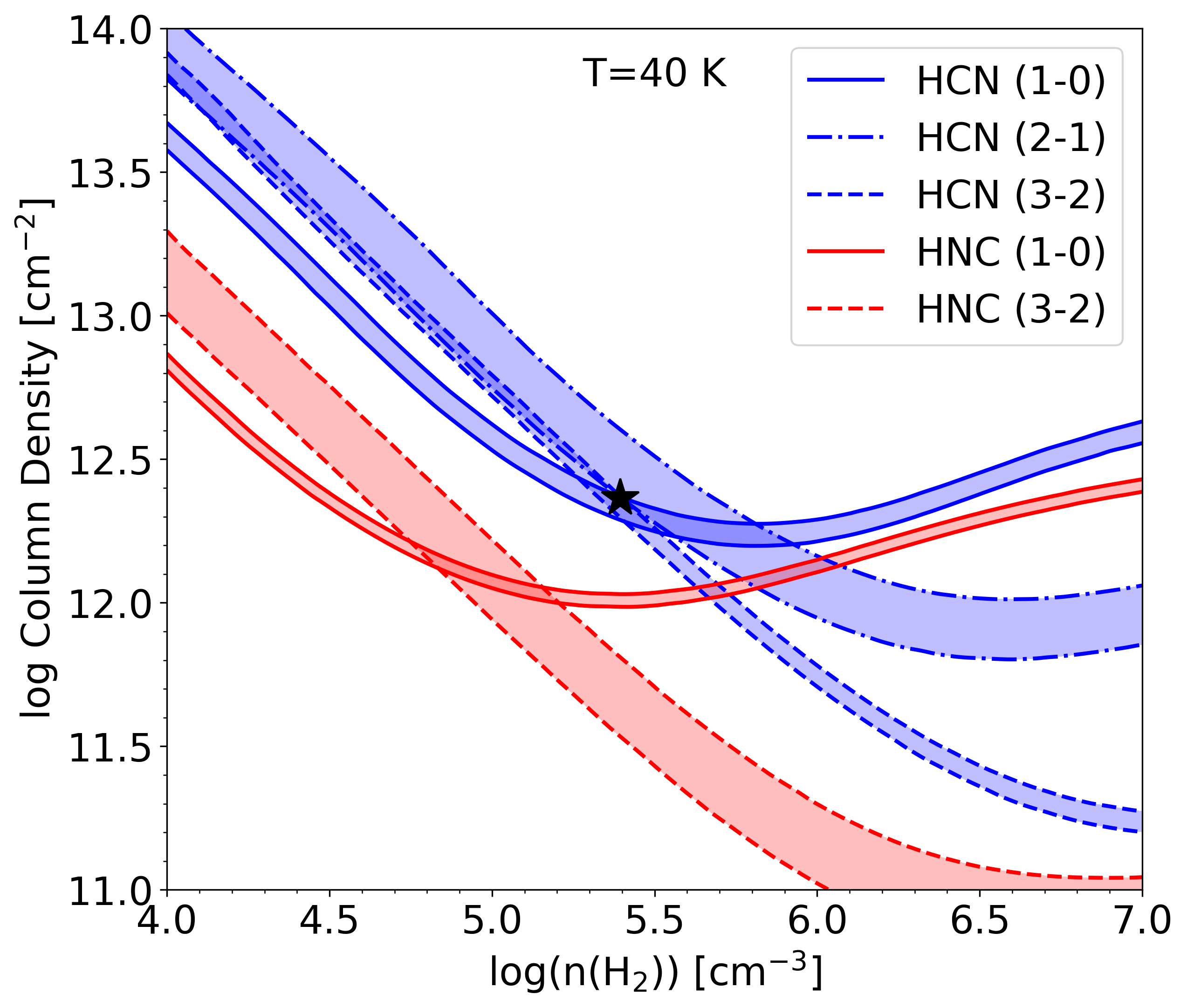}
		\includegraphics[width=0.32\linewidth]{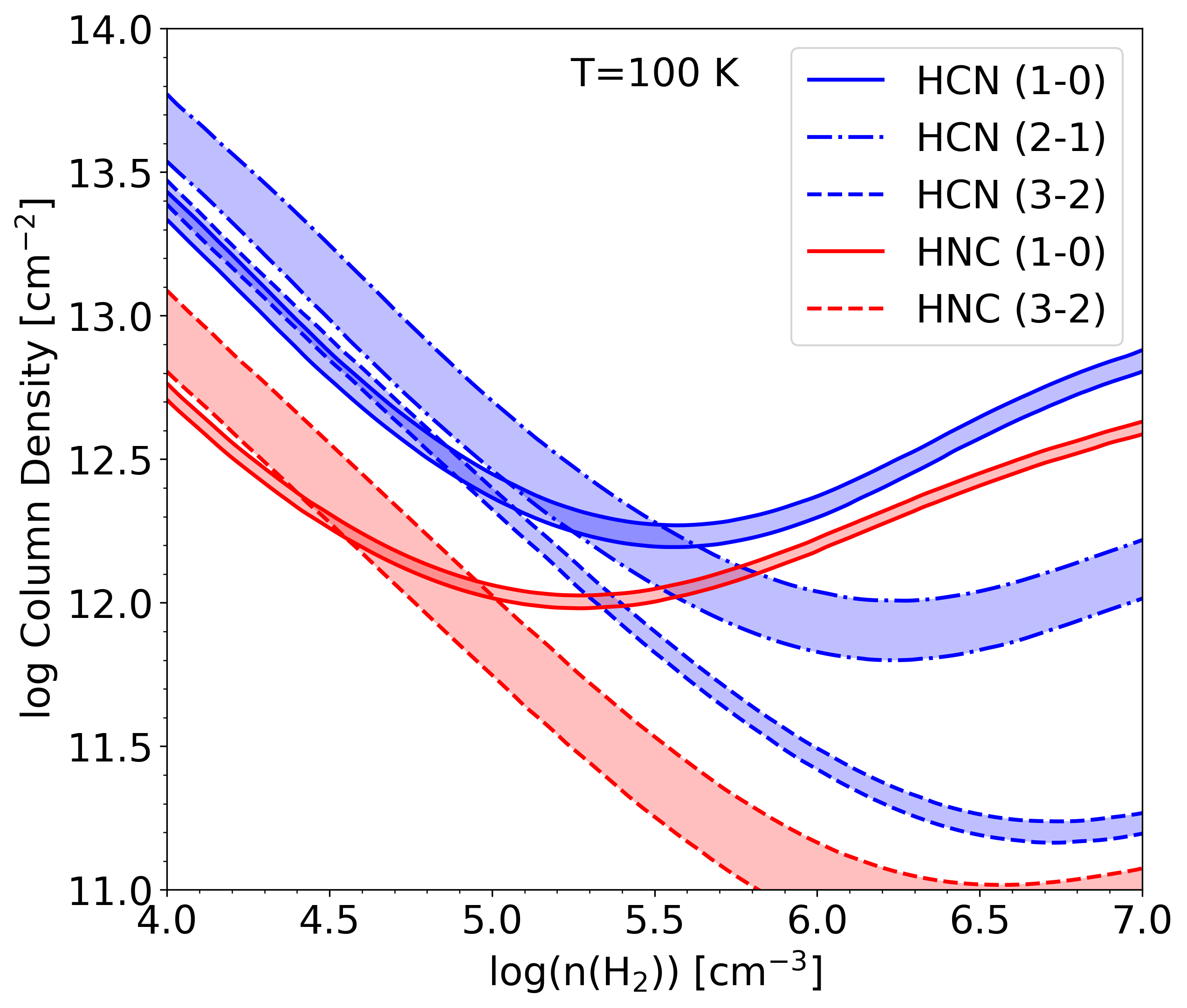}
	\end{tabular}
		\caption{Loci of density and kinetic temperatures compatible with the $J=1\rightarrow0, 2\rightarrow1$, and $3\rightarrow2$ HCN and HNC line intensities towards Globule B. The shaded regions indicate 1$\sigma$ uncertainties. A black star indicates highly constrained regions of overlap across all three transitions of HCN.}
		\label{Radex}
	\end{figure*}

\subsection{PDR Modeling}
\label{PDR}

To interpret the RADEX results for HCN and HNC column densities in terms of a physical model of the globules, we employed the publicly available 
Meudon PDR code \citep[http://ism.obspm.fr]{LePetit06}. As a first approximation, we modeled the globule as a plane-parallel slab of gas and dust, irradiated by a fixed UV radiation field, and computed the radiative transfer equation for each point through an iterative process. Heating and cooling processes (such as the photoelectric effect and cosmic rays, and infrared and millimeter emission, respectively) are incorporated into the model, which utilizes a large chemical network for the purposes of predicting abundances. The model code then computes the abundances of relevant species for an assumed set of physical conditions. A model spectrum representing a 100,000~K white dwarf star \citep{Levenhagen17}, scaled to the flux of the Helix CSPN \citep{Oke90}, provided the ionization source to one side of the slab, while the standard interstellar radiation field was assumed to irradiate the other side. The ionized region of the Helix that surrounds a globule is assumed to result in constant pressure across the ionized-neutral zones.

The density of the slab of gas was initially set to 10$^3$ cm$^{-3}$, with a gas temperature of 100~K. Using the analysis of \cite{ODell05} and neutral gas properties in PNe \citep{Natta98}, we can constrain the expected pressure outside the globules. Assuming a constant pressure boundary is necessary to maintain the globules for the age of the nebula, we estimate a representative pressure of $\sim$10$^7$~K~cm$^{-3}$ for the Meudon models. This is consistent with an extinction (due to dust) within the globule of A$_V \geq$0.5 \citep{ODell05}. Given a typical globule diameter of $\sim$2$''$ \citep{ODell05, Meaburn98} or a thickness of 400~AU, we then varied the pressure in our computed models until they achieved consistent depth for extinctions of A$_V$=0.5--1.0. 

\begin{figure*}
		\includegraphics[width=0.49\linewidth, trim={0 0 0 0}, clip]{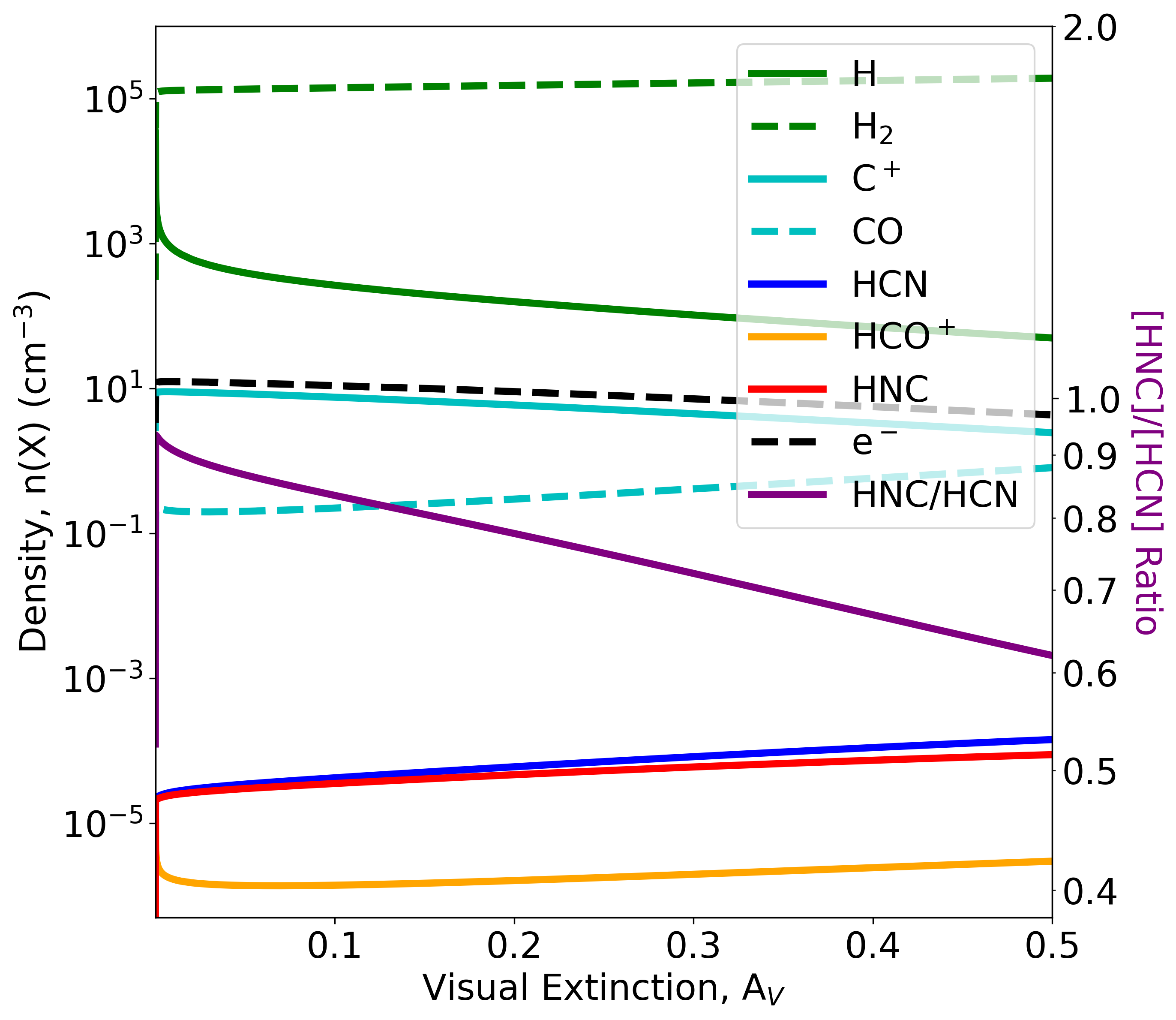}
		\includegraphics[width=0.49\linewidth, trim={0 0 0 0}, clip]{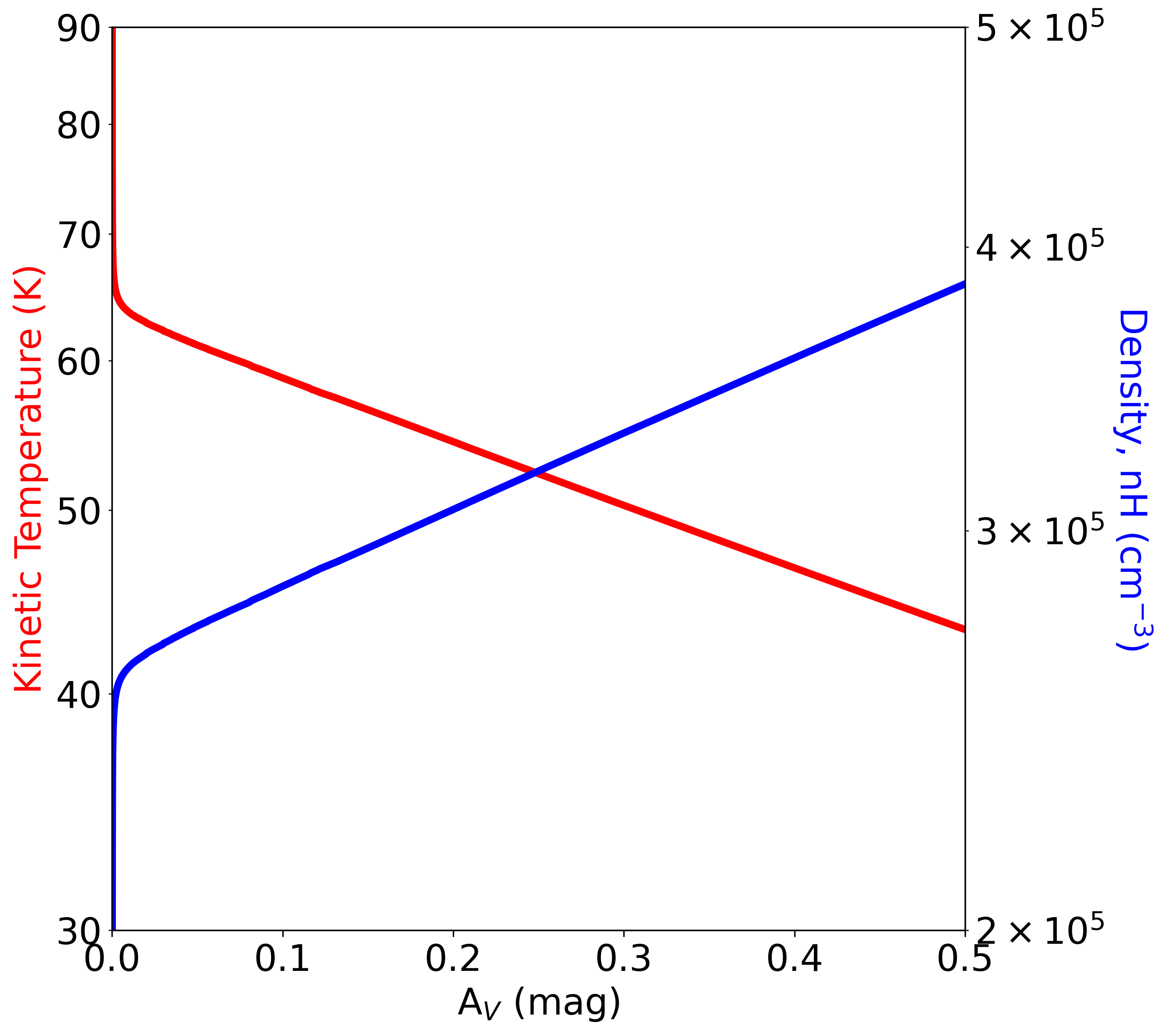} \\
		\includegraphics[width=0.49\linewidth, trim={0 0 0 0}, clip]{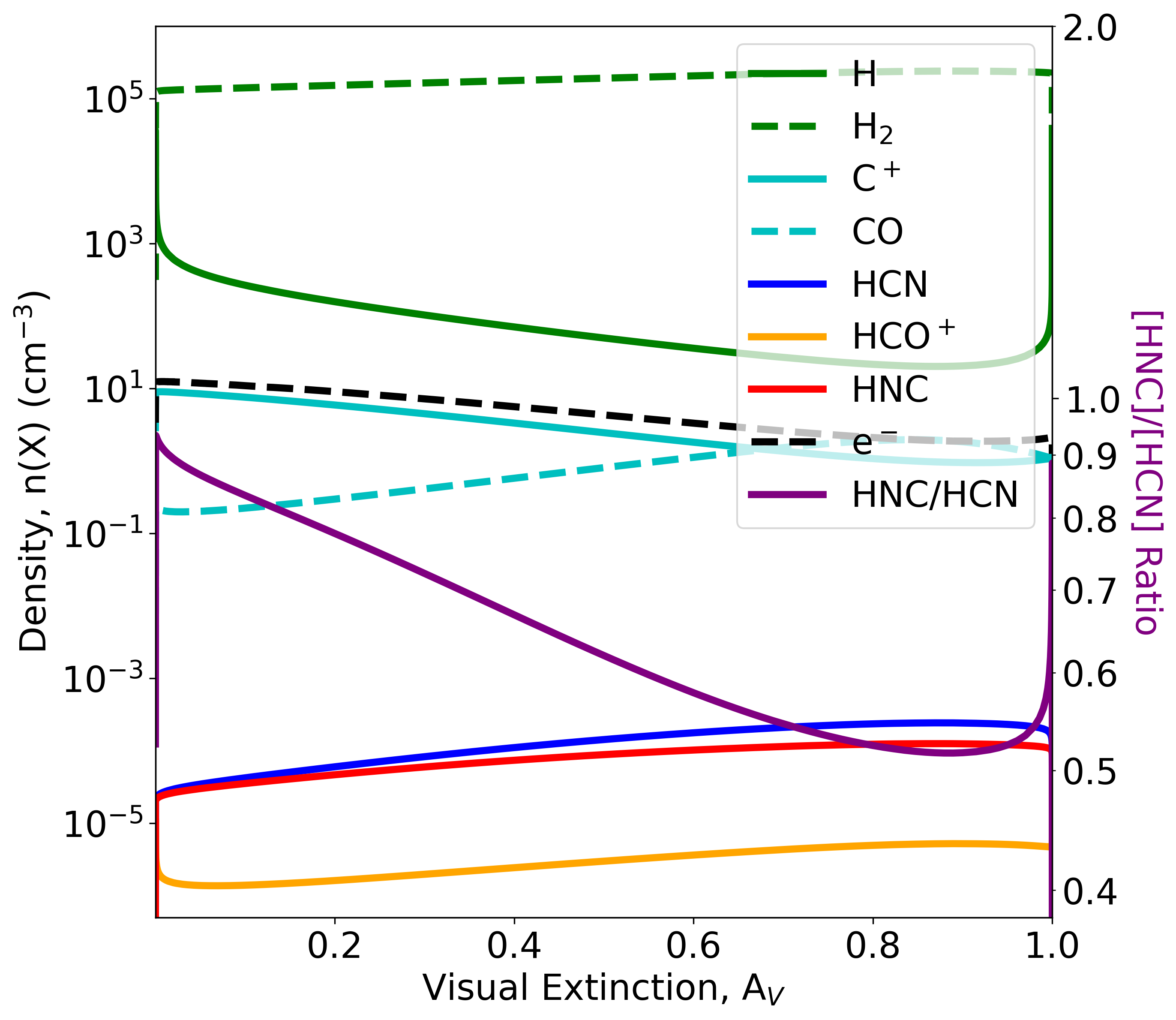}
		\includegraphics[width=0.49\linewidth, trim={0 0 0 0}, clip]{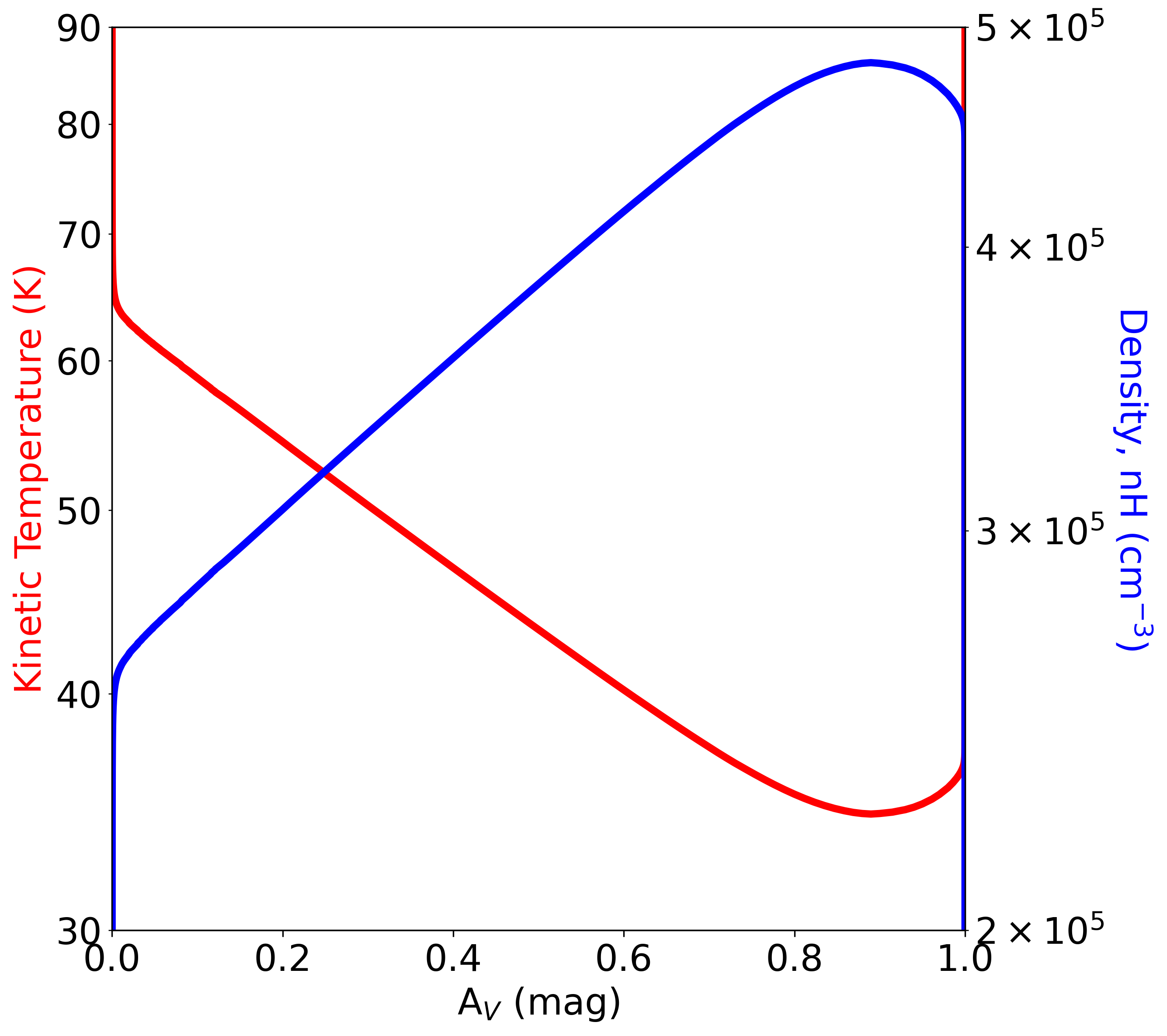}
		\caption{
		Molecular and atomic densities and gas parameters across a gas slab of extinction A$_V$=0.5 (top) and A$_V$=1.0 (bottom), with fixed pressure P=$10^7$~K~cm$^{-3}$ and density $\rho$=$10^5$~cm$^{-3}$, irradiated by a WD model spectrum of T=100,000~K \citep{Levenhagen17}. {\em Left:} Density of selected species across the visual extinction of the slab. The abundance ratio of HNC/HCN is plotted as a solid purple line, with scale indicated on the right-hand y-axis. {\em Right:} Temperature and density across the model slab.}
		\label{PDR_AV10_1}
		\label{PDR_AV05_1}
	\end{figure*}

The resulting chemical abundances of several key molecular and atomic species are displayed in Figure \ref{PDR_AV10_1}. In these models, we find the HNC/HCN ratio is below unity across the slab. 
A fixed initial pressure results in heightened density of the gas inside the model globule and a gradient in temperatures across $\sim$40--60~K, consistent with the range predicted by \cite{Graninger14}. Curiously, the models show that the HNC/HCN ratio declines with decreasing temperature inside an individual globule, contrary to what is observed for samples of cloud cores \citep{Schilke92, Graninger14, Riaz18}.

\begin{figure*}
		\includegraphics[width=0.48\textwidth]{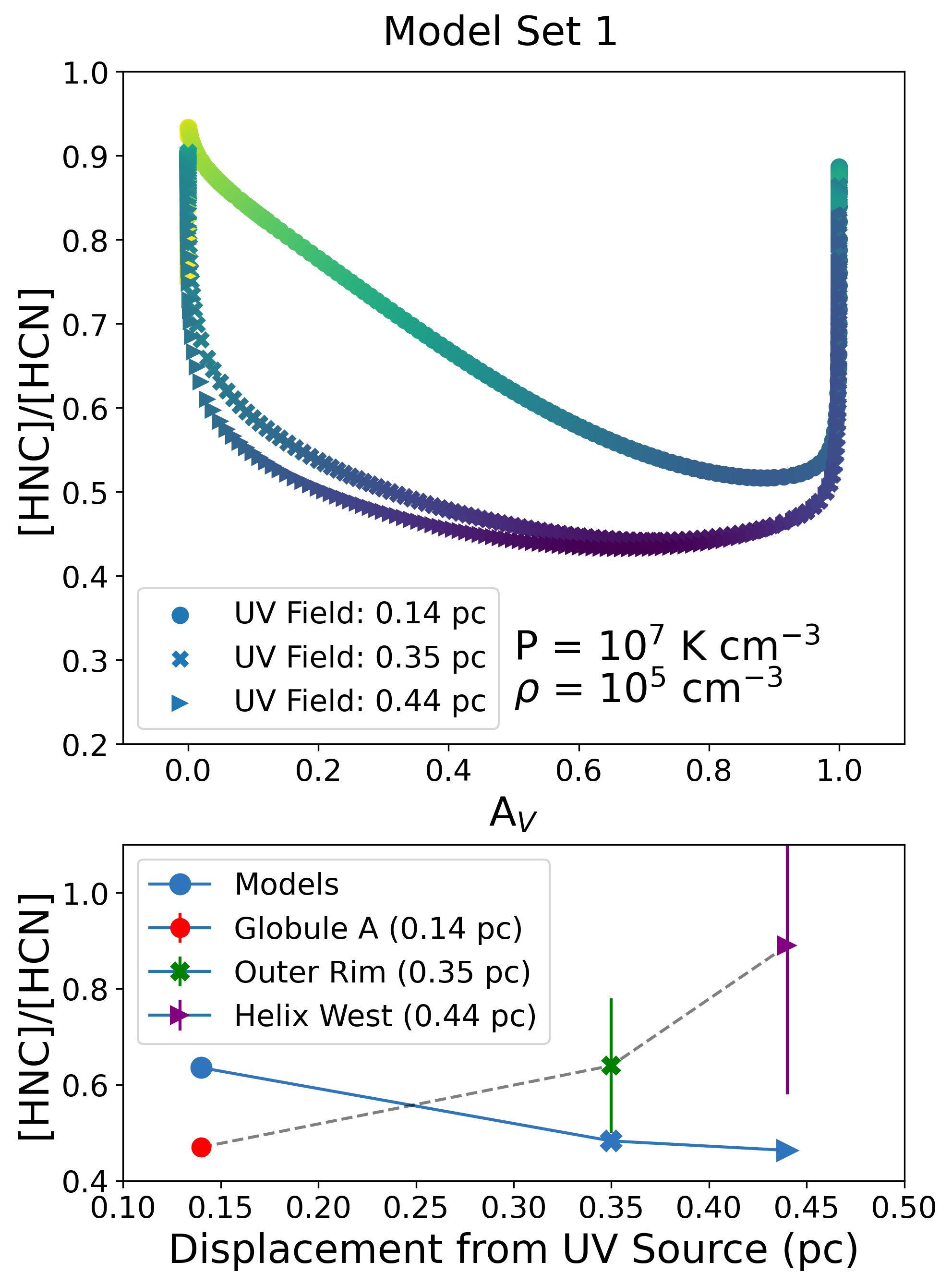}
		\includegraphics[width=0.515\textwidth]{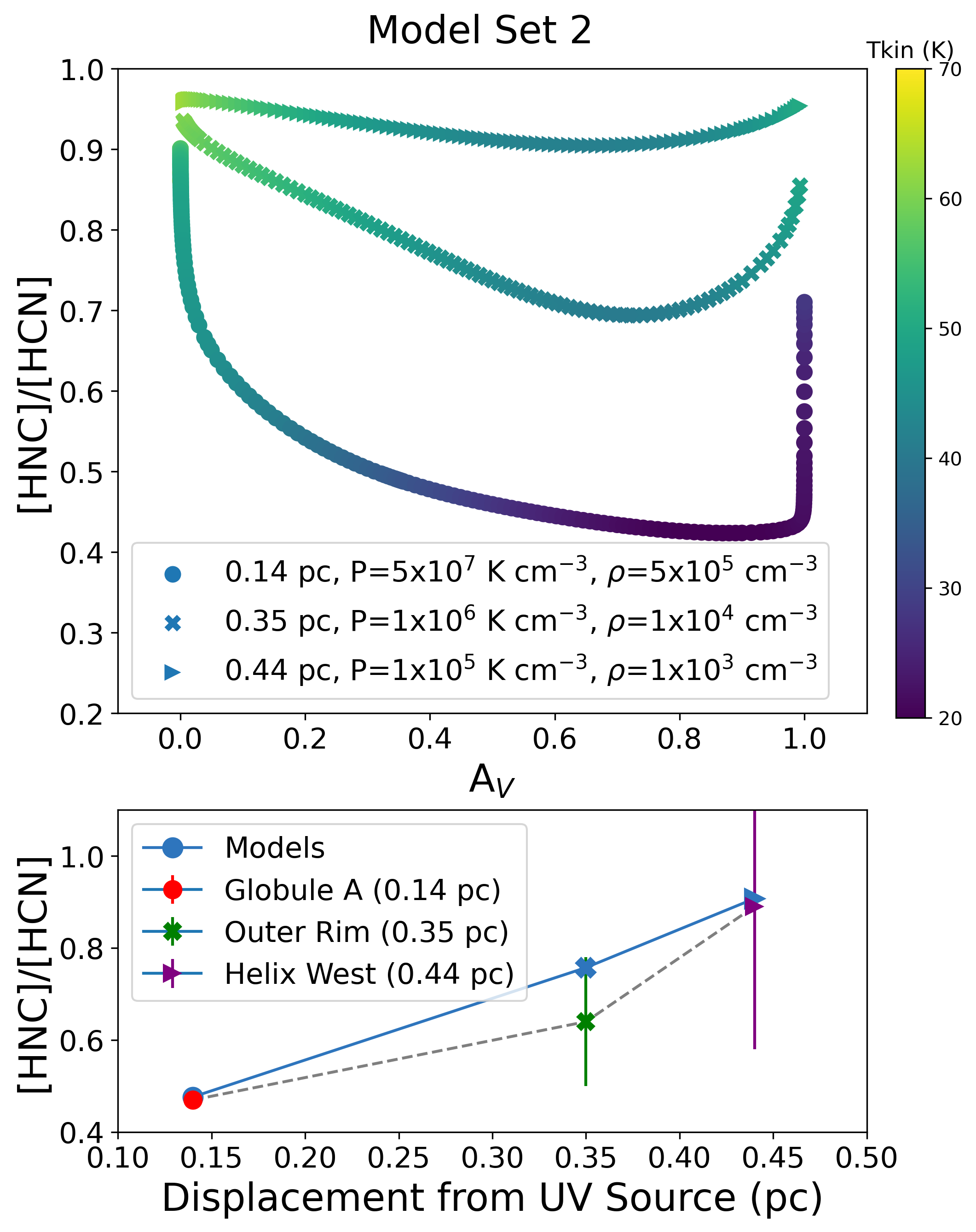}
		\caption{
		\textit{Top row.} Curves of the [HNC]/[HCN] abundance ratio output by the Meudon PDR modeling code are plotted against extinction, for a representative slab of gas with integrated A$_V$=1.0 of a representative slab of gas. Points are colored based on kinetic gas temperature, which peaks toward the edges of the slab. \textit{Bottom row.} The average abundance ratio for each model curve is marked with the corresponding symbol (light blue), outlining the predicted trend with respect to displacement from the UV flux source. Additional points (red, green, blue) mark the IRAM 30m observed line ratios for the three modeled positions (Globules, Rim, Plumes), respectively. \textit{Left:} Model Set 1. HNC/HCN of three model slabs with increasing distances from the UV source, thus decreasing incident UV flux. Pressure and density remain fixed. 
		\textit{Right:} Model Set 2. UV flux, pressure, and density are all decreased with increasing distance from the CSPN. See Section \ref{PDR}. }
		\label{Fig_PDR_HNCHCN}
		\label{Fig_PDR_HNCHCN_PD}
	\end{figure*}

We used the same fundamental PDR model to simulate molecular gas parcels at three positions of the Helix Nebula as observed with the 30m telescope: a globule (A or C), the outer Rim, and the plume positions (East and West). In one set of models (hereafter Model Set 1), we fixed gas pressure and density across the three positions, with values of 10$^7$ K~cm$^{-3}$ and 10$^5$ cm$^{-3}$, respectively, and varied UV flux according to the relative distances of three positions from the CSPN. This set of models is illustrated in the left panel of Figure \ref{Fig_PDR_HNCHCN_PD}. The right panel of Figure \ref{Fig_PDR_HNCHCN_PD} represents a set of models in which density is progressively decreased across the three modeled positions (Model Set 2). 
In addition to the detailed results of the modeling of HNC/HCN as a function of A$_V$, we have calculated an A$_V$-weighted mean ratio for each model, for comparison with the observed HNC/HCN ratios (bottom panels of Figure \ref{Fig_PDR_HNCHCN} and Table \ref{table_PDR_Values}).

The model curves demonstrate that, in each case, the HNC/HCN ratio declines toward the center of the modeled region, reaching a minimum near A$_V \sim$0.7. However, in Model Set 1, an increase in distance from the UV source leads to a decrease in the weighted mean HNC/HCN, counter to observations (Figure \ref{Fig_PDR_HNCHCN}, bottom left). 
In contrast, we find Model Set 2 --- in which pressure, density, and UV irradiation are all simultaneously decreased, to mimic increasing distance from an irradiated central star --- appears to yield good agreement with the observed variation in HNC/HCN line ratios (Figure \ref{Fig_PDR_HNCHCN}, bottom right).

The preceding comparison between the models and observations suggests that the combination of changing pressure (density) and UV irradiation, as opposed to one or the other ``ingredient'', may best explain the anticorrelation of HNC/HCN with distance from CSPN that is observed in the Helix. 
Model Set 2 should not be considered a `best fit' given that we have not explored the full parameter space, but rather an indication that the combination of well-constrained variation in UV flux and variation in pressure can reproduce the observed trend in HNC/HCN.


\begin{table}
	\caption{Average Line Ratios from PDR Meudon Code}
	\label{table_PDR_Values}
\resizebox{\linewidth}{!}{%
	\begin{tabular}{l | c | c c c | c}
\hline\hline
				& Model Set 1 & \multicolumn{3}{c |}{Model Set 2} &	Observed Ratio \\
\hline
Pressure (K cm$^{-3}$) & 10$^7$	& 5$\times$10$^7$ & 10$^6$ & 10$^5$ & \\
\hline
UV Field: 0.14 pc	& 	0.64		& 0.48 &	& 		& 0.46-0.48 \\
UV Field: 0.35 pc	&	0.48		&	& 0.55 &		& 0.64 \\
UV Field: 0.44 pc	&	0.46		&	&	& 0.73	& 0.66-1.11 \\
\hline
	\end{tabular}
	}
	\tablefoot{Average value of the HNC/HCN line ratio from Meudon PDR models (Figure \ref{Fig_PDR_HNCHCN}). The observed ratios for Helix positions with similar UV field strengths (see Table \ref{Ratio_Table}) are presented in the right-most column.
}
\end{table}

\subsection{Comparison of RADEX and PDR Model Results}

Contours of HCN and HNC column density in Figure \ref{Radex} indicate that HCN preferentially resides somewhat deeper within the globules. 
If the conversion of HNC into HCN does scale with local gas temperature \citep{Schilke92, Graninger14, Hacar19}, we would expect to find a shell of low ratio HNC/HCN surrounding an increasing ratio in the densest regions, where the gas is coldest and HNC is better shielded from UV that could drive chemical reprocessing. 
However, the RADEX models suggest that HNC/HCN decreases towards the core of the globule.

Modeling of photodissociation with a gas slab using the Meudon code produced similar results. A notable increase in HCN can be seen at the core of the PDR model slab, even with respect to the rise in HNC density (Figure \ref{PDR_AV10_1}, left and right). HNC/HCN drops to 0.43-0.52 at the core compared to the globule surface value of 0.93 (Figure \ref{Fig_PDR_HNCHCN}, left). By decreasing the pressure, and thus the density, of the model slabs, we achieve good agreement with the observed HNC/HCN line ratios across the Helix (Figure \ref{Fig_PDR_HNCHCN}, right). The fact that this model well reproduces the observational results suggests that both UV irradiation and the local gas physical conditions -- in particular, gas pressure (or density) -- play a role in determining the HNC/HCN ratio.

Nevertheless, the process that connects UV irradiation to the relative production and destruction rates of HCN and HNC remains poorly constrained. The HNC/HCN ratio has been found to be well correlated with gas heating in PDR environments \citep[e.g.,][]{Graninger14, Hacar19}, while selective photodissociation directly relates UV irradiation with the conversion of HNC into HCN  \citep{Aguado17}. The results presented here suggest that UV irradiation likely plays a key role via one or both of these mechanisms. That is, UV photons may be heating the gas via photoionization and the photoelectric effect on dust grains, or may be preferentially suppressing the HNC abundance via selective photodissociation; either or both processes would enhance the abundance of HCN relative to HNC as UV flux increases.

The molecule C$_2$H is a strong tracer of UV ionization in PDR environments and comparisons with HNC/HCN would establish the significance ionization plays on the ratio. As a byproduct of isocyanide dissociation, the presence of CN in regions of suppressed HNC/HCN would alternatively suggest the effect of selective photodissociation on the ratio abundance. Thus, additional observations of such photodissociation-generated species could help disentangle the effects of UV-driven heating vs. photodissociation in driving HNC/HCN.


\section{Conclusions} \label{sect_conclusions}

In this study, we present the results of a 1--3~mm line survey of the Helix Nebula using the IRAM 30~m and APEX 12~m telescopes. The targeted regions consist of three prominent dense molecular knots within the ionized zone, a position that has been kinematically resolved into two locations along the molecular rim of the nebula, and two diametrically opposite positions at the edges of the extended molecular envelope. Transitions from five molecules were targeted and detected across the regions: the $J = 1 \rightarrow 0$ and $J = 3 \rightarrow 2$ lines of HCN, HCO$^+$, and HNC, the $J = 2 \rightarrow 1$ lines of HCN and HCO$^+$, as well as $^{12}$CO and $^{13}$CO in all globules. These observations have yielded the first detections of HCN, HNC, HCO$^+$, C$_2$H, and CN in the globules of NGC 7293.

The results support previous indications that HNC is more readily destroyed than its isomer HCN in regions of higher UV irradiation in PNe. Targeting positions of variable distance to the Helix Nebula central star as a means to test the anticorrelation between the HNC/HCN line ratio and CSPN UV luminosity in PNe \citep{Bublitz19}, we have determined that the HNC/HCN ratio is indeed anticorrelated with distance from the CSPN, as expected if the HNC/HCN ratio decreases as irradiating CSPN UV flux increases. 
This suggests that significant UV-driven modification of the HNC/HCN ratio likely occurs during the PN stage, wherein the ratio may not be fixed during earlier (late AGB and pre-PN) evolutionary stages \citep[as asserted by][]{Schmidt17}. 
These Helix Nebula HNC/HCN line ratio data, combined with data from previous PN studies, demonstrate that this anticorrelation between HNC/HCN and the strength of UV irradiation spans 0.03--1.11 in HNC/HCN integrated line intensity ratio and a factor 10$^4$ in UV flux.

Radiative transfer (RADEX) models fit to the Globule B observed line intensities indicate column densities of $\sim$2.5$\times$10$^{12}$~cm$^{-2}$ and $\sim$1.5$\times$10$^{12}$~cm$^{-2}$ for HCN and HNC, respectively.  
Although the observational results within the Helix Nebula and across a larger sample of PNe clearly indicate that UV irradiation is important in determining the HNC/HCN ratio in PN molecular gas, our PDR modeling indicates that the UV flux gradient alone cannot reproduce the observed variation of HNC/HCN across the Helix. Instead, HNC/HCN appears to be dependent on both UV irradiation and gas pressure and density.

Additionally, the HCO$^+$/HCN and HCO$^+$/HNC ratios are both found to increase with distance from the CSPN. 
The increase in relative abundance of HCO$^+$ suggests that X-rays, with their superior penetrating power relative to UV, may be increasingly important in driving PN molecular chemistry as distance from the CSPN increases.


%
\longtab{
\begin{longtable}{lllcccc}
\caption{Helix Spectral Line Data \label{table_longtable}}\\
\hline\hline
Position & Molecule & Transition & T$_{mb}$ & $V_{LSR}$ & $\Delta V_{1/2}$ & $\int T_{mb} dV$ \\ 
 &  &  & (K) & (km s$^{-1}$) & (km s$^{-1}$) & (K km s$^{-1}$) \\
\hline
\endfirsthead
\caption{continued.}\\
\hline\hline
Position & Molecule & Transition & T$_{mb}$ & $V_{LSR}$ & $\Delta V_{1/2}$ & $\int T_{mb} dV$ \\ 
 &  &  & (K) & (km s$^{-1}$) & (km s$^{-1}$) & (K km s$^{-1}$) \\
\hline
\endhead
\hline
\endfoot
Globule A 		& HCN 		& $J = 1 \rightarrow 0, F = 0 \rightarrow 1$ & 0.028(0.003) & -22.68$\pm$0.09 & 1.59$\pm$0.23 & 0.046$\pm$0.005 \\
			&			& \hspace{18mm} $F = 2 \rightarrow 1$	& 0.101(0.003) & -22.52$\pm$0.02 & 1.34$\pm$0.05 & 0.139$\pm$0.005 \\
			&			& \hspace{18mm} $F = 1 \rightarrow 1$	& 0.053(0.003) & -22.52$\pm$0.04 & 1.13$\pm$0.08 & 0.062$\pm$0.004 \\
			& HCN 		& $J = 3 \rightarrow 2$				& 0.07(0.04) & -22.01$\pm$0.20 &  $\sim$0.68	& 0.050$\pm$0.027 \\
			& HCO$^+$ 	& $J = 1 \rightarrow 0$				& 0.107(0.003) & -22.56$\pm$0.02 & 1.12$\pm$0.06 & 0.123$\pm$0.005 \\
			& HCO$^+$	& $J = 3 \rightarrow 2$				& $<$0.03 & - & - & - \\
			& HNC 		& $J = 1 \rightarrow 0$				& 0.084(0.004) & -22.53$\pm$0.03 & 1.32$\pm$0.07 & 0.114$\pm$0.005 \\
			&HNC	 	& $J = 3 \rightarrow 2$				& 0.18(0.06) & -22.45$\pm$0.08 & 0.56$\pm$0.012 & 0.109$\pm$0.027 \\
			& $^{13}$CO 	& $J = 1 \rightarrow 0$				& 0.26(0.03) & -22.71$\pm$0.04 & 0.61$\pm$0.97 & 0.170$\pm$0.025 \\
			& $^{12}$CO 	& $J = 1 \rightarrow 0$				& 0.88(0.08) & -22.44$\pm$0.04 & 0.89$\pm$0.14 & 0.833$\pm$0.092 \\
			& $^{12}$CO 	& $J = 2 \rightarrow 1$				& 1.21(0.10) & -22.42$\pm$0.03 & 0.92$\pm$0.07 & 1.194$\pm$0.078  \\
\hline
Globule B 		& HCN 		& $J = 1 \rightarrow 0, F = 0 \rightarrow 1$ & 0.033(0.004) & -16.98$\pm$0.05 & 0.66$\pm$0.78 & 0.024$\pm$0.005 \\
			& 	 		& \hspace{18mm} $F = 2 \rightarrow 1$	& 0.083(0.004) & -17.75$\pm$0.04 & 1.67$\pm$0.11 & 0.145$\pm$0.008 \\
			& 	 		& \hspace{18mm} $F = 1 \rightarrow 1$	& 0.053(0.004) & -16.95$\pm$0.07 & 1.37$\pm$0.13 & 0.077$\pm$0.007 \\
			& HCN 		& $J = 3 \rightarrow 2$				& 0.24(0.03) & -17.13$\pm$0.06 & 1.53$\pm$0.16 & 0.399$\pm$0.035 \\
			& HCO$^+$ 	& $J = 1 \rightarrow 0$				& 0.110(0.006) & -17.04$\pm$0.04 & 1.78$\pm$0.08 & 0.198$\pm$0.008 \\
			& HCO$^+$	& $J = 3 \rightarrow 2$				& 0.43 (0.04) & -17.32$\pm$0.04 & 0.94$\pm$0.09 & 0.433$\pm$0.032 \\
			& HNC 		& $J = 1 \rightarrow 0$				& 0.082(0.005) & -16.99$\pm$0.04 & 1.84$\pm$0.11 & 0.158$\pm$0.008 \\
			& HNC 		& $J = 3 \rightarrow 2$				& 0.14(0.06) & -17.18$\pm$0.21 & 1.51$\pm$0.60 & 0.226$\pm$0.065 \\
			& $^{13}$CO 	& $J = 1 \rightarrow 0$				& 0.09(0.02) & -17.39$\pm$0.24 & 2.00$\pm$0.62 & 0.200$\pm$0.047 \\
			& $^{12}$CO 	& $J = 1 \rightarrow 0$				& 0.60(0.10) & -17.09$\pm$0.08 & 1.19$\pm$0.15 & 0.763$\pm$0.099 \\
			& $^{12}$CO 	& $J = 2 \rightarrow 1$				& 1.13(0.13) & -17.33$\pm$0.04 & 0.83$\pm$0.09 & 0.997$\pm$0.088 \\
\hline
Globule C 	& HCN 		& $J = 1 \rightarrow 0, F = 0 \rightarrow 1$ & 0.020(0.002) & -27.85$\pm$0.12 & 1.10$\pm$0.20 & 0.024$\pm$0.004 \\
			& 	 		& \hspace{18mm} $F = 2 \rightarrow 1$	& 0.075(0.002) & -27.94$\pm$0.03 & 1.21$\pm$0.08 & 0.094$\pm$0.005 \\
			& 	 		& \hspace{18mm} $F = 1 \rightarrow 1$	& 0.044(0.002) & -27.90$\pm$0.05 & 1.28$\pm$0.12 & 0.059$\pm$0.005 \\
			& HCN 		& $J = 3 \rightarrow 2$				& $<$0.3 & -27.70$\pm$0.23 & $\sim$0.23 &  - \\
			& HCO$^+$ 	& $J = 1 \rightarrow 0$				& 0.072(0.003) & -27.93$\pm$0.03 & 1.20$\pm$0.07 & 0.090$\pm$0.004 \\
			& HCO$^+$	& $J = 3 \rightarrow 2$				& 0.25 (0.07) & -28.00$\pm$0.09 & 0.93$\pm$0.18 & 0.247$\pm$0.045 \\
			& HNC 		& $J = 1 \rightarrow 0$				& 0.063(0.003) & -27.86$\pm$0.03 & 1.32$\pm$0.07 & 0.085$\pm$0.004 \\
			& HNC 		& $J = 3 \rightarrow 2$				& 0.24(0.08) & -27.72$\pm$0.09 & 0.58$\pm$0.18 & 0.145$\pm$0.044 \\
			& $^{13}$CO 	& $J = 1 \rightarrow 0$				& 0.18(0.03) & -27.90$\pm$0.07 & 0.84$\pm$0.17 & 0.165$\pm$0.026 \\
			& $^{12}$CO 	& $J = 1 \rightarrow 0$				& 0.60(0.07) & -28.00$\pm$0.07 & 1.11$\pm$0.17 & 0.712$\pm$0.088 \\
			& $^{12}$CO 	& $J = 2 \rightarrow 1$				& 2.09(0.09) & -27.91$\pm$0.01 & 0.79$\pm$0.03 & 1.747$\pm$0.064 \\
\hline		
Rim (Outer) 	& HCN 		& $J = 1 \rightarrow 0, F = 0 \rightarrow 1$ & 0.03(0.02) & -47.06$\pm$0.36 & 1.67$\pm$0.85 & 0.057$\pm$0.026 \\ 
			& 	 		& \hspace{18mm} $F = 2 \rightarrow 1$	& 0.12(0.02) & -47.16$\pm$0.07 & 1.84$\pm$0.35 & 0.239$\pm$0.032 \\
			& 	 		& \hspace{18mm} $F = 1 \rightarrow 1$	& 0.09(0.02) & -47.11$\pm$0.10 & 1.89$\pm$0.55 & 0.174$\pm$0.035 \\
			& HCN 		& $J = 3 \rightarrow 2$				& $<$0.14 & -47.30$\pm$0.12 & - & - \\
			& HCO$^+$ 	& $J = 1 \rightarrow 0$				& 0.16(0.01) & -47.65$\pm$0.06 & 3.25$\pm$0.17 & 0.555$\pm$0.023 \\
			& HCO$^+$	& $J = 3 \rightarrow 2$				& $<$0.05	 & - 				& - & - \\
			& HNC 		& $J = 1 \rightarrow 0$				& 0.13(0.02) & -47.37$\pm$0.09 & 2.18$\pm$0.23 & 0.301$\pm$0.025 \\
			& HNC 		& $J = 3 \rightarrow 2$				& $<$0.25 & -47.10$\pm$0.04 & - & - \\
\hline			
Rim (Inner) 	& HCN 		& $J = 1 \rightarrow 0, F = 0 \rightarrow 1$ & 0.04(0.02) & -10.26$\pm$0.24 & $\sim$0.66  & 0.031$\pm$0.021 \\ 
			& 	 		& \hspace{18mm} $F = 2 \rightarrow 1$	& 0.11(0.02) & -9.66$\pm$0.08 & 2.05$\pm$0.32 & 0.234$\pm$0.033 \\
			& 	 		& \hspace{18mm} $F = 1 \rightarrow 1$	& 0.06(0.02) & -9.73$\pm$0.16 & 2.04$\pm$0.64 & 0.125$\pm$0.034  \\
			& HCN 		& $J = 3 \rightarrow 2$				& $<$0.17 & -10.83$\pm$0.06 & - & - \\
			& HCO$^+$ 	& $J = 1 \rightarrow 0$				& 0.17(0.01) & -9.68$\pm$0.05 & 2.44$\pm$0.12 & 0.453$\pm$0.019 \\
			& HCO$^+$	& $J = 3 \rightarrow 2$				& $<$0.040 & - & - & - \\
			& HNC 		& $J = 1 \rightarrow 0$				& 0.13(0.02) & -9.67$\pm$0.08 & 1.70$\pm$0.17 & 0.230$\pm$0.021 \\
			& HNC 		& $J = 3 \rightarrow 2$				& $<$0.11 & -9.30$\pm$0.18 & - & - \\
\hline			
East$^*$ 		& HCN 		& $J = 1 \rightarrow 0$ & 0.05(0.01) & -30.41$\pm$0.12 & 1.88$\pm$0.46 & 0.101$\pm$0.016 \\
			& HCN 		& $J = 3 \rightarrow 2$ & $<$0.10	& - 				& - 			& - \\
			& HCO$^+$ 	& $J = 1 \rightarrow 0$ & 0.11(0.01) & -29.53$\pm$0.05 & 1.50$\pm$0.14 & 0.172$\pm$0.019 \\
			& HCO$^+$	& $J = 3 \rightarrow 2$ & 	0.09 (0.05) & -30.81$\pm$0.38 & 2.17$\pm$0.63 & 0.196$\pm$0.061 \\
			& HNC 		& $J = 1 \rightarrow 0$ & 0.11(0.01) & -29.53$\pm$0.05 & 1.50$\pm$0.14 & 0.110$\pm$0.018 \\
			& HNC 		& $J = 3 \rightarrow 2$ & $<$0.14 & - 				& 				- & - \\
\hline			
West$^*$ 		& HCN 		& $J = 1 \rightarrow 0$ & 0.03(0.01) & -21.76$\pm$0.66 & 5.67$\pm$0.66 & 0.177$\pm$0.014 \\
			& HCN 		& $J = 3 \rightarrow 2$ & $<$0.27	& - & - & - \\
			& HCO$^+$ 	& $J = 1 \rightarrow 0$ & 0.11(0.01) & -20.69$\pm$0.16 & 4.19$\pm$0.33 & 0.473$\pm$0.023  \\
			& HCO$^+$	& $J = 3 \rightarrow 2$ & 0.08$^\dagger$ & -21.46$\pm$0.27 & 1.24$\pm$0.64 & 0.111$\pm$0.049 \\
			& HNC 		& $J = 1 \rightarrow 0$ & 0.06(0.01) & -21.23$\pm$0.30 & 2.77$\pm$0.69 & 0.166$\pm$0.088 \\
			& HNC 		& $J = 3 \rightarrow 2$ & $<$0.11 & - & - & - \\

\multicolumn{7}{l}{
\tablefoot{$^\dagger$: Uncertain \\
	$^*$: Only central line peaks for Helix positions East and West presented due to the significant number of overlaid, but resolvable velocity profiles detected \\
	Peak line intensities are in units of the main beam temperature (T$_{mb}$). Velocities of the lines and their FWHM line width ($\Delta V_{1/2}$) are with respect to the LSR. Integrated line intensity ($\int T_{mb}dV$) is computed from the intensity over the width of the line. The 1$\sigma$ formal errors are included for each measurement.
}}
\end{longtable}
}



%
\longtab{
\begin{longtable}{lllcccc}
\caption{Additional Helix Spectral Lines \label{table_bonuslines}}\\
\hline\hline
Position & Molecule & Transition & T$_{mb}$ & $V_{LSR}$ & $\Delta V_{1/2}$ & $\int T_{mb} dV$ \\ 
 &  &  & (K) & (km s$^{-1}$) & (km s$^{-1}$) & (K km s$^{-1}$) \\
\hline
\endfirsthead
\caption{continued.}\\
\hline\hline
Position & Molecule & Transition & T$_{mb}$ & $V_{LSR}$ & $\Delta V_{1/2}$ & $\int T_{mb} dV$ \\ 
 &  &  & (K) & (km s$^{-1}$) & (km s$^{-1}$) & (K km s$^{-1}$) \\
\hline
\endhead
\hline
\endfoot
Globule A		& C$_2$H		& $N = 1 \rightarrow 0$ &			& & & \\ 
			&			& $J = 3/2 \rightarrow 1/2, F = 1 \rightarrow 0$ & 0.016 (0.003) & -22.22$\pm$0.26 & 1.73$\pm$0.38 & 0.030$\pm$0.007 \\
			& 			& $J = 1/2 \rightarrow 1/2, F = 1 \rightarrow 1$ & 0.011 (0.004) & -22.54$\pm$0.21 &1.15$\pm$0.55 & 0.013$\pm$0.005 \\
			& CN		& $N = 1 \rightarrow 0, J= 1/2 \rightarrow 1/2$ & 0.23 (0.05) & -22.52$\pm$0.12 & 1.03$\pm$0.36 & 0.25$\pm$0.06 \\
			& CN		& $N = 1 \rightarrow 0, J= 3/2 \rightarrow 1/2$ & 0.25 (0.05) & -22.24$\pm$0.22 & 1.97$\pm$0.44 & 0.52$\pm$0.11 \\
\hline
Globule B		& C$_2$H		& $N = 1 \rightarrow 0$ &			& & & \\ 
			&			& $J = 3/2 \rightarrow 1/2, F = 1 \rightarrow 0$ & 0.020 (0.006) & -16.91$\pm$0.23 & 1.32$\pm$0.35 & 0.03$\pm$0.008 \\
			& 			& $J = 1/2 \rightarrow 1/2, F = 1 \rightarrow 1$ & 0.018 (0.006) & -16.91$\pm$0.16 & 0.67$\pm$0.36 & 0.013$\pm$0.007 \\
			& CN		& $N = 1 \rightarrow 0, J= 1/2 \rightarrow 1/2$ & 0.11 (0.06)& -17.11$\pm$0.61& 3.31$\pm$1.19 & 0.38$\pm$0.11 \\
			& CN		& $N = 1 \rightarrow 0, J= 3/2 \rightarrow 1/2$ & 0.15 (0.06)$^\dagger$ & -17.34$\pm$0.26 & 0.90$\pm$0.57 & 0.14$\pm$0.07$^\dagger$ \\
\hline
Globule C		& C$_2$H		& $N = 1 \rightarrow 0$ &			& & & \\ 
			&			& $J = 3/2 \rightarrow 1/2, F = 1 \rightarrow 0$ & 0.020 (0.003)$^\dagger$ & -27.71$\pm$0.11 & 0.87$\pm$0.16 & 0.02$\pm$0.003$^\dagger$ \\
			& 			& $J = 1/2 \rightarrow 1/2, F = 1 \rightarrow 1$ & 0.012 (0.003) & -27.66$\pm$0.14 & 1.28$\pm$0.30 & 0.02$\pm$0.004 \\
			& CN		& $N = 1 \rightarrow 0, J= 1/2 \rightarrow 1/2$ & 0.112 (0.045) & -28.16$\pm$0.37 & 1.90$\pm$0.64 & 0.23$\pm$0.07 \\
			& CN		& $N = 1 \rightarrow 0, J= 3/2 \rightarrow 1/2$ & 0.20 (0.05) & -27.88$\pm$0.13& 1.01$\pm$0.32 & 0.22$\pm$0.06 \\
\hline
Rim (Outer)	& C$_2$H		& $N = 1 \rightarrow 0$ &			& & & \\ 
			&			& $J = 3/2 \rightarrow 1/2, F = 1 \rightarrow 0$ & 0.046 (0.008) & -47.87$\pm$0.26 & 3.18$\pm$0.52 & 0.15$\pm$0.02 \\
			& 			& $J = 1/2 \rightarrow 1/2, F = 1 \rightarrow 1$ & 0.020 (0.009) & -47.82$\pm$0.58 & 2.19$\pm$1.14 & 0.05$\pm$0.02 \\
\hline
Rim (Inner)	& C$_2$H		& $N = 1 \rightarrow 0$ &			& & & \\ 
			&			& $J = 3/2 \rightarrow 1/2, F = 1 \rightarrow 0$ & 0.013 (0.009) & -11.20$\pm$2.39 & 15.67$\pm$4.99 & 0.21$\pm$0.06 \\
			& 			& $J = 1/2 \rightarrow 1/2, F = 1 \rightarrow 1$ &0.046 (0.008) & -7.71$\pm$0.25 & 3.23$\pm$0.50 & 0.16$\pm$0.02 \\
\hline
East			& C$_2$H		& $N = 1 \rightarrow 0$ &			& & & \\ 
			&			& $J = 3/2 \rightarrow 1/2, F = 1 \rightarrow 0$ & $<$0.016 & - & - & - \\
			& 			& $J = 1/2 \rightarrow 1/2, F = 1 \rightarrow 1$ & $<$0.009 & - & - & - \\
\hline
West			& C$_2$H		& $N = 1 \rightarrow 0$ &			& & & \\ 
			&			& $J = 3/2 \rightarrow 1/2, F = 1 \rightarrow 0$ & $<$0.01	 & - & - & - \\
			& 			& $J = 1/2 \rightarrow 1/2, F = 1 \rightarrow 1$ & $<$0.02	 & - & - & - \\

\multicolumn{7}{l}{
\tablefoot{$^\dagger$: Uncertain \\
	$^*$: Only central line peaks for Helix positions East and West presented due to the significant number of overlaid, but resolvable velocity profiles detected \\
	Line properties identified in the same manner as those in Table \ref{table_longtable}.
}}
\end{longtable}
}


\begin{acknowledgements}
J.B. wishes to acknowledge useful discussion with E. van Dischoeck.
This publication is based on observations carried out under project number 109-17 with the IRAM 30m telescope. IRAM is supported by INSU/CNRS (France), MPG (Germany) and IGN (Spain). 
This work is also based on data acquired with the Atacama Pathfinder Experiment (APEX) under program ID 0103.D-0682. APEX is a collaboration between the Max-Planck-Institut fur Radioastronomie, the European Southern Observatory, and the Onsala Space Observatory. 
The work of T.F. is supported in part by the French National Research Agency in the framework of the Investissements d'Avenir program (ANR-15-IDEX-02), through the funding of the ``Origin of Life'' project of the Universit\'{e} Grenoble-Alpes.
M.S.G., J.A., and V.B. acknowledge support of the AxiN and EVENTs / NEBULAE WEB research programs supported by the Spanish MICINN (grants AYA2016-78994-P and PID2019-105203GB-C21).
The authors also acknowledge IPAG and the support of the Chateaubriand Fellowship of the Office for Science \& Technology of the Embassy of France in the United States, by which much of this work was funded.
\end{acknowledgements}

\bibliographystyle{aa-package/bibtex/aa}
\bibliography{Helix_References.bib}

\begin{appendix}
\section{}

The following tables comprise a detailed list of the properties of all emission line features obtained from this study, as well as spectral line figures that are not substantially addressed in this work.



	\begin{figure}[!h]
		\includegraphics[width=0.32\linewidth, trim={2cm 1.25cm 3.5cm 11cm}, clip]{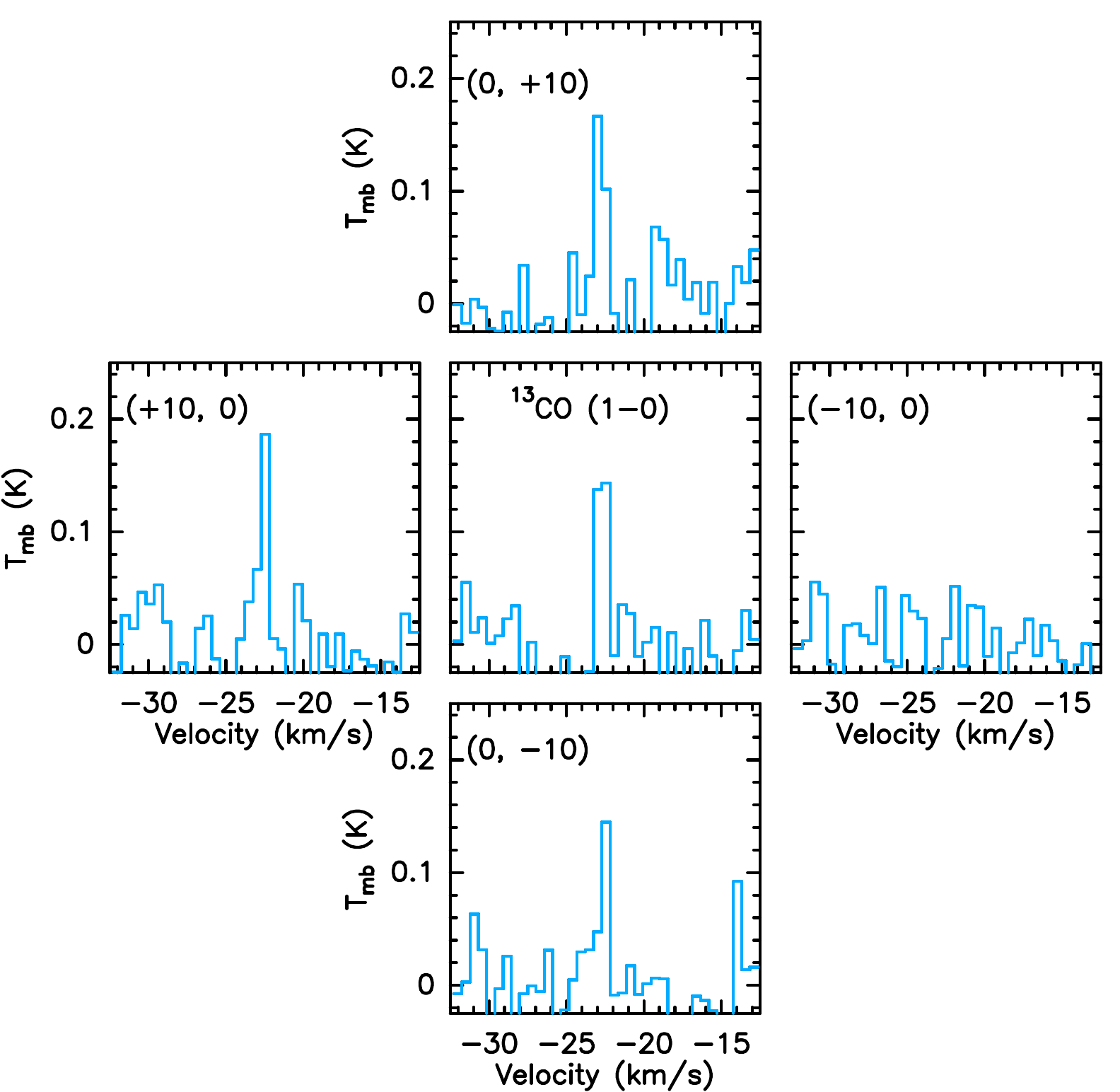}
		\includegraphics[width=0.32\linewidth, trim={2cm 1.25cm 3.5cm 11cm}, clip]{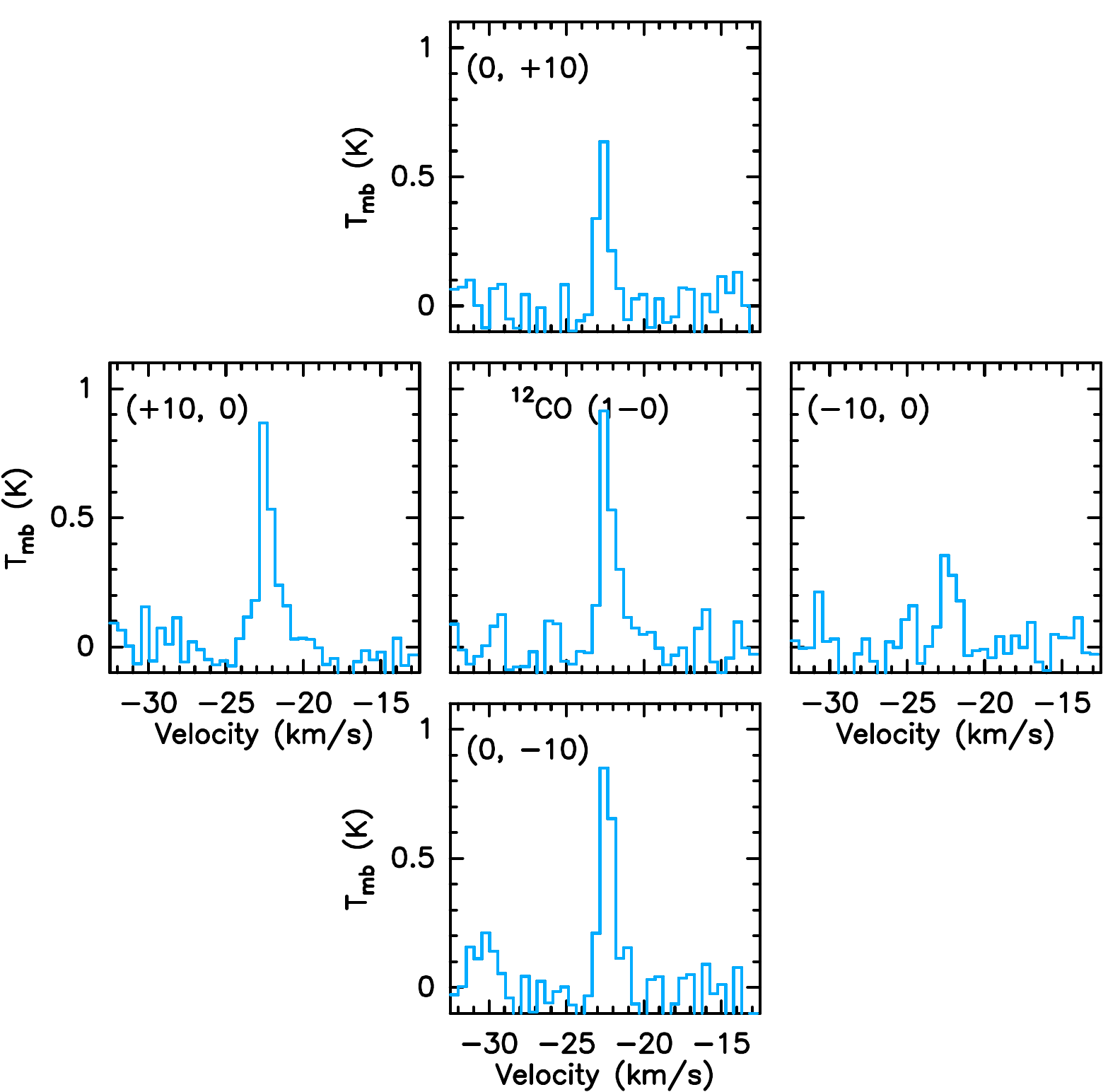}
		\includegraphics[width=0.32\linewidth, trim={2cm 1.25cm 3.5cm 11cm}, clip]{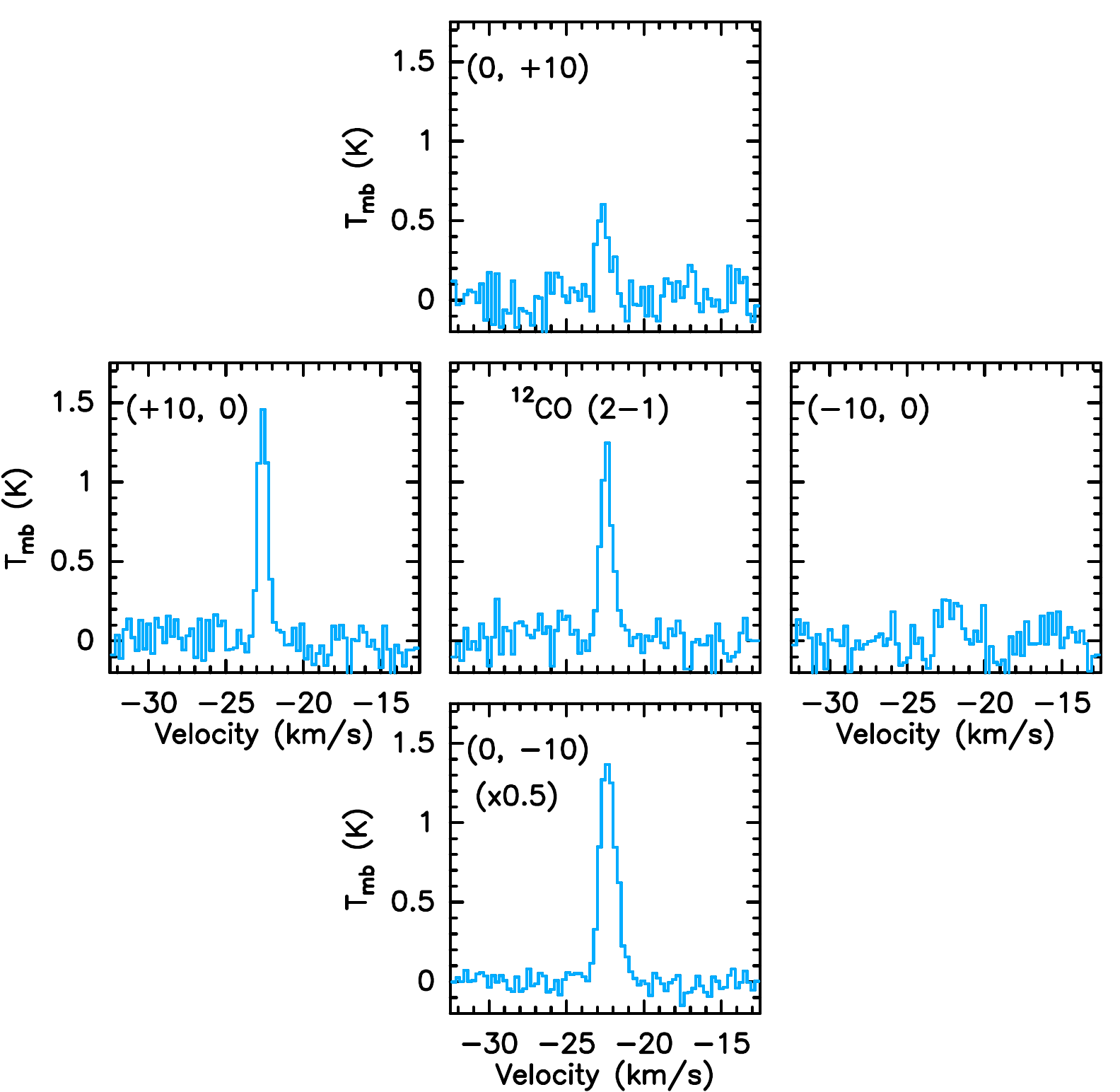}
		\caption{
			Spectra of Globule A in the $J=1 \rightarrow 0$ lines of $^{13}$CO (left) and $^{12}$CO (center), as well as the $J=2 \rightarrow 1$ line of $^{12}$CO (right). Offset positions from the central pointing are indicated within each spectral window by their R.A. and dec. The spacing of 10$''$ between pointings is roughly half the 30m beam at half-power at 110 GHz, and 1 beamwidth at 230 GHz. The CSPN is to the NE.
			}
		\label{Glob_5ptA}
	\end{figure}

	\begin{figure}[!h]
		\includegraphics[width=0.32\linewidth, trim={2cm 1.25cm 3.5cm 11cm}, clip]{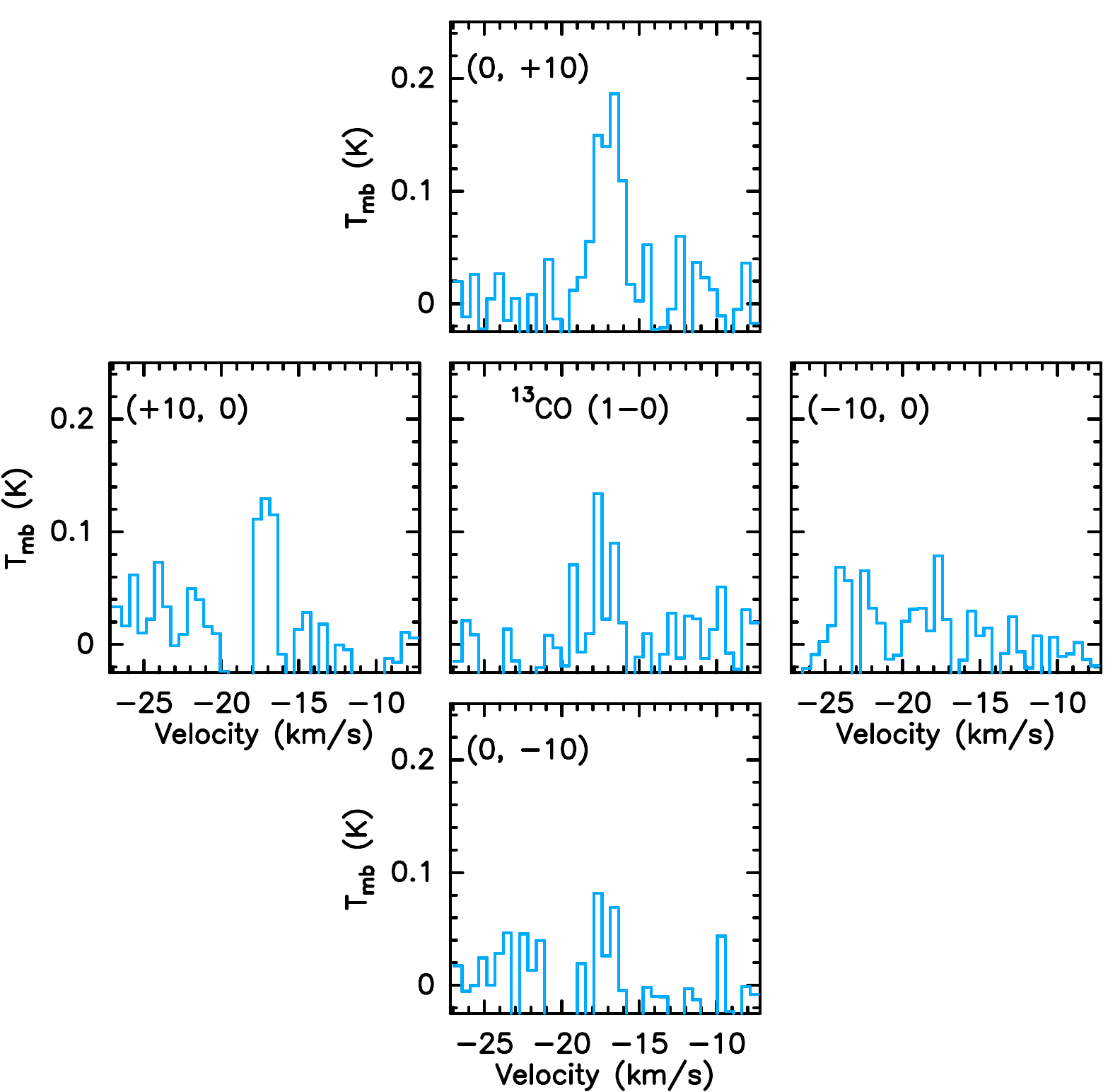}
		\includegraphics[width=0.32\linewidth, trim={2cm 1.25cm 3.5cm 11cm}, clip]{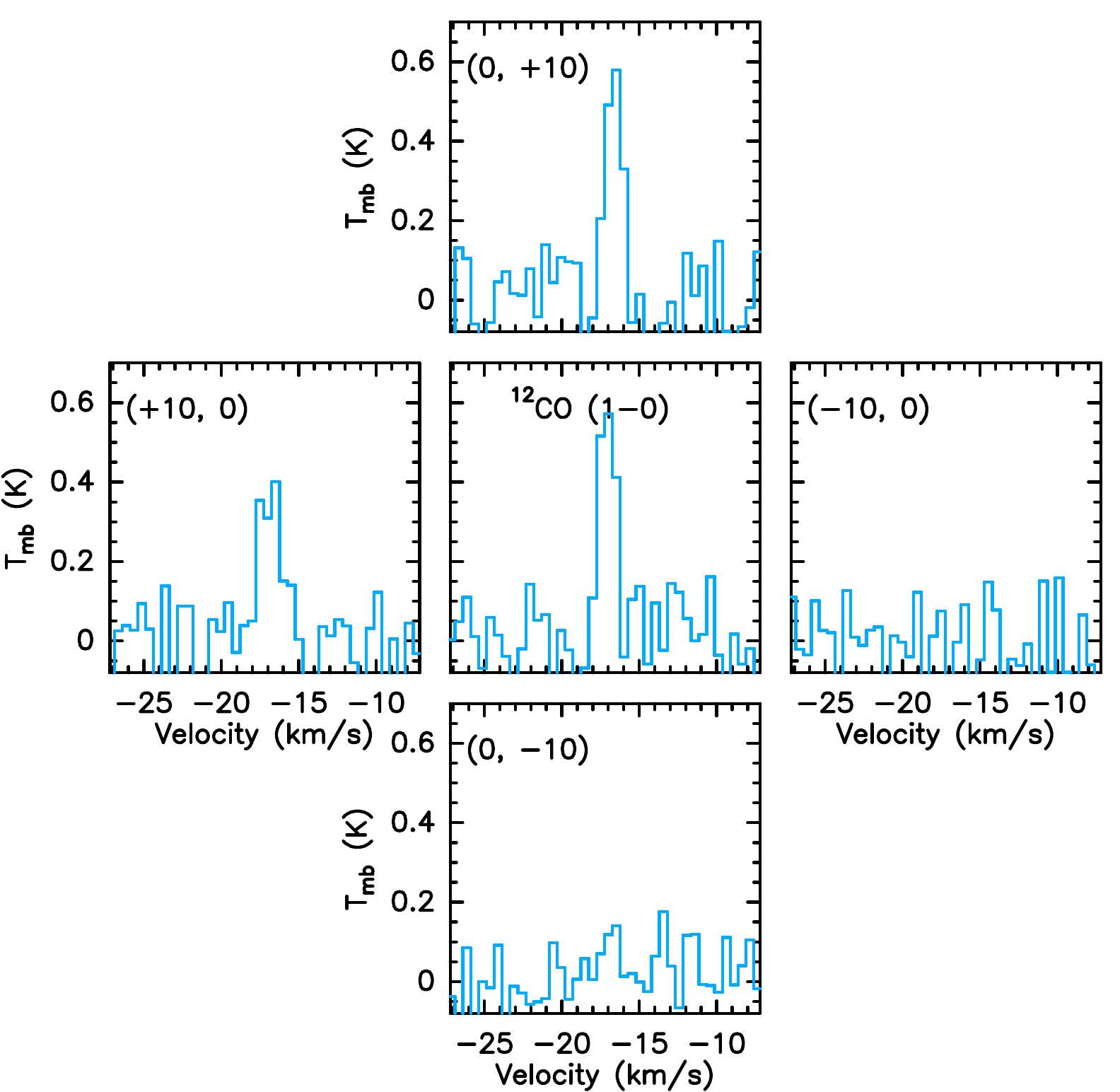}
		\includegraphics[width=0.32\linewidth, trim={2cm 1.25cm 3.5cm 11cm}, clip]{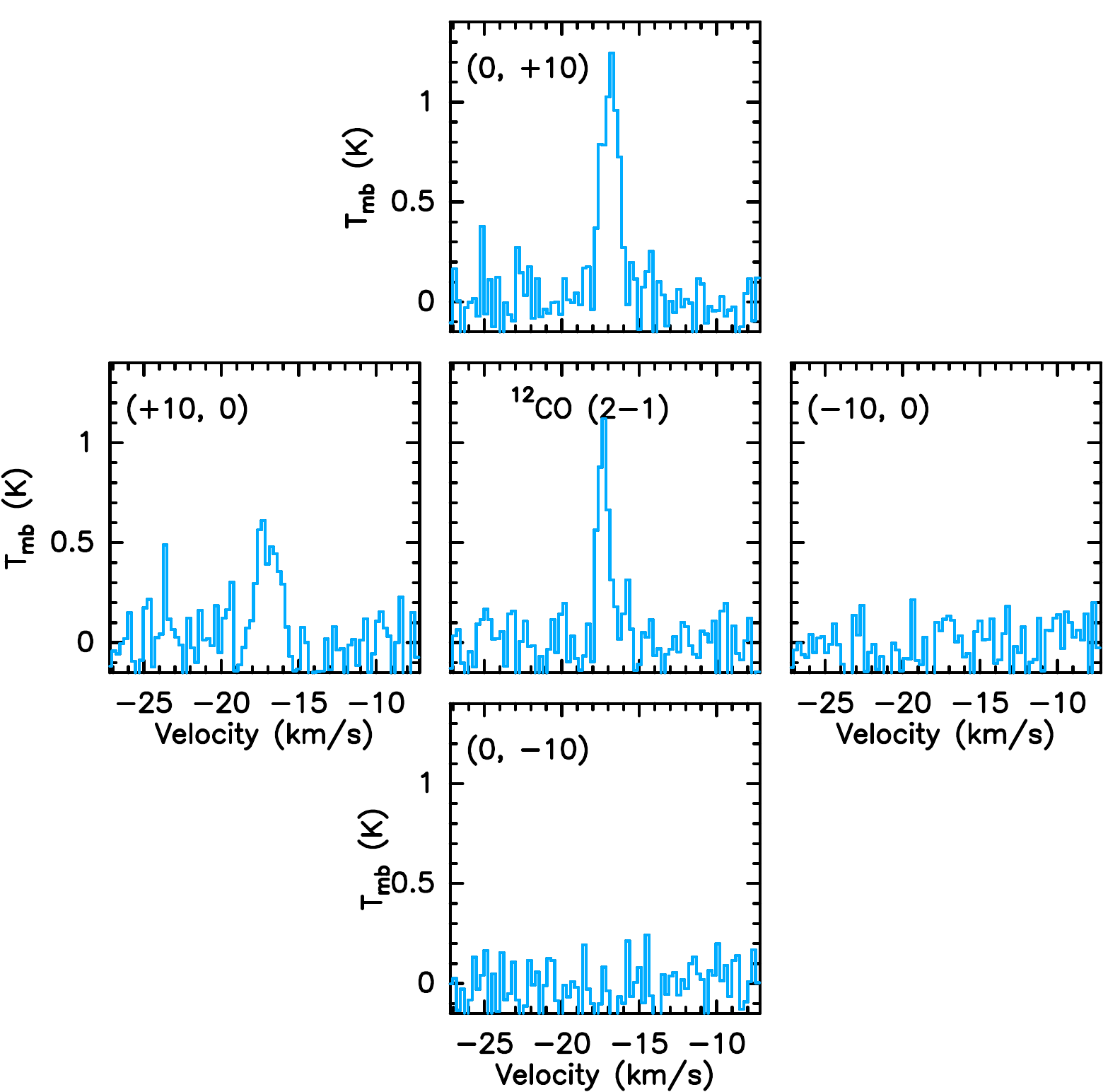}
		\caption{
			Spectra of Globule B in the $J=1 \rightarrow 0$ lines of $^{13}$CO (left) and $^{12}$CO (center), as well as the $J=2 \rightarrow 1$ line of $^{12}$CO (right). Offset positions are indicated as in Figure \ref{Glob_5ptA}. The CSPN is to the SW, coinciding with the pointings that display decreased CO emission.}
		\label{Glob_5ptB}
	\end{figure}

	\begin{figure}[!h]
		\includegraphics[width=0.32\linewidth, trim={2cm 1.25cm 3.5cm 11cm}, clip]{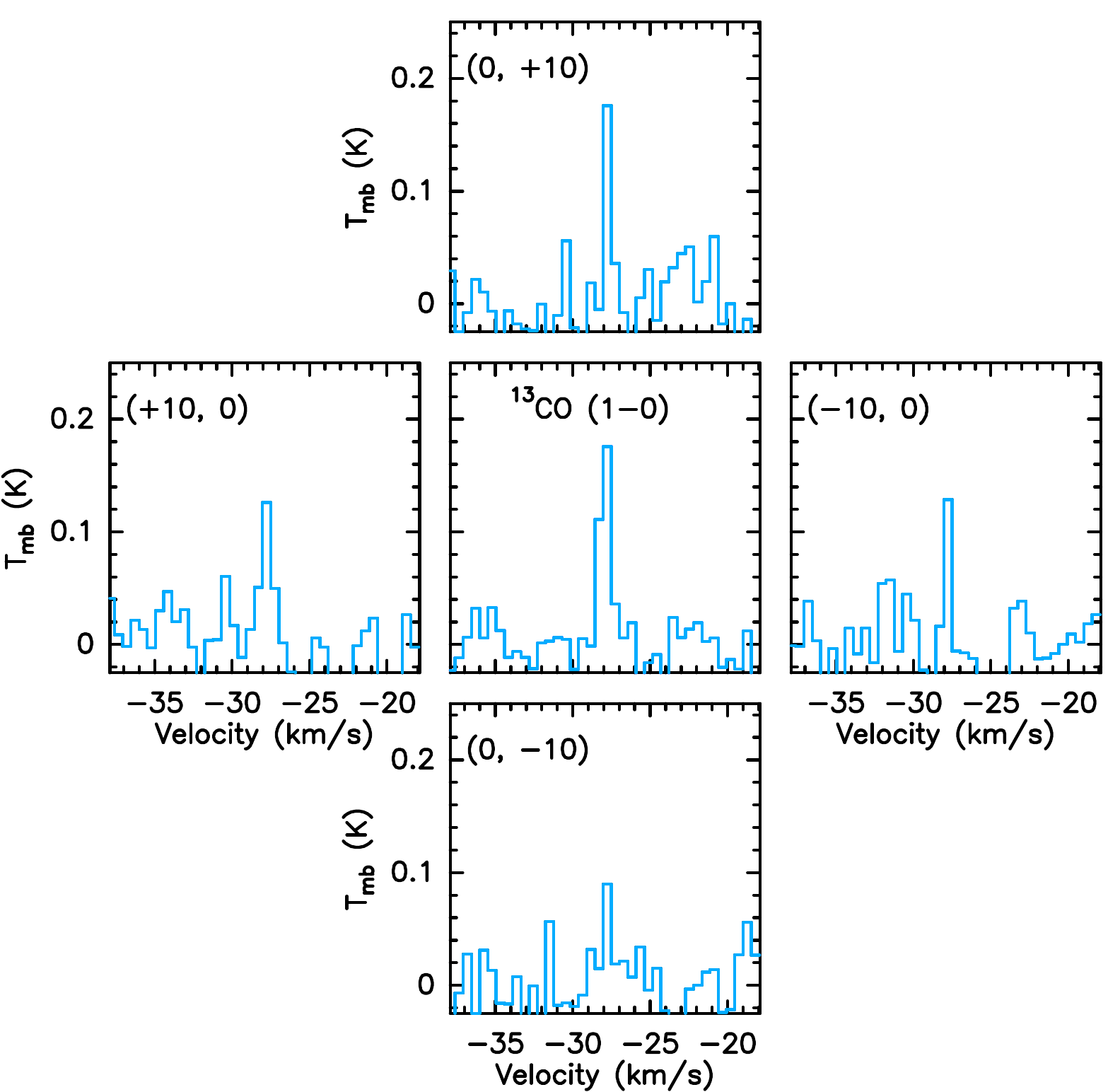}
		\includegraphics[width=0.32\linewidth, trim={2cm 1.25cm 3.5cm 11cm}, clip]{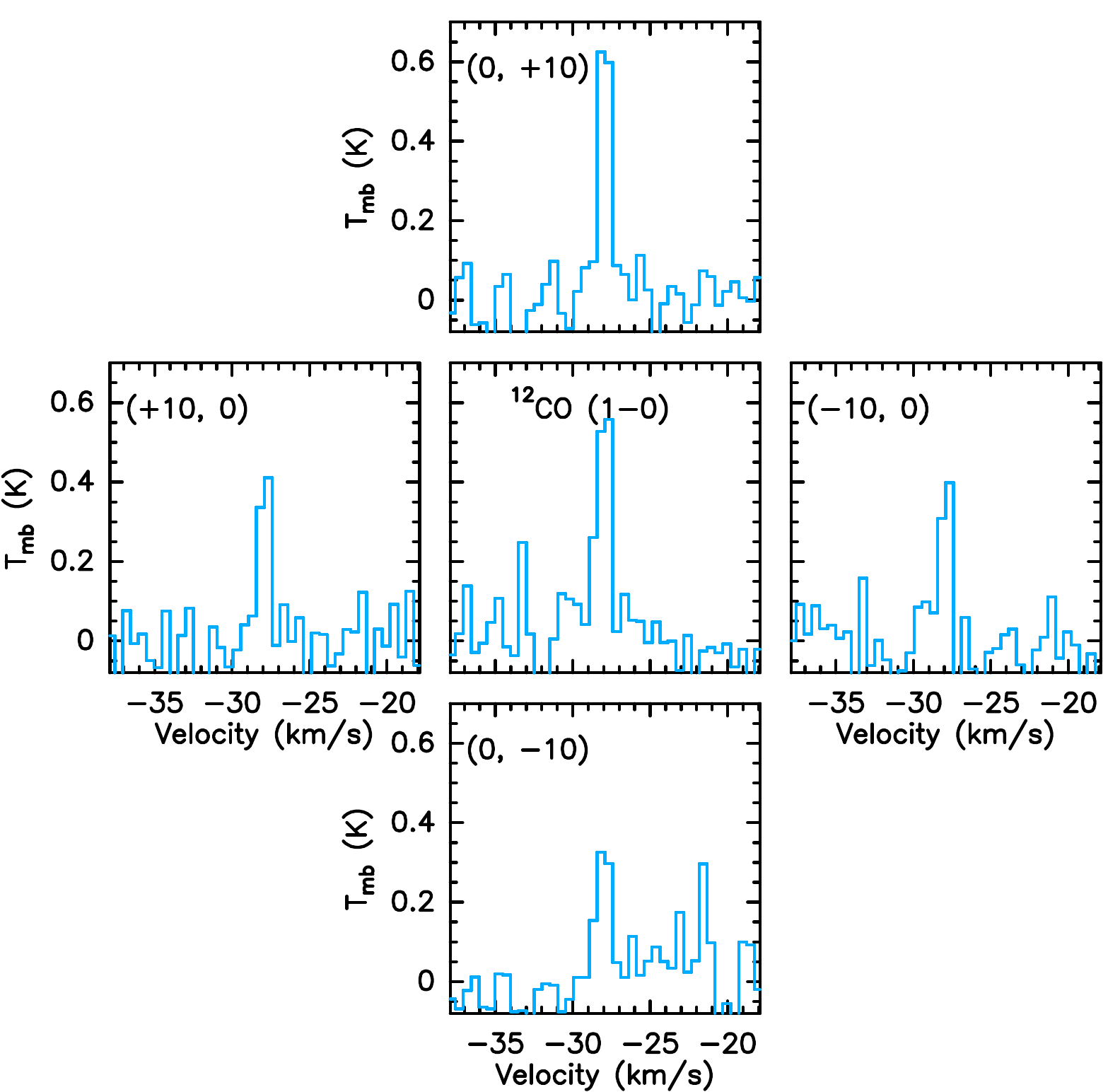}
		\includegraphics[width=0.32\linewidth, trim={2cm 1.25cm 3.5cm 11cm}, clip]{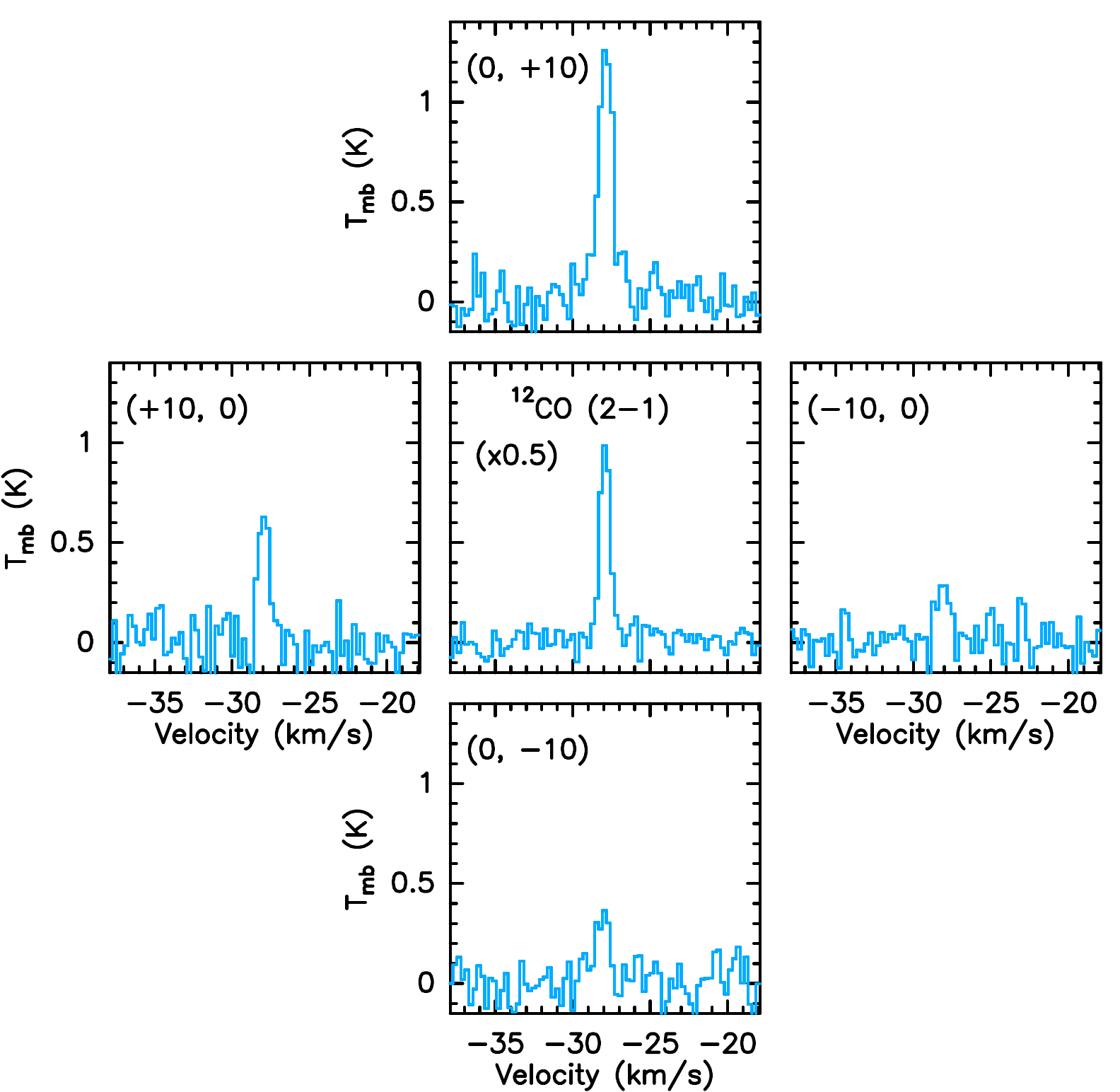}
		\caption{
			Spectra of Globule C in the $J=1 \rightarrow 0$ lines of $^{13}$CO (left) and $^{12}$CO (center), as well as the $J=2 \rightarrow 1$ line of $^{12}$CO (right). Offset positions are indicated as in Figure \ref{Glob_5ptA}. The CSPN is to the south, coinciding with decreased CO emission at (0, --10).}
		\label{Glob_5ptC}
	\end{figure}

	\begin{figure*}
		\includegraphics[width=0.49\linewidth, trim={2.4cm 5cm 3.5cm 4cm}, clip]{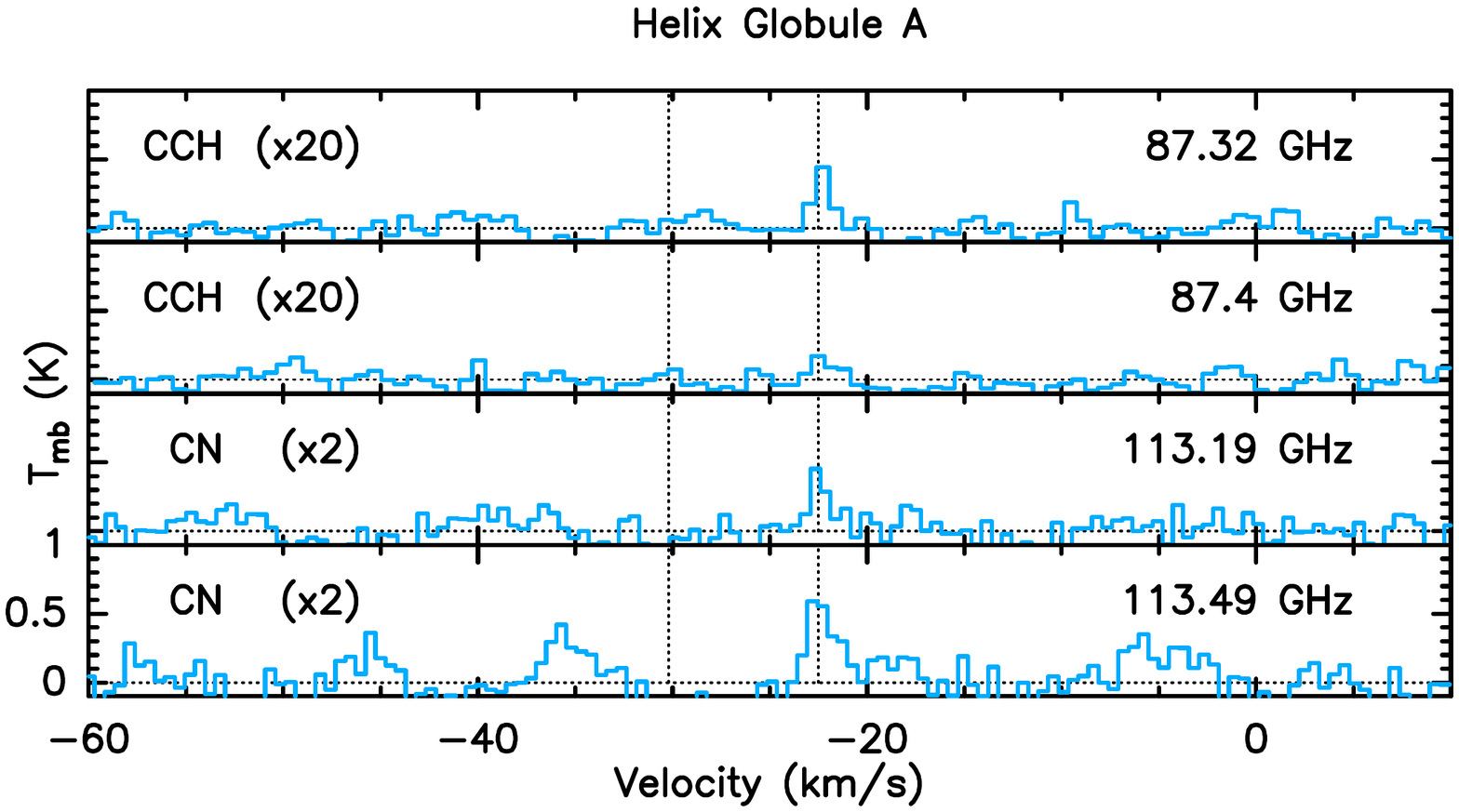}
		\includegraphics[width=0.49\linewidth, trim={2.4cm 5cm 3.5cm 4cm}, clip]{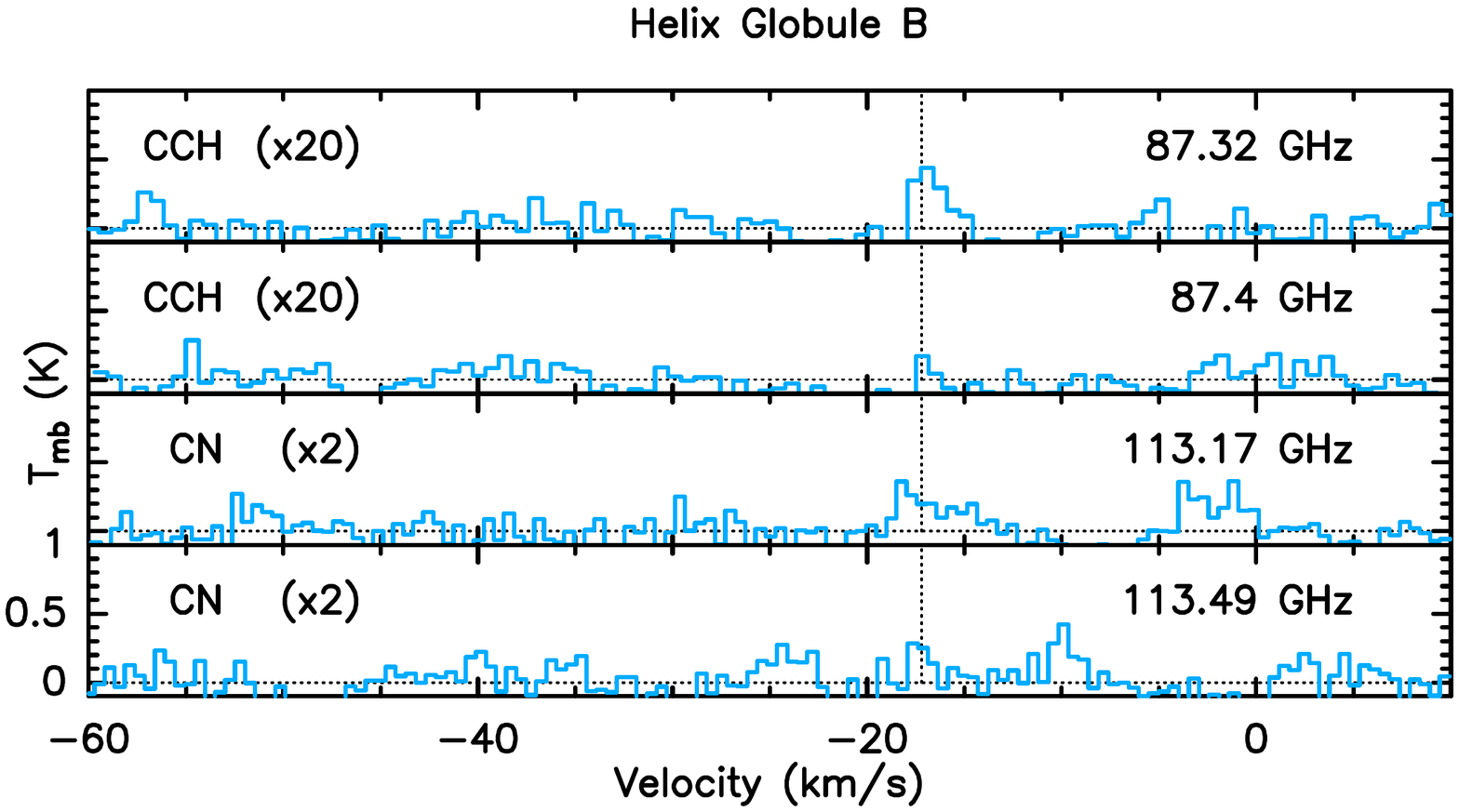} \\
		\includegraphics[width=0.49\linewidth, trim={2.4cm 5cm 3.5cm 4cm}, clip]{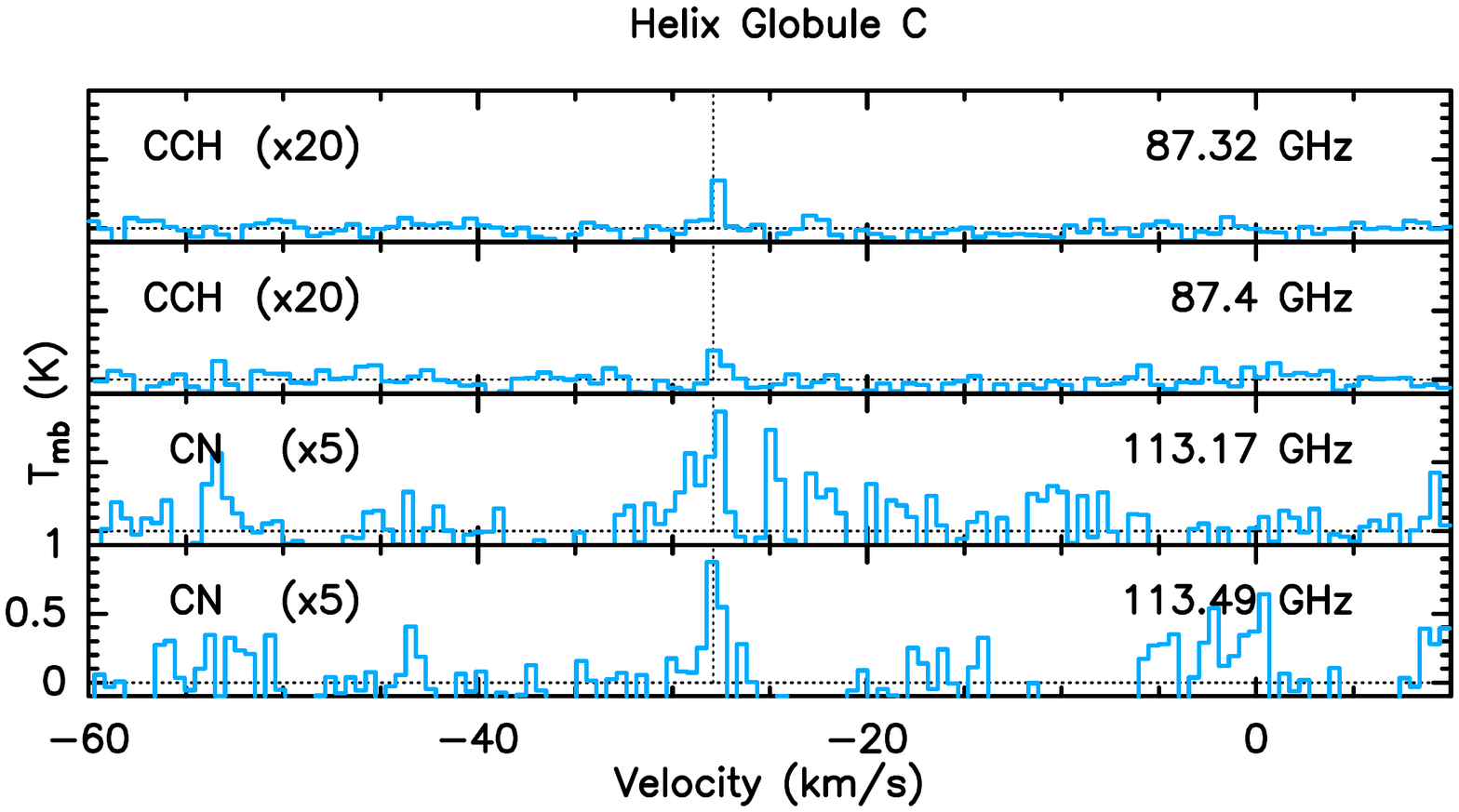}
		\includegraphics[width=0.49\linewidth, trim={2.4cm 7.5cm 3.5cm 7cm}, clip]{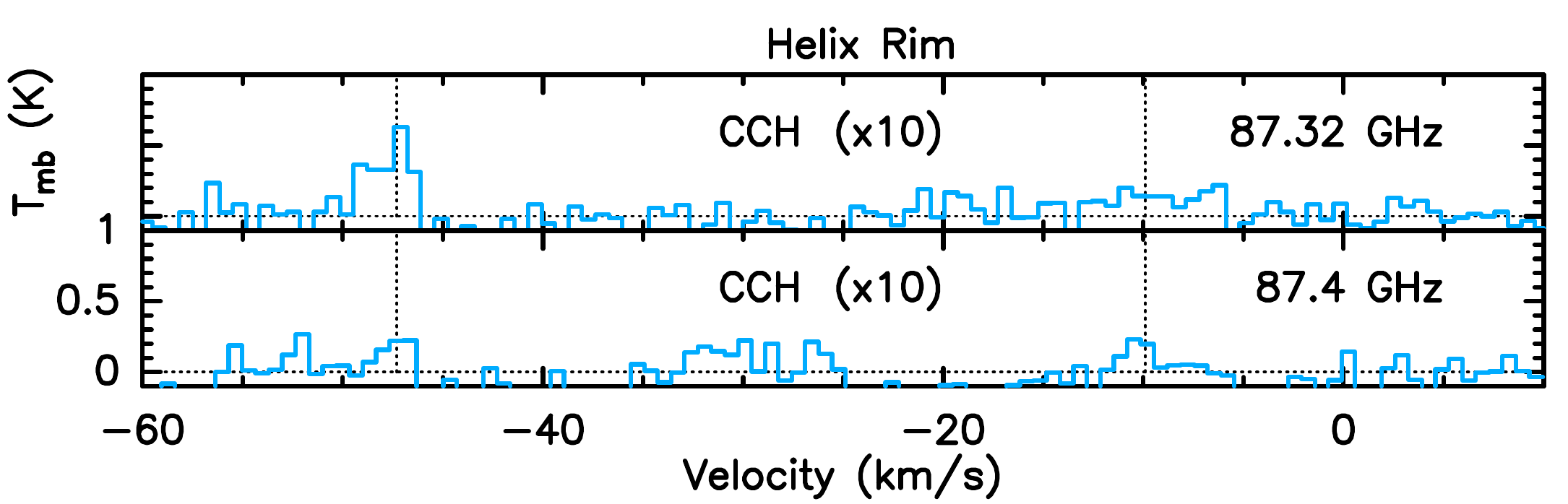} \\
		\includegraphics[width=0.49\linewidth, trim={2.4cm 7.5cm 3.5cm 7cm}, clip]{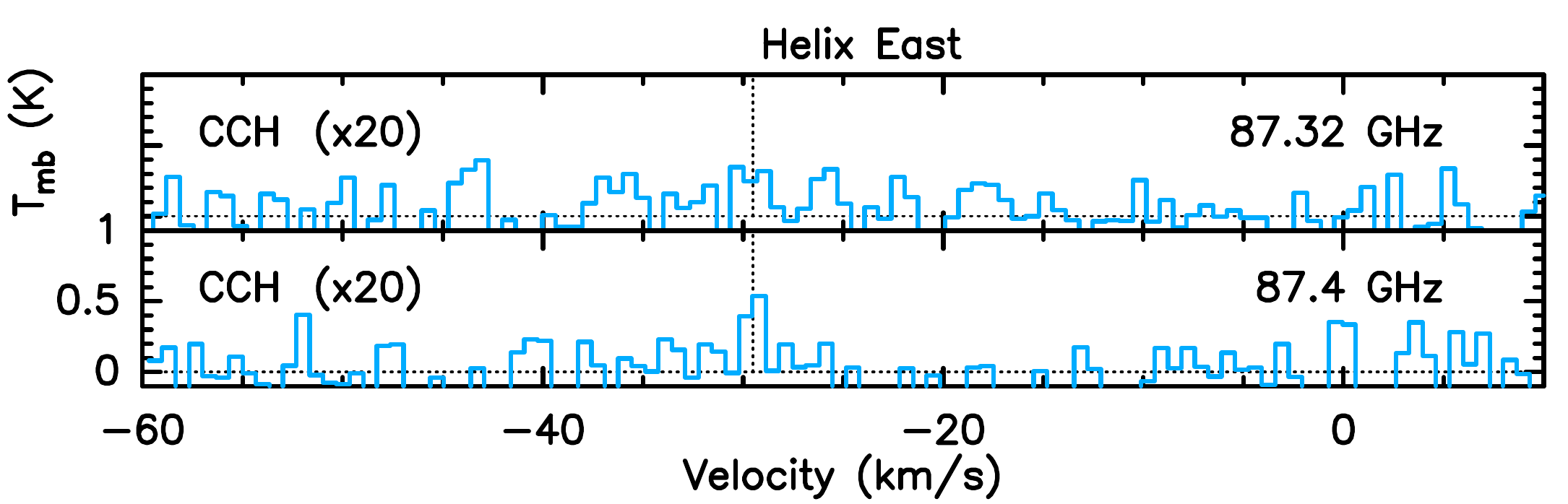}
		\includegraphics[width=0.49\linewidth, trim={2.4cm 7.5cm 3.5cm 7cm}, clip]{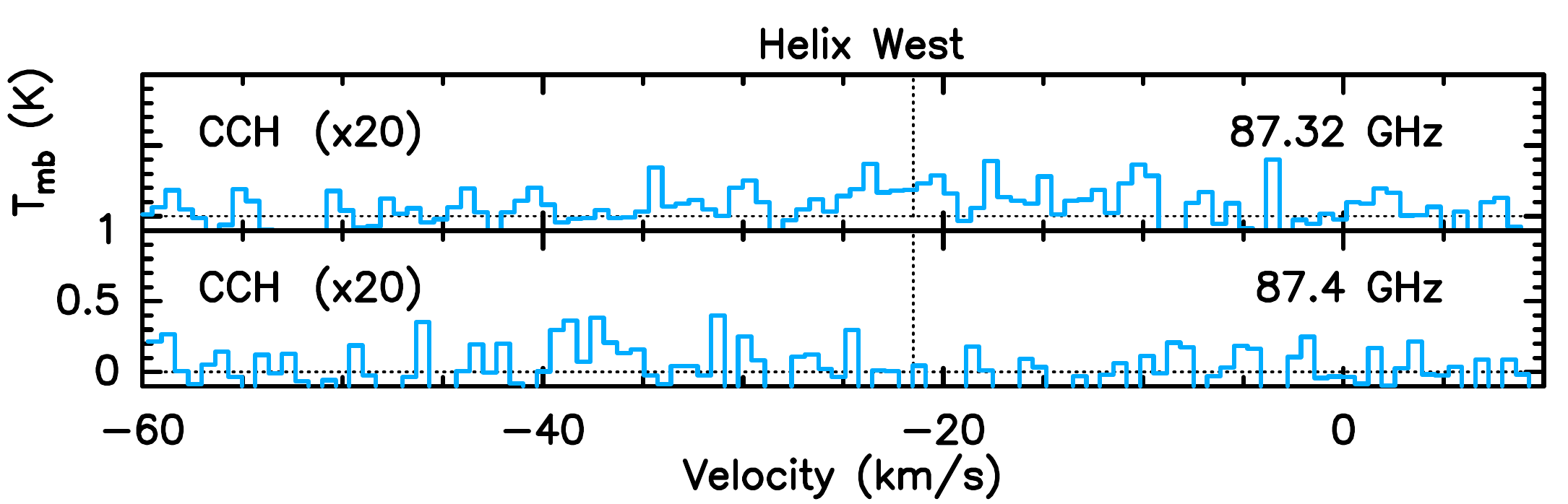}
		\caption{Additional spectral lines observed in each of the six targeted positions of the Helix Nebula.}
		\label{BonusSpectra}
	\end{figure*}

\end{appendix}
\end{document}